\documentclass[11pt]{article}
\usepackage[T1]{fontenc}
\usepackage{lmodern}
\usepackage[margin=1in]{geometry}
\usepackage{amsmath,amssymb,amsthm,mathtools}
\usepackage{microtype,enumitem,booktabs,array}
\usepackage[numbers,sort&compress]{natbib}
\usepackage{xcolor}
\usepackage{tikz}
\usetikzlibrary{arrows.meta,positioning,calc}
\usepackage[colorlinks=true,linkcolor=blue!45!black,citecolor=blue!45!black,urlcolor=blue!45!black]{hyperref}
\hypersetup{pdftitle={Graph Isomorphism via Inherited Decompositions},pdfauthor={Mohammad T. Hajiaghayi and Mahdi JafariRaviz}}
\setlist{topsep=4pt,itemsep=3pt,leftmargin=1.65em}
\newtheorem{theorem}{Theorem}[section]
\newtheorem{lemma}[theorem]{Lemma}
\newtheorem{proposition}[theorem]{Proposition}
\newtheorem{corollary}[theorem]{Corollary}
\theoremstyle{definition}
\newtheorem{definition}[theorem]{Definition}

\theoremstyle{remark}
\newtheorem{remark}[theorem]{Remark}
\DeclareMathOperator{\Alt}{Alt}
\DeclareMathOperator{\Sym}{Sym}
\DeclareMathOperator{\Aut}{Aut}
\DeclareMathOperator{\Iso}{Iso}
\DeclareMathOperator{\GL}{GL}

\DeclareMathOperator{\Soc}{Soc}

\DeclareMathOperator{\Out}{Out}
\DeclareMathOperator{\sgn}{sgn}

\DeclareMathOperator{\polylog}{polylog}

\newcommand{\F}{\mathbb F}

\newcommand{\R}{\mathbb R}
\newcommand{\one}{\mathbf 1}

\newcommand{\restr}[2]{#1\big|_{#2}}
\newcommand{\poly}{\operatorname{poly}}
\newcommand{\kcert}{\kappa}
\newcommand{\Solve}{\mbox{\normalfont\scshape Isomorphism}}
\newcommand{\Continue}{\mbox{\normalfont\scshape Continue}}

\newcommand{\Gd}{\widehat\Gamma_d}
\newcommand{\proc}[1]{\mbox{\normalfont\scshape #1}}
\newcounter{algorithm}
\newcounter{algline}

\newcommand{\pc}{\stepcounter{algline}\makebox[1.4em][r]{\scriptsize\arabic{algline}}\>}
\newenvironment{pseudocode}[2]{%
  \begin{figure}[htbp]\centering
  \begin{minipage}{\linewidth}\small
  \refstepcounter{algorithm}\label{#2}%
  \hrule\medskip
  \noindent\textbf{Algorithm \thealgorithm. #1}\par\smallskip
  \setcounter{algline}{0}%
}{\smallskip\hrule\end{minipage}\end{figure}}

\title{Deterministic $n^{O(\log n)}$ Graph Isomorphism Algorithm \\ via Inherited Decompositions}
\author{MohammadTaghi Hajiaghayi\thanks{University of Maryland. Email: \texttt{hajiaghayi@gmail.com}.}
\and Mahdi JafariRaviz\thanks{University of Maryland. Email: \texttt{mahdijafariraviz@gmail.com}.}}
\date{}

\begin{document}
\maketitle

\begin{abstract}
We give a deterministic $n^{O(\log d)}$-time algorithm for
isomorphism of graphs with $n$ vertices and maximum degree $d$.
Since $d<n$, this is an $n^{O(\log n)}$-time algorithm for graph
isomorphism. To our knowledge, this is the first proof of such a
worst-case bound for general graphs. It improves the explicit cubic
logarithmic exponent in the analysis of the corrected Luks--Babai
algorithm. For graphs of maximum degree
$d$, the best previous bound known to us, due to Grohe, Neuen, and
Schweitzer, is $n^{O((\log d)^c)}$ for an absolute constant $c$.
The algorithm uses the
local-certificate and color-refinement steps of Babai's
quasipolynomial algorithm, but retains the decomposition of each
recursive instance when it solves an auxiliary coloring problem.
There are two reasons for retaining this information. First, a local
test can make progress without removing many vertices: it can
eliminate a layer of permutation symmetry. A logarithmic bound on
the total weight of these layers allows tests of constant size.
Second, the auxiliary bipartite graphs arising in the symmetry-breaking
step can be represented inside the current instance, after retaining
a row decomposition when necessary. These recursive calls inherit
the progress already made. Two further observations give the
dependence on the degree. The recursive calls made by a local test
check disjoint parts of the instance, so together they cost about
as much as one call on the whole instance. And for graphs of maximum
degree $d$, each large symmetric action that the algorithm must
break acts on at most $d$ points. We present the
algorithm through colorings, partitions, and bipartite graphs, and
collect the permutation-group ingredients in one appendix. The result
extends to string isomorphism and returns one isomorphism and
generators for the full automorphism group.
\end{abstract}

\section{Introduction}
\label{sec:introduction}

Graph isomorphism asks whether two graphs differ only in the names of
their vertices. Given graphs $X$ and $Y$ on $n$ vertices, we seek a
bijection $\pi\colon V(X)\to V(Y)$ such that
\begin{equation*}
 uv\in E(X)\quad\Longleftrightarrow\quad
 \pi(u)\pi(v)\in E(Y).
\end{equation*}
The definition is elementary, but the worst-case complexity of finding
such a bijection has resisted a general polynomial-time algorithm.
The difficulty is not usually a lack of local information. Degree
sequences, neighborhood colors, and repeated refinement often identify
many vertices. The difficult instances are those in which large sets of
vertices remain indistinguishable under the available information.

Luks's algorithm for bounded-degree graphs introduced a way to describe
and restrict the remaining vertex permutations without listing them
individually~\cite{Luks82}. For maximum degree $d$ it runs in time
$n^{O(d)}$, and Babai, Kantor, and Luks improved the exponent to
$O(d/\log d)$~\cite{BKL83}. Babai's quasipolynomial algorithm for
general graphs combines Luks's approach with local certificates and a
combinatorial analysis of highly symmetric auxiliary
structures~\cite{Babai16,Babai18}. Helfgott's
detailed exposition of the corrected algorithm gives the explicit
bound
\begin{equation}
 \exp\bigl(O((\log n)^3)\bigr).
 \label{eq:previous-bound}
\end{equation}
The running-time analysis, including the treatment of auxiliary
recursive calls, is part of the difficulty: Babai's correction of
Split-or-Johnson was necessary to obtain the quasipolynomial
bound~\cite{BabaiFix17,H}. Grohe, Neuen, and Schweitzer adapted these
methods to graphs of maximum degree $d$ and obtained the bound
$n^{O((\log d)^c)}$ for an absolute constant
$c$~\cite[Theorem~1.1]{GNS18}.

We refine both the recursive reductions and their analysis to obtain,
for graphs of maximum degree $d\geq2$, the bound
\begin{equation}
 n^{O(\log d)}=\exp\bigl(O(\log n\log d)\bigr).
 \label{eq:target-bound}
\end{equation}
A graph on $n$ vertices has maximum degree less than $n$, so for all
graphs this gives the bound
\begin{equation}
 \exp\bigl(O((\log n)^2)\bigr)=n^{O(\log n)}.
 \label{eq:general-bound}
\end{equation}
To our knowledge, this is the first proof of an $n^{O(\log n)}$
worst-case bound for general graph isomorphism, and the first bound
for graphs of maximum degree $d$ whose exponent is linear in
$\log d$. The improvement comes
from changing which auxiliary instances are solved, which
decompositions they inherit, and what is charged when a recursive
call keeps almost all of the current domain. For the dependence on
the degree we also show that the many recursive calls made by a
constant-size local test check disjoint parts of the domain, so they
cost about as much as one call on the whole domain. Du~\cite{Du17}
previously stated the bound~\eqref{eq:general-bound};
Appendix~\ref{app:du-gaps}
explains why that argument does not establish it.

\subsection{Main result}

The algorithm computes the complete set of isomorphisms in a compact
form. If $X$ and $Y$ are isomorphic and $\pi$ is one isomorphism, every
other isomorphism is obtained by composing $\pi$ with an automorphism
of $X$. Thus it suffices to return $\pi$ and a generating set of
$\Aut(X)$; there is no need to list a possibly factorial number of
isomorphisms.

\begin{theorem}[Graphs of maximum degree $d$]
\label{thm:main-graph}
There is a deterministic algorithm that, given two graphs on $n$
vertices of maximum degree at most $d$, decides whether they are
isomorphic in time $\exp(O(\log(n+2)\log(d+2)))$. If they are
isomorphic, it returns one isomorphism and generators for the full
automorphism group of the first graph within the same bound. The
constants do not depend on $d$.
\end{theorem}

No vertex of a graph on $n$ vertices has more than $n-1$ neighbors.
Taking $d=n-1$ in Theorem~\ref{thm:main-graph}, and using
$\log(d+2)\leq\log(n+2)$, gives the bound for general graphs.

\begin{corollary}[Graph isomorphism]
\label{cor:main-general}
There is a deterministic algorithm that, given two graphs on $n$
vertices, decides whether they are isomorphic in time
$\exp(O((\log(n+2))^2))$. If they are isomorphic, it returns one
isomorphism and generators for the full automorphism group of the
first graph within the same bound.
\end{corollary}

The proof of Theorem~\ref{thm:main-graph} uses a standard intermediate
problem. A coloring is a map
$x\colon\Omega\to\Sigma$, where $\Omega$ is a finite set of positions.
A generating set specifies a family $G$ of allowed permutations of
$\Omega$, closed under composition and inversion. Given two colorings
$x,y$, \emph{string isomorphism} asks for all $g\in G$ with $x^g=y$.
Luks showed that for graphs of maximum degree $d$ it suffices to solve
string isomorphism for groups in a restricted class~\cite{Luks82}.
Let $\Gd$ be the class of finite groups whose composition factors are
all isomorphic to subgroups of $\Sym(d)$. This is Luks's class, in the
notation of Grohe, Neuen, and Schweitzer~\cite[Definition~2.3]{GNS18}.
It is smaller than the class, also often denoted $\Gamma_d$, that
allows simple groups of Lie type of bounded rank. The class $\Gd$ is
closed under subgroups and homomorphic images~\cite[Lemma~2.4]{GNS18}.
A polynomial-time reduction, whose running time does not depend on
$d$, reduces isomorphism of graphs of maximum degree $d$ to string
isomorphism for groups in $\Gd$~\cite[Theorem~8.1]{GNS18}.
Section~\ref{sec:degree-graphs} gives this reduction in a form that
also returns an isomorphism and the automorphism generators.

\begin{theorem}[String isomorphism]
\label{thm:main-string}
Let $d\geq2$. String isomorphism on $N$ positions under a group in
$\Gd$, given by generators, can be solved deterministically in time
$\exp(O(\log(N+2)\log(d+2)))$. The algorithm returns either the empty
set or one isomorphism and generators for the entire color-preserving
subgroup. The running time has an additional polynomial dependence on
the explicit input description, and the constants do not depend on
$d$.
\end{theorem}

The statement uses the usual generator representation of permutation
groups. Canonization is a separate task; Babai obtained a
quasipolynomial algorithm for that task as well~\cite{BabaiCanon19}.
The theorem here concerns isomorphism and automorphism generators.

The algorithm has two layers. The inner layer is a general procedure
for string isomorphism, with no restriction on the group. It is built
from local certificates and inherited decompositions in
Sections~\ref{sec:model}--\ref{sec:analysis}, and it runs in time
$\exp(O((\log(N+2))^2))$ on $N$ positions
(Proposition~\ref{prop:general-procedure}). The
outer layer is a recursion for groups in $\Gd$
(Section~\ref{sec:degree}). It keeps the decompositions and the local
certificates of the inner layer, but it calls the inner layer only on
instances with at most $d^2$ positions. The
group-theoretic structural input, including the
classification-dependent primitive-group results, is stated and used
in Appendix~\ref{app:groups} and Section~\ref{sec:degree-primitive}.

\subsection{Relation to previous work}

Early computational work on graph isomorphism includes Corneil and
Gotlieb's heuristic treatment~\cite{CorneilGotlieb70}; Read and Corneil's
survey~\cite{ReadCorneil77} describes the development of the problem,
its applications, and its unsettled complexity. Algorithms for
restricted graph classes supplied several of the first structural
methods. In particular, Hopcroft and Wong obtained a linear-time
isomorphism test for planar graphs~\cite{HopcroftWong74}. These algorithms use structural restrictions to control the
relabelings that must be considered.

The group-theoretic approach represents families of possible
relabelings by generators and restricts those families recursively.
Babai gave a randomized polynomial-time test for vertex-colored
graphs with bounded color multiplicities, using a chain of
permutation groups~\cite{Babai79}.
Furst, Hopcroft, and Luks established polynomial-time algorithms for
basic operations on generator-specified permutation
groups~\cite{FHL80}. Luks reduced bounded-degree graph isomorphism to
color-preserving subgroup problems for groups with restricted
composition factors, obtaining a polynomial-time algorithm for every
fixed maximum degree~\cite{Luks82}. Babai and Luks extended this
approach to canonical labeling, including polynomial-time canonization
for bounded-degree graphs. Their general
canonization algorithm runs in time $\exp(n^{1/2+o(1)})$, using
Zemlyachenko's color-valence reduction together with the group-theoretic
canonical-labeling machinery~\cite{BabaiLuks83}. Combining group methods
with degree reduction also gave general subexponential algorithms:
Babai, Kantor, and Luks obtained the bound
$\exp(O(\sqrt{n\log n}))$~\cite{BKL83}.

Combinatorial refinement provides another source of structure. The
Weisfeiler--Leman method refines colors using incidence information
and relates the resulting partitions to graph
symmetry~\cite{WeisfeilerLeman68}. Its limitations are substantial:
Cai, F\"urer, and Immerman constructed nonisomorphic graphs requiring
linear Weisfeiler--Leman dimension to distinguish, even among
degree-three colored graphs. Their examples admit linear-time
canonical labeling: failure of low-dimensional refinement can coexist
with efficient isomorphism testing on the same graph family~\cite{CFI92}.
Neuen and Schweitzer proved
exponential lower bounds for individualization--refinement algorithms
whose refinement operators, cell selectors, and node invariants are
captured by a fixed dimension of Weisfeiler--Leman
refinement~\cite{NeuenSchweitzer18}. These are lower bounds for the
specified methods. They help explain the role of additional
permutation-group information in algorithms that combine refinement
with algebraic reductions.

Babai combined Luks's string-isomorphism framework with local
certificates and combinatorial partitioning to obtain a general
quasipolynomial-time algorithm~\cite{Babai16,Babai18}. The corrected
algorithm replaces a recursive step in the Split-or-Johnson
procedure~\cite{BabaiFix17}. Helfgott's exposition of the corrected
algorithm gives the explicit running-time estimate
$\exp(O((\log n)^3))$ in Appendix~A~\cite{H}; this is the quantitative
benchmark for general graphs used here. The local certificates connect information about
small test sets in an auxiliary action to restrictions on the
original permutation group. Babai subsequently obtained
quasipolynomial-time canonization by modifying the symmetry-breaking
construction~\cite{BabaiCanon19}. Canonization remains a separate
output requirement from the isomorphism cosets computed in this paper.

Later work adapted these methods to additional graph structure.
Grohe, Neuen, and Schweitzer obtained an $n^{O((\log d)^c)}$
isomorphism test for graphs of maximum degree $d$, for an absolute
constant $c$~\cite{GNS18}, improving the $n^{O(d/\log d)}$ bound of
Babai, Kantor, and Luks~\cite{BKL83}. Their main structural tool is a
dichotomy for primitive groups in $\Gd$: such a group is small, or it
is a product of Johnson actions on at most $d$
labels~\cite[Theorem~3.13]{GNS18}. We use this dichotomy directly.
Their algorithm first passes to a larger domain with an almost
$d$-ary sequence of partitions~\cite[Theorem~4.6]{GNS18}; we keep
the original domain. Wiebking developed
multiple-coset isomorphism to combine graph decompositions with
Babai's framework, obtaining an $n^{\polylog k}$ bound, where $k$ is
the smaller treewidth of the two input graphs~\cite{Wiebking20}.
Neuen obtained a $(n+m)^{\polylog d}$ algorithm for hypergraph
isomorphism under a supplied permutation group whose composition
factors embed into the symmetric group on $d$ points; here $m$ is
the number of hyperedges~\cite{Neuen20}. These bounds exploit the
degree, treewidth, or supplied group restriction. Their exponents are
polylogarithmic in the parameter, while the exponent
in~\eqref{eq:target-bound} is linear in $\log d$. In particular, they
do not give the exponent in~\eqref{eq:general-bound} for unrestricted
graph isomorphism.

The present algorithm builds on the Luks--Babai framework. It uses
local certificates and coherent refinement while retaining the
decomposition of the current permutation action through auxiliary
calls. Two ingredients support this change in the general procedure:
a hereditary bound on visible alternating quotients, used to permit
constant-size local tests, and realizations of auxiliary bipartite
instances inside the current domain. Retained row decompositions
supply such realizations when the large action is on single labels.
For groups in $\Gd$, two further facts make the exponent linear in
$\log d$. The recursive calls of the local tests have disjoint windows, and
every alternating action has at most $d$ labels. The analysis charges
recursive restrictions and auxiliary calls to the same inherited
decomposition. Section~\ref{sec:overview} explains these reductions
and their role in the running-time bound.

Du~\cite{Du17} previously stated an $n^{O(\log n)}$ bound for
general graph isomorphism. Appendix~\ref{app:du-gaps} gives an infinite
family of graph inputs on which an incidence test used by that
algorithm fails, and identifies independent gaps in its recursion
and assembly of partial answers. That preprint therefore does not
establish the stated bound. Our quantitative comparison for general
graphs is with the
explicit bound~\eqref{eq:previous-bound} for the corrected
Luks--Babai algorithm. The present approach retains permutation
actions and complete isomorphism cosets throughout the recursion;
it does not use Du's spectral decomposition or exposure argument.

\paragraph{Organization.}
Section~\ref{sec:overview} gives an overview of the approach and
the running-time argument. Sections~\ref{sec:model}--\ref{sec:graph-rectangles}
define the recursive objects and prove the auxiliary graph reductions.
Section~\ref{sec:graph-algorithm} gives the general procedure and its
correctness argument. Section~\ref{sec:analysis} gives its
running-time analysis and fixes all numerical parameters.
Section~\ref{sec:degree} gives the recursion for groups in $\Gd$ and
proves Theorem~\ref{thm:main-string}. Section~\ref{sec:degree-graphs}
gives the reduction from graphs to strings and proves
Theorem~\ref{thm:main-graph}. Appendix~\ref{app:groups}
contains the structural statements and their detailed
permutation-group proofs.
Appendix~\ref{app:du-gaps} discusses the major gaps in Du's preprint.

\section{Our approach and techniques}
\label{sec:overview}

Our algorithm builds on the Luks--Babai framework: represent the
remaining relabelings by generators, test their compatibility with
local information, and use refinement to expose a partition or an
auxiliary graph. Two difficulties determine the running time. A local
test can eliminate symmetry while keeping almost every position.
An auxiliary graph can require substantial recursive work before the
original instance can continue. We modify the local-test endpoint and
stop the symmetry-breaking procedure at its first bipartite graph.
Both changes use a decomposition that remains attached to the instance
throughout these operations.

We first describe the general procedure, which places no restriction
on the group. Its exponent is quadratic in $\log N$. The recursion for
groups in $\Gd$ reuses its parts, and the last part of this section
explains what changes to make the exponent $O(\log N\log d)$.

We distinguish the graph's $n$ vertices from the positions on which
the recursive algorithm stores colors. In our application these
positions come from the reduction from graphs to strings in
Section~\ref{sec:degree-graphs}, and there are polynomially many of
them. Write $s$ for
the current number of positions, $q$ for the number of auxiliary
labels, and $N$ for the number of positions at the start of a phase.
A phase keeps this starting degree fixed; its calls inherit the
decomposition until their domains are small enough to start a new phase.
Auxiliary labels may name blocks or coordinates; they need not be
graph vertices. In particular, a graph on $q$ labels may describe
constraints on many more than $q$ positions.

\paragraph{Keeping the decomposition and its restrictions.}
For a transitive active action, the state retains a hierarchy of
invariant partitions. Its successive ratios $b_1,\ldots,b_t$ satisfy
$\prod_i b_i=s$. Each level also carries a cap $c_i$: even after
further restrictions and quotient actions, its entire action cannot
induce all even permutations on more than $c_i$ labels. Initially
$c_i=b_i$ suffices.
A restriction can lower a cap without creating a new partition.
This records progress even when the blocks themselves remain unchanged.

When a call selects an orbit, it intersects the old partitions with
that orbit and inherits their caps, replacing each cap by its minimum
with the new interval ratio. A change from row coordinates to column
coordinates splits or reorders old factors and preserves their
restrictions. Thus a later auxiliary problem cannot regain an
allowance for symmetry that an earlier step has already removed.

On an actual $b$-block quotient of the whole current group with no
further invariant partition, the standard primitive-group reduction
either enumerates its $b^{O(\log b)}$ relabelings or exposes a large
alternating action on auxiliary labels. The latter case requires the
two reductions below.

\paragraph{Why constant-size tests suffice.}
A local certificate examines a fixed number $\kcert$ of auxiliary
labels and asks whether color-preserving relabelings can realize
every even permutation of them. We examine all such tests and
comparisons between ordered tests in the two inputs; for fixed
$\kcert$, there are polynomially many.
The test set is small, but answering its question can require exact
recursive isomorphism calls throughout the current domain.
The procedure grows a window of checked positions, stopping as
nonfull if the remaining relabelings no longer realize every even
test permutation. Otherwise, if fixing a position prevents some even permutation of
the test labels, then fixing all test labels splits that position's
orbit into pieces of size at most $s/\kcert$.

The difficult endpoint occurs when no such position remains; fixing
the test labels can then leave an entire orbit intact.
First ask whether the current relabelings fixing the unchecked region
pointwise still realize all even test permutations. If so, they
preserve the checked colors and fix the remaining positions, giving
automorphisms of the whole coloring: a full certificate.
Otherwise, we find an unchecked orbit on
which the test action is visible: the induced relabeling of that
orbit determines the permutation of the test labels. Restricting
to relabelings that fix every test label removes this symmetry,
although the orbit may keep its size.

Assign weight $\kcert$ to removing the test action.
Theorem~\ref{thm:hereditary-budget} bounds the total weight along
every sequence of restrictions and orbit selections by $O(\log s)$.
Its measure $\sigma$ never
increases under such operations, and the endpoint call lowers it by
$\kcert$. Local calls therefore have either
$s'\leq s/\kcert$ or $\sigma'\leq\sigma-\kcert$.
A bound on each individual symmetry would not give this conclusion;
the bound must persist through the entire recursive chain.

\paragraph{The first bipartite graph gives enough progress.}
Babai's aggregation and refinement steps either yield a useful
partition, exploit already certified symmetry, or produce an
nontrivial homogeneous coherent configuration---a stable coloring of ordered
label pairs---on a subset $C$ containing a fixed fraction of the
original $q$ labels. The value $q$ refers to that original label
set throughout the reduction. For configurations above the label
threshold specified below, we follow Split-or-Johnson only to its
first bipartite graph. After fixing one source label and trying
its target images, either the resulting color classes are balanced
or there is a dominant class $W$ and a smaller class $R$.
At this first step, both sides are still subsets of the old
label set. For a fixed $2/3<\eta<1$, they satisfy
\[
 |W|>\eta|C|,\qquad |R|\leq |C|-|W|,
 \qquad |W||R|\leq\eta(1-\eta)q^2.
\]

The bipartite graph is nontrivial and semiregular: degrees are
constant on each side. Every class of
left vertices with identical neighborhoods has size at most
$|W|/2$. Once the right side is fixed pointwise, an automorphism can only
permute within these classes; the remaining choices act on the
smaller right side. Consequently its complete isomorphism answer,
projected to $W$ and extended across the other labels with the
required parity, reduces the parent's selected cap to $\beta q$
for a fixed $\beta<1$.
The extension may admit extra maps. A subsequent check of the
original colors removes them while preserving all parent
isomorphisms covered by this branch.

\paragraph{Representing the rectangle on existing positions.}
Fix $1/3<\alpha<1/2$ and put $\lambda=N^\alpha$ throughout the
phase. An auxiliary configuration with $q\leq\lambda$ can be
solved on its ordered pairs, using at most $N^{2\alpha}$ positions.
For $q>\lambda$ and sufficiently large $N$, we have
$\binom q3>N$, and the action also exceeds the logarithmic
threshold of Corollary~\ref{cor:physical-support-thresholds}.
Applied to the transitive action on the current positions and its
supplied alternating label action, that result identifies the
kernel orbits with single labels, ordered pairs, or unordered pairs
of distinct labels. Each label or pair represents the same number of positions.

In a pair representation, pull each matrix-entry color back to all
positions over that pair. The positions over $W\times R$ form
an actual subgroup orbit $\Delta$, and
\[
 \frac{|\Delta|}{s}\leq
 \frac{2|W||R|}{q(q-1)}\leq\theta<1
\]
for a fixed $\theta$; the factor two is needed for unordered pairs.
The inherited call on $\Delta$ returns exactly the permitted
matrix isomorphisms after projection. Its physical size decreases,
and it keeps the old hierarchy, caps, and symmetry measure.

\paragraph{Retaining rows to supply a missing coordinate.}
For a single-label action, let $m$ be its degree and regard its
fibers as rows. The row-normal-form theorem chooses a common
coordinate system inside the rows, so that every even permutation
of the rows is available while leaving the within-row coordinates
unchanged. It also specifies the remaining local variation inside a row.
If that variation is transitive inside a row, the row terminal
reduces to polynomially many exact isomorphism calls on individual
rows. Combining their answers must impose one residual action
shared by all rows, a sum constraint on their local corrections,
and the required parity of the row matching. These conditions are
part of the terminal's
$\exp(O((\log N)^2))$ additional work.

If the local variation is intransitive, its orbits define columns
across all rows. We continue on the column action, lifting each
restriction to all compatible relabelings of the positions. These
full preimages retain the independent row action.
A later alternating action of degree $q>\lambda$ on the columns
must again have a single-label representation: there are at most
$N^{1-\alpha}$ columns, too few for a pair representation at this scale.
When $q\leq m$, couple its even permutations with the same permutations
of $q$ selected rows. This supplies a grid of equal cells on which
the rectangle can be represented. Its inherited query occupies
the fraction
\[
 \frac{|W||R|}{mq}\leq\frac{q}{4m}\leq\frac14
\]
of the current positions. When needed, a second query with permuted
target colors handles the other parity class.

If instead $q>m$, construct new independent rows of degree $q$.
There can be only one such enlargement while the row action is
retained. The two degrees $m,q>\lambda$ belong to distinct
intervals of the same hierarchy; all other ratios have product
at most $N/(mq)<N^{1-2\alpha}<\lambda$. The entire remaining
action therefore has no alternating section of degree greater than $q$.
The coordinate change retains the old caps, so later restrictions
cannot produce another larger row requirement.

We process a largest original orbit first, preserving the row action
on the only child that can exceed half the parent domain. A child
that loses the retained rows already has a domain decrease, a symmetry
decrease, or a genuine hierarchy refinement, which pays for their
later reconstruction.

\paragraph{Complete answers prevent repeated continuations.}
Every recursive call returns either the empty set or all compatible
relabelings, represented by one isomorphism and generators for the
source coloring's stabilizer. This is
essential when processing different orbits: each later query must
retain every choice consistent with the preceding ones.

Complete output also determines the cost of an auxiliary graph. We finish its
recursive solution and assemble its full coset before resuming
the parent under the projected answer. The work spent in the
auxiliary recursion and the work of the parent continuation add;
the auxiliary leaves do not each initiate a new continuation.

The parent resumes under the full preimage of the extended label
coset in its original action, saved before the aggregation choices.
In particular, choosing the diagonal subgroup to represent a
rectangle incurs no enumeration of its index. The rectangle call
is paid by its smaller physical domain, and the parent continuation
by the smaller cap. Caps learned inside the rectangle call belong
to that call; the parent's decrease follows from the rectangle's
twin-class argument.

\paragraph{The running-time analysis.}
The analysis combines domain size, the hereditary measure $\sigma$,
the partition hierarchy, and the cap sum
\[
 Q=\sum_i(\log b_i)\log\max\{c_i,2\}\leq(\log s)^2.
\]
A constant-factor domain decrease pays polynomially many inherited
calls. The extra visible-orbit check pays through $\sigma$; genuine interval
refinements consume a separate $O(\log s)$ hierarchy measure.
Enumerating the primitive quotient leaves kernel-orbit calls on at
most $s/b$ positions, which pays its $b^{O(\log b)}$ cost.
Reducing a cap
from at least $q$ to $\beta q$ lowers $Q$ by $\Omega(\log q)$,
which pays the $q^{O(1)}$ parent continuations. These resources
give a total phase cost of $\exp(O((\log N)^2))$ times its
fresh-call cost.

Choose $2\alpha<\gamma<1$. Ordinary inherited calls start a fresh
phase only at $N^\gamma$ positions or fewer. Small configurations use
at most $N^{2\alpha}$ positions, and row calls use at most
$N^{1-\alpha}$; both are below the same threshold.
Neither a new auxiliary description nor a change of coordinates
by itself restarts the phase. We obtain
\[
 F(N)\leq\exp(O((\log N)^2))\bigl(1+F(N^\gamma)\bigr).
\]
Successive quadratic logarithmic costs form a geometric series, so
the general procedure takes time $\exp(O((\log N)^2))$. The degree
recursion below uses this bound only on instances with at most $d^2$
positions.

\paragraph{Groups with restricted composition factors.}
Now let the group lie in $\Gd$. The analysis above pays $N^a$ for the
local calls of one giant step. A decrease of order $x$ per call is
needed to pay for them, and this is why the potential has the term
$x^2$. Two facts let us replace $x^2$ by $x\log d$.

First, every alternating section of a group in $\Gd$ has degree at
most $d$. So every label count satisfies $q\leq d$, every cap can
start at $\min\{d,b_i\}$, and the cap sum is at most
$(\log s)\log d$. A primitive quotient of degree $b$ either has order
$b^{O(\log d)}$ or is a product of Johnson actions on at most $d$
labels~\cite[Theorem~3.13]{GNS18}. Its enumeration is then paid by a
decrease of order $\log b\log d$.

Second, the local calls have disjoint windows. One endpoint makes a
call at each window update, and there can be $s$ updates. But the new
windows are disjoint. For a fixed coset of the test quotient, the calls
of all stages therefore have disjoint domains. So the local calls split
into $q^{O(1)}$ families, and each family weighs at most as much as one
call on the whole domain. A decrease of order $\log d$ per call pays
for $q^{O(1)}\leq d^{O(1)}$ families.

The measure $\sigma$ itself can still be of order $\log N$ when $d$
is fixed (Remark~\ref{rem:degree-example}). What changes is the
price of one unit of $\sigma$: it costs $O(\log d)$ in the exponent,
not $O(\log N)$. The potential of the degree recursion is
\begin{equation*}
 A\bigl[(x+\sigma+\rho)\log(d+2)+Q\bigr]=O(\log N\log d).
\end{equation*}
One local call may contain both orbits of the two kinds above, so
$x+\sigma$ is taken as a maximum over points of the domain.

Auxiliary configurations have at most $d$ labels. The degree recursion
solves each of them by the general procedure on at most $d^2$
positions, in time $\exp(O((\log d)^2))$. This is an additive cost
of the giant step, and there are $q^{O(1)}$ of them. The degree
recursion never uses rectangles, rows, or phases. For $d<N$ the
additive cost is at most $\exp(O(\log N\log d))$. For $d\geq N$ the
general procedure already runs within the claimed bound, and the
degree recursion is not needed.

\section{Colorings, partitions, and allowed relabelings}
\label{sec:model}

Unless a base is displayed, logarithms are natural.
We distinguish the \emph{active domain} $\Omega$, on which colors are
stored, from an \emph{auxiliary label set} $\Gamma$. Their sizes are
$s=|\Omega|$ and $q=|\Gamma|$, respectively. An auxiliary label may
name a block of positions or a coordinate of a subset representation.
It need not be a vertex of the input graph. This distinction matters:
a small auxiliary label set can describe a large number of positions.

The allowed relabelings form a group $G\leq\Sym(\Omega)$. We use
right actions and write $x^g$ for the coloring obtained by relabeling
$x$ with $g$. A restriction is represented by generators, together
with one representative when it is a coset. Basic operations such as
finding orbits, fixing specified points, computing an explicitly
given image, and lifting image generators take polynomial time.
An arbitrary intersection of two permutation groups is not treated
as a polynomial-time operation.
An \emph{overcoset} is a coset containing all isomorphisms of the
current branch, possibly together with extra relabelings. Those
extra relabelings are removed when the original colors are checked.

Every fresh call uses the faithful permutation image on its own active
domain, with a polynomial-size strong generating set. Its caller retains
the homomorphism used to lift the returned generators and representative;
operations in a larger covering action are charged to that caller.
Before the call, compress the two input colorings jointly: equal color
records receive the same integer name, distinct records receive distinct
names, and the common palette has at most $2|\Omega|$ entries. Colors
are never renamed independently in the two inputs. Reading and comparing
the original records contributes the polynomial dependence on the explicit
input description in Theorem~\ref{thm:main-string} and
Proposition~\ref{prop:general-procedure}. Subsequent calls use
the compressed palette, including any shared tags introduced by a
construction. The general recursive routine $\Solve$ is specified in
Algorithm~\ref{alg:main-recursion}, and the degree-sensitive routine
in Algorithm~\ref{alg:degree-recursion}.

\subsection{Exact answers and the orbit chain rule}

Every recursive call returns the complete set of compatible
relabelings. This requirement is essential even when the original
question asks only for one isomorphism. Suppose $G$ has several
orbits on $\Omega$. After solving the first orbit, later orbits must
be processed under precisely the relabelings compatible with the
first answer. Returning only one choice would discard solutions.

If the answer on the first orbit is $H\pi$, absorb $\pi$ into the
target coloring and continue under $H$. Repeat for the remaining
orbits. This is the orbit chain rule. The recursive domains are
disjoint, but their allowed relabelings are processed sequentially.
For an exhaustive family of branches, the successful answers can
again be represented by one coset: take one successful representative,
adjoin the stabilizer generators in each successful answer, and
adjoin the ratios of successful representatives. This is valid
because the exhaustive union is the full isomorphism coset.

We process a largest orbit of the group at the start of a chain
first. This order will preserve a retained row action on the only
child that can contain more than half of the current domain. Every
later conditional query lies in a different original orbit, and
therefore has size at most half of the parent domain.

\subsection{The partition hierarchy}

A block system is a partition of $\Omega$ that every allowed
relabeling permutes. In a transitive action all blocks have the same
size. We keep a hierarchy of such partitions down to singletons.
If its successive ratios are $b_1,\ldots,b_t$, then
\[
 b_1\cdots b_t=s.
\]
When a recursive call restricts the allowed relabelings and selects
one orbit, it intersects the old partitions with that orbit. The
old hierarchy is retained; new invariant partitions only refine it.

Each interval of the hierarchy also carries an upper bound $c_i$
on the alternating sections that can occur in that interval. A
section means a quotient of a subgroup. The definition is used only
for the analysis: it ensures that the bound remains valid after
further restrictions and quotient actions. The initial bound
$c_i=b_i$ is always available. A symmetry-breaking step can reduce
$c_i$ even when the partition itself does not change. We will charge
the two kinds of progress separately.
When an interval of new ratio $b'$ inherits a bound $c$, we store
$\min\{c,b'\}$. Thus every stored bound is at most its current
interval ratio, including after an orbit restriction.
Formally, the \emph{alternating width} of a finite group is the
largest $k\geq5$ for which it has a section isomorphic to $\Alt(k)$,
or zero if there is no such section. An interval bound $c_i$ is
a bound on this quantity for its entire constituent action.

The standard primitive-group reduction either enumerates a group of
order $b^{O(\log b)}$ on an actual $b$-block quotient, or exposes a
large alternating action on an explicit auxiliary label set. Product
actions refine the actual hierarchy. Enumeration is performed only
on a quotient of the whole current group; a local constituent
deeper in a hierarchy is not independently enumerated.

\subsection{The three structural interfaces}

The main algorithm uses the following facts. Their proofs are in
Appendix~\ref{app:groups}.

\begin{enumerate}[label=(\roman*)]
\item \textbf{Inherited symmetry measure.} Every transitive active
action has a nonnegative measure
$\sigma(G,\Omega)=O(\log(s+2))$, nonincreasing under a
subgroup restriction and passage to one faithful orbit action.
When the induced action has a visible quotient $\Alt(k)$, $k\geq\kcert$, with a
transitive kernel, passage to that kernel reduces $\sigma$ by at
least $k$ (Theorem~\ref{thm:hereditary-budget}). The algorithm does
not need to compute $\sigma$.
\item \textbf{Large actions have one or two coordinates.} If a
supplied alternating action on $q$ labels is sufficiently large
relative to $s$, its kernel orbits are indexed either by one label,
by an ordered pair, or by an unordered pair of distinct labels
(Corollary~\ref{cor:physical-support-thresholds}). The quotient map
and its equal-size fibers are computable.
\item \textbf{Natural rows can be made independent.} If the action
is on single labels and the kernel is transitive in every row,
then above a logarithmic row-size threshold we can choose coordinates
so that all even row permutations act without changing the within-row
coordinate (Theorem~\ref{thm:logarithmic-clean-complement}). The
full normal form also describes the remaining within-row variation
(Theorem~\ref{thm:logarithmic-gat-normal-form}).
\end{enumerate}

The first interface explains progress without a large domain
decrease. The other two identify auxiliary graph problems that can
be solved on a subset of the current domain. These are statements
about the actual permutation action, not merely about an abstract
group isomorphic to it.

\section{Local tests of constant size}
\label{sec:graph-tests}

Let $\Gamma$ be an auxiliary label set on which the current action
induces all even permutations. Fix a constant $\kcert$ and consider
every $\kcert$-element test set $T\subseteq\Gamma$. A local
certificate asks whether the color-preserving relabelings can still
realize every even permutation of $T$. It either supplies such
relabelings or records an obstruction. The number of test sets and
their ordered comparisons is polynomial because $\kcert$ is fixed.

The obstacle is that a constant-size test may reveal a large orbit
on which no position becomes distinguishable. The ordinary domain-
size analysis gives no useful decrease there. Our endpoint procedure
explicitly looks for an orbit on which the test action is visible.
If it finds one, the recursive call consumes the corresponding
symmetry measure. It does not assume that an unaffected region can
be ignored.

\begin{lemma}[Local-test interface]
\label{lem:main-local-test}
Suppose the physical parent action $G\leq\Sym(\Omega)$ is transitive,
with $s=|\Omega|$.
For a fixed sufficiently large $\kcert$, all test certificates and
ordered comparisons can be computed with polynomial structural work
and polynomially many inherited recursive calls of the following
two types:
\[
 (s,\sigma)\longmapsto(s/\kcert,\sigma),
 \qquad
 (s,\sigma)\longmapsto(s,\sigma-\kcert).
\]
Every comparison returns the complete isomorphism coset. The
certificates and the equality classes of ordered comparisons are
preserved by all parent isomorphisms.
\end{lemma}

\begin{proof}
Maintain a window of positions whose colors have already been
checked. The current allowed relabelings preserve those colors.
If their image on $T$ is proper, the certificate is nonfull.
Otherwise, add the affected positions: those whose point
stabilizers do not map onto $\Alt(T)$. The kernel orbits within
each affected orbit have size at most a $1/\kcert$ fraction.

If no affected positions remain, test whether the pointwise
stabilizer of the remaining region still maps onto $\Alt(T)$.
If it does, this gives an actual color-preserving full certificate.
If it does not, simplicity guarantees a remaining orbit on which
the test quotient is visible. Its kernel is transitive, so the
first structural interface decreases $\sigma$ by $\kcert$.
Add all such orbits, an equivariant choice, and continue.

Each update enlarges the window. There are at most $s$ updates,
and enumerating the constant-size test quotient costs only a
constant factor. Ordered comparisons run the two endpoint
procedures in lockstep, under the full ordered-tuple transporter.
The prefix-recovery argument in Section~\ref{sec:certificates}
shows that a final marked-window isomorphism preserves every
earlier window. Thus discarded comparisons cannot reappear, and
the comparison relation is exact. That section gives the full
construction and the visibility test.
\end{proof}

Algorithm~\ref{alg:local-certificate} spells out one endpoint procedure.
Here $P_T$ consists of the allowed relabelings that preserve $T$ and
induce an even permutation on it; $\phi_T$ is this induced action.
Relabelings outside $T$ remain unrestricted. For a set $U$ of positions,
$A_{(U)}$ denotes the members of $A$ fixing every position in $U$.
The window $W$ records colors already checked. Returning \textsc{Full}
requires actual automorphisms of the entire coloring, whereas
\textsc{Nonfull} records that even the current partial automorphisms
cannot realize the whole test action.

\begin{pseudocode}{A constant-size local certificate}{alg:local-certificate}
\textbf{Input:} A transitive inherited state $\mathcal S$, a coloring $x$, and a
test set $T\subseteq\Gamma$ of size $\kcert$.\\
\textbf{Output:} A full or nonfull certificate, with its final window
and exact endpoint group.
\begin{tabbing}
\hspace{1.6em}\=\hspace{1.4em}\=\hspace{1.4em}\=\hspace{1.4em}\=\kill
\pc \textbf{procedure} $\proc{LocalCertificate}(\mathcal S,x,T)$\\
\pc \>$W\gets\varnothing$; $A\gets P_T$\\
\pc \>\textbf{loop}\\
\pc \>\>\textbf{if} $\phi_T(A)\ne\Alt(T)$ \textbf{then return} $(\textsc{Nonfull},W,A)$\\
\pc \>\>$E\gets\{v\in\Omega\setminus W:\phi_T(A_v)\ne\Alt(T)\}$\\
\pc \>\>\textbf{if} $E\ne\varnothing$ \textbf{then} $V\gets E$\\
\pc \>\>\textbf{else}\\
\pc \>\>\>$U\gets\Omega\setminus W$; $K\gets A_{(U)}$\\
\pc \>\>\>\textbf{if} $\phi_T(K)=\Alt(T)$ \textbf{then return} $(\textsc{Full},W,A,K)$\\
\pc \>\>\>$V\gets\bigcup\{\Delta:\Delta\text{ is an }A\text{-orbit in }U,
                      \ \phi_T(A_{(\Delta)})=1\}$\\
\pc \>\>$A\gets\proc{CheckNewWindow}(\mathcal S,A,V,x)$\\
\pc \>\>$W\gets W\cup V$
\end{tabbing}
\end{pseudocode}

The operation $\proc{CheckNewWindow}$ returns the full subgroup of
$A$ preserving $x$ on $V$. It enumerates the constantly many cosets
of $\ker(\phi_T|_A)$ in $A$ and uses $\Continue$ on the kernel
orbits in $V$, with the inherited state and the orbit chain rule.
Affected orbits give the domain decrease in
Lemma~\ref{lem:main-local-test}; the orbits selected in the other
case give its symmetry-measure decrease. No update is a fresh call
on the whole marked coloring.

For ordered comparisons, run two copies in lockstep under the full
ordered-tuple transporter, maintaining the complete comparison coset
after each new window. Each update uses the same inherited kernel-orbit
queries. A mismatch or empty coset ends that comparison. The
prefix-recovery argument in the proof above shows that a discarded
comparison cannot become feasible later.

\subsection{From the tests to an auxiliary graph}

Aggregate the full certificates, and use the comparison classes of
the nonfull certificates as tuple colors. Babai's Design Lemma and
the aggregation procedure then expose either useful symmetry or
a nontrivial colored auxiliary graph. Two points require care in
using this step with constant-size tests.

First, the full certificates must be actual automorphisms of the
entire coloring. This is why the endpoint procedure fixes the
unexamined region at full termination. Second, orbital colors
cannot be assigned arbitrary names independently in the two
instances. We fix a bounded source tuple, branch over its target
images and the image of a source orbital, and refine the resulting
binary directed graphs. The number of these choices is polynomial
in $q$. The detailed argument, including the large-symmetry branch,
is in Section~\ref{subsec:aggregation-constant-threshold}.

The resulting interface has three outcomes, with an absolute
constant $\delta>0$:
\begin{enumerate}
\item A complete answer reduces to inherited calls on domains of
size at most $s/2$.
\item A colored partition gives a family of permitted relabelings
whose alternating-section bound is a fixed fraction of $q$.
\item We obtain a nontrivial homogeneous coherent configuration
on $C\subseteq\Gamma$, with $|C|\geq\delta q$.
\end{enumerate}

A coherent configuration is an edge coloring of a complete directed
graph whose intersection counts depend only on edge colors. It can
be obtained by stable pair refinement. Homogeneity means all vertices
have the same diagonal color. The algorithm uses the configuration
to choose a partition or a bipartite subgraph; the main recursion
does not require the reader to manipulate its intersection numbers.

\section{Solving the first auxiliary bipartite graph}
\label{sec:graph-rectangles}

The usual Split-or-Johnson step continues to analyze the auxiliary
configuration. Our recursion stops earlier. It solves the first
bipartite graph reached by that procedure, using the old domain
whenever the graph has an inherited representation.

\subsection{The first rectangle and what its answer gives}

Choose an absolute balance parameter $\eta>2/3$. In a homogeneous
coherent configuration on $C$, individualize a source vertex and
branch over its target image. If every resulting color class has
size at most $\eta|C|$, the coloring already gives a balanced
restriction. Otherwise let $W$ be the unique larger class.
The first step of Split-or-Johnson supplies another class $R$
and a nonempty, noncomplete semiregular bipartite graph between
$W$ and $R$. Both sides are still actual subsets of $C$; no
artificial vertices have been introduced.

\begin{lemma}[First-rectangle interface]
\label{lem:main-rectangle}
With polynomially many choices, we obtain either a fixed-fraction
restriction on the auxiliary action, or a bipartite graph on
disjoint $W,R\subseteq C$ satisfying
\[
 |W|>\eta|C|,\qquad |R|\leq |C|-|W|,
 \qquad |W||R|\leq\eta(1-\eta)q^2.
\]
Every class of left vertices having identical neighborhoods has
size at most $|W|/2$. The complete isomorphism coset of this
bipartite graph under the relabelings induced by the current
auxiliary action gives an overcoset containing every parent
isomorphism in the current branch, with an alternating-section
bound at most $\beta q$ on all of $\Gamma$, for some fixed
$\beta<1$.
\end{lemma}

The intuition behind the last conclusion is direct. An automorphism
that fixes the right side can only permute left vertices with the
same neighborhood. These twin classes are small. Every other
automorphism acts on the smaller right side. Thus large symmetric
choices on the left are bounded by either a twin class or the
number of right vertices. We may forget the constraints on the
complement of $W$ after recording this bound, since $W$ occupies
a fixed fraction of the original label set.

\begin{proof}
The construction is the first entrance described in
Lemma~\ref{lem:first-soj-rectangle}. For $w=|W|$ and $r=|R|$,
the function $w(|C|-w)$ decreases on the relevant range, giving
the displayed product bound. Semiregularity gives the twin bound:
complement the bipartite graph if necessary so that every right
vertex has degree at most $w/2$. Each nonempty left twin class
is contained in one such neighborhood.

The kernel of the action on $R$ permutes only left twin classes;
the quotient acts on $R$. Hence the left projection has alternating
width at most $\max\{w/2,r\}$. Extend its full coset by arbitrary
permutations of $\Gamma\setminus W$, and impose the required
parity. Its width is at most
\[
 \max\{w/2,r,q-w\}
 \leq\max\{1/2,1-\eta\delta\}\,q.
\]
Choose $\beta$ larger than this constant and than the constants
for the balanced-partition cases. Lifting the overcoset and
checking the original colors removes any extra maps admitted
by the extension. It preserves every genuine parent isomorphism.
\end{proof}

\subsection{A rectangle represented by pairs of labels}

Suppose the kernel orbits of the current action are indexed by
ordered or unordered pairs of distinct labels of $\Gamma$.
Every pair has the same number of positions above it. The
positions above $W\times R$ form one orbit of an appropriate
subgroup of the allowed relabelings. If this orbit is $\Delta$,
then
\begin{equation}
 \frac{|\Delta|}{s}=
 \begin{cases}
 |W||R|/(q(q-1)),&\text{ordered pairs},\\
 2|W||R|/(q(q-1)),&\text{unordered pairs}.
 \end{cases}
 \label{eq:graph-rectangle-size}
\end{equation}
The factor of two in the second line reflects the unordered-pair
domain. This count, rather than the number of newly introduced
matrix entries, is what pays for the recursive call.

Pull the color of a matrix entry back to every position above it.
An allowed relabeling preserves these pulled-back colors exactly
when its action on the rectangle preserves the matrix. A complete
answer on $\Delta$ therefore projects to the complete permitted
matrix answer. The query is an actual subgroup-orbit restriction,
so it inherits the hierarchy and the symmetry measure. Equations
\eqref{eq:graph-rectangle-size} and the product bound in
Lemma~\ref{lem:main-rectangle} give $|\Delta|\leq\theta s$ for
some fixed $\theta<1$ and sufficiently large $q$.

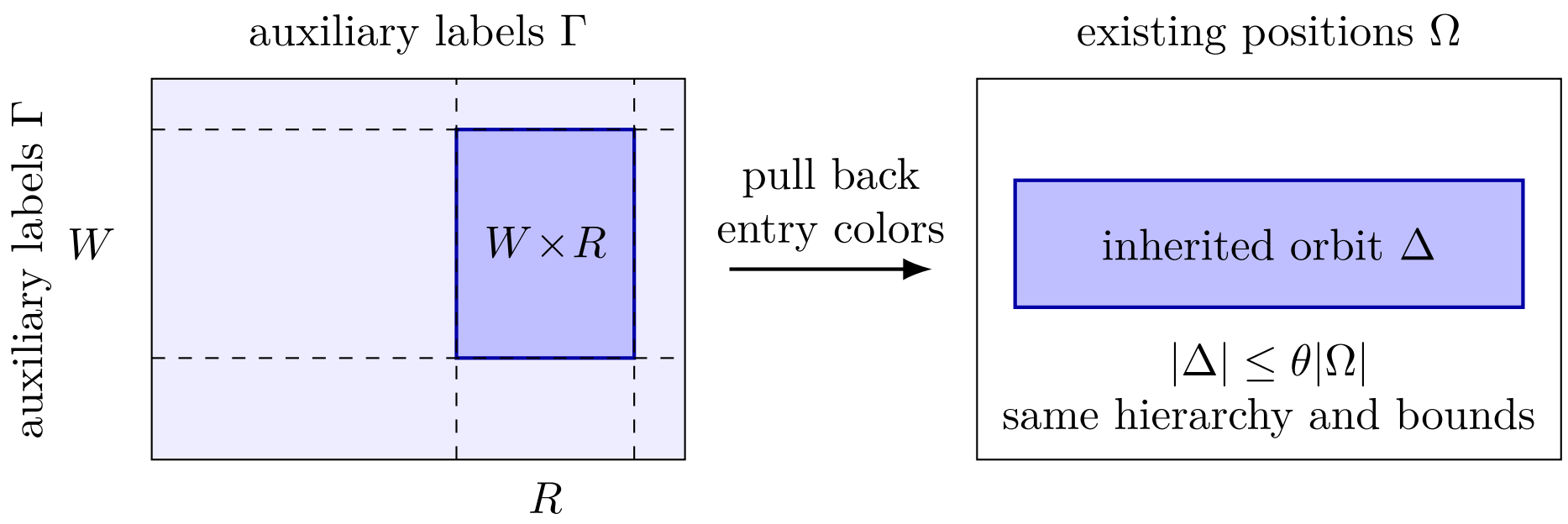
\begin{figure}[t]
\centering
\begin{tikzpicture}[font=\small,>=Latex]
 \fill[blue!7] (0,0) rectangle (4.2,3.0);
 \draw (0,0) rectangle (4.2,3.0);
 \fill[blue!25] (2.4,0.8) rectangle (3.8,2.6);
 \draw[blue!65!black,thick] (2.4,0.8) rectangle (3.8,2.6);
 \draw[dashed] (0,0.8)--(4.2,0.8);
 \draw[dashed] (0,2.6)--(4.2,2.6);
 \draw[dashed] (2.4,0)--(2.4,3.0);
 \draw[dashed] (3.8,0)--(3.8,3.0);
 \node at (2.1,3.35) {auxiliary labels $\Gamma$};
 \node[rotate=90] at (-0.95,1.5) {auxiliary labels $\Gamma$};
 \node at (3.1,-0.3) {$R$};
 \node[anchor=east] at (-0.15,1.7) {$W$};
 \node at (3.1,1.7) {$W\!\times\!R$};
 \draw[->,thick] (4.55,1.5)--(6.15,1.5)
   node[midway,above,align=center] {pull back\\entry colors};
 \draw (6.5,0) rectangle (11.1,3.0);
 \fill[blue!25] (6.8,1.2) rectangle (10.8,2.2);
 \draw[blue!65!black,thick] (6.8,1.2) rectangle (10.8,2.2);
 \node at (8.8,3.35) {existing positions $\Omega$};
 \node at (8.8,1.7) {inherited orbit $\Delta$};
 \node[align=center] at (8.8,0.55)
   {$|\Delta|\leq\theta|\Omega|$\\same hierarchy and bounds};
\end{tikzpicture}
\caption{The matrix is solved inside an existing action. The picture
is schematic; the exact fiber count is~\eqref{eq:graph-rectangle-size}.
The matrix answer is projected back before the parent continues.}
\label{fig:rectangle}
\end{figure}

\subsection{A rectangle supplied by retained rows}

Now suppose the large action is on single labels. Think of the
positions as $m$ rows, each with the same set of within-row
coordinates. The row-normal-form theorem supplies independent
even row permutations. It also describes the remaining local
variation on the coordinates inside a row.

If this local variation is transitive, the row subroutine of
Theorem~\ref{thm:transitive-gat-terminal} uses polynomially many
calls on one row, together with $\exp(O((\log s)^2))$ additional
work. The terminal must match entire families of row types and
their common residual choices; merely sorting the rows would
not suffice.

If the local variation is intransitive, its orbits define columns. We keep the
row permutations and recurse on the column action. Suppose a
later column action has $q$ natural blocks and $q\leq m$. Couple
its even permutations diagonally to the same permutations of
$q$ selected rows. Above a pair of distinct labels this gives a
grid of equal cells. The rectangle uses $|W||R|$ cells, whereas
the current grid contains $mq$ such cells, up to a common fiber
factor. Thus the corresponding query has relative size
\[
 \frac{|W||R|}{mq}\leq\frac{q}{4m}\leq\frac14.
\]
Lemma~\ref{lem:diagonal-carrier} proves that these are actual
subgroup-orbit calls and that their answers project to the full
required rectangle coset. A bounded number of parity cases is
sufficient.

The diagonal subgroup is used only to solve the matrix. Its index
in the parent group is not enumerated: the parent continues under
the lifted matrix answer, not under this auxiliary subgroup.

\subsection{The auxiliary query and the parent continuation}

Algorithm~\ref{alg:auxiliary} gives the order of these operations.
Its input includes a frozen copy $\mathcal S_0$ of the parent state
before the aggregation choices. In particular, $\Gamma$ and $q$ refer
to the original label set, even when the coherent configuration uses
only $C\subseteq\Gamma$. The phase parameters $N$ and $\lambda$ are
defined in Section~\ref{sec:graph-algorithm}.

The procedure first computes auxiliary answers and extends them to
overcosets on all of $\Gamma$. It then checks the original colors
under their lifts. This order separates two kinds of progress: the
auxiliary query uses a smaller inherited domain, and the parent
continuation uses a smaller interval bound. Bounds learned inside
the auxiliary query are not transferred to the parent.

\begin{pseudocode}{Solve an auxiliary graph, then resume its parent}{alg:auxiliary}
\textbf{Input:} The frozen parent $\mathcal S_0$, its $q$ labels
$\Gamma$, and corresponding coherent configurations $C_X,C_Y$.\\
\textbf{Output:} A list of exact parent continuation cosets covering
this aggregation branch.
\begin{tabbing}
\hspace{1.6em}\=\hspace{1.4em}\=\hspace{1.4em}\=\hspace{1.4em}\=\hspace{1.4em}\=\kill
\pc \textbf{procedure} $\proc{AuxiliaryReduction}(\mathcal S_0,C_X,C_Y)$\\
\pc \>\textbf{if} $|C_X|\ne|C_Y|$ \textbf{then return} the empty list\\
\pc \>$\mathcal E\gets$ an empty list of overcosets on $\Gamma$\\
\pc \>\textbf{if} $q\leq\lambda$ \textbf{then}\\
\pc \>\>$\mathcal I\gets\Solve$ on the full colored-pair structures $C_X^2,C_Y^2$\\
\pc \>\>Extend $\mathcal I$ using the outer color classes, fixed anchors, and parity\\
\pc \>\>Append the resulting nonempty overcoset to $\mathcal E$\\
\pc \>\textbf{else}\\
\pc \>\>\textbf{for each} source--target choice in the first-rectangle construction \textbf{do}\\
\pc \>\>\>\textbf{if} it gives a balanced partition \textbf{then}\\
\pc \>\>\>\>Append its full-$\Gamma$ overcoset to $\mathcal E$; \textbf{continue}\\
\pc \>\>\>Align the two rectangles; \textbf{skip} this choice if side sizes differ\\
\pc \>\>\>Represent the rectangle on an actual inherited orbit $\Delta$\\
\pc \>\>\>Pull matrix-entry colors back to $\Delta$ in both instances\\
\pc \>\>\>Use $\Continue$ to obtain its complete permitted rectangle answer\\
\pc \>\>\>\textbf{if} this answer is empty \textbf{then continue}\\
\pc \>\>\>Project to the left side; extend over $\Gamma\setminus W$ and impose parity\\
\pc \>\>\>Append the resulting full-$\Gamma$ overcoset to $\mathcal E$\\
\pc \>$\mathcal A\gets$ an empty list of exact parent answers\\
\pc \>\textbf{for each} $E\in\mathcal E$ \textbf{do}\\
\pc \>\>Append $\proc{LiftAndContinue}(\mathcal S_0,E)$ to $\mathcal A$\\
\pc \>\textbf{return} the list $\mathcal A$
\end{tabbing}
\end{pseudocode}

The small-configuration call uses the action induced by all
permutations of $C_X$ on ordered pairs; it has at most $q^2$ positions.
Lemma~\ref{lem:exact-configuration} proves the bound for its extension.
For a large configuration, the inherited representation is either the
cross-pair orbit or the row--column diagonal orbit described above.
If the diagonal action supplies only one parity class of rectangle
maps, query both classes by twisting the target matrix, as in
Lemma~\ref{lem:diagonal-carrier}. This uses at most two inherited calls
and does not require an odd physical lift. Project the entire returned
coset, including its representative and generators, and retain the
source--target alignment.

Here $\proc{LiftAndContinue}(\mathcal S_0,E)$ lifts $E$ to a coset
$H\tau$ in the original parent action. It replaces the selected
interval bound by the minimum of its old value and $\beta q$,
invokes $\Continue$ under $H$ on $(x,y^{\tau^{-1}})$, and
right-multiplies the returned answer by $\tau$. Thus it checks the
original coloring constraints while retaining the full lifted coset.
All other parent bounds remain unchanged.
The rectangle subgroup is used only for the query on $\Delta$.
The resulting continuation cosets are kept as a list: their union
for one aggregation branch need not itself be a coset. Only after
all aggregation branches have been processed does
$\proc{CombineExact}$ combine the exhaustive family by the coset
rule in Section~\ref{sec:model}. It returns $\varnothing$ if every
answer is empty.

\section{The algorithm}
\label{sec:graph-algorithm}

This section gives the general procedure, which places no restriction
on the group. The recursion for groups in $\Gd$ in
Section~\ref{sec:degree} reuses its steps.
We organize the recursion into phases. A phase starts on $N$
positions. Choose fixed constants
\begin{equation}
 \frac13<\alpha<\frac12,\qquad 2\alpha<\gamma<1,
 \qquad \lambda=N^\alpha.
 \label{eq:scale-constraints}
\end{equation}
An inherited call remains in this phase until its active domain
has size at most $N^\gamma$. Only then may it start a fresh phase.
The values of $N$ and $\lambda$ remain fixed throughout the phase.
An auxiliary configuration on $q\leq\lambda$ labels is an allowed
exception: it starts a fresh call on at most $q^2\leq N^{2\alpha}$
positions, already below the same threshold.

\subsection{The recursive state}

The state contains the active action, the two colorings, the
retained partition hierarchy with its interval bounds, and,
when present, the independent row action and its invariant
column partition. Origin labels on intervals record which old
interval is refined by a new one. A counter $h\in\{0,1,2\}$ records
the remaining row-construction events before a charged reset.
The analysis also follows
$\sigma$; no maximization over subgroup chains is performed by
the algorithm.

At a fresh intransitive instance, initialize the hierarchy and
interval bounds on each original orbit before imposing any
conditional coloring constraints. Process a largest original
orbit first. Every later call inherits the state of the original
orbit containing it. Initializing a new hierarchy after each
conditional restriction would not justify the running-time bound.

Algorithm~\ref{alg:main-recursion} distinguishes the initialization
of a fresh phase from an inherited call. All state changes in
$\Continue$ retain the phase size $N$. The only new phases are its
threshold call, the small-configuration call in
Algorithm~\ref{alg:auxiliary}, and the row calls described below.
The faithful-action and joint-palette convention of
Section~\ref{sec:model} applies at each of these boundaries.

\begin{pseudocode}{The phase recursion}{alg:main-recursion}
\textbf{Input:} Two colorings and their allowed relabelings; inherited
calls also receive the state described above.\\
\textbf{Output:} The complete permitted isomorphism coset.
\begin{tabbing}
\hspace{1.6em}\=\hspace{1.4em}\=\hspace{1.4em}\=\hspace{1.4em}\=\kill
\pc \textbf{procedure} $\Solve(\Omega,G,x,y)$\\
\pc \>Use the faithful action and jointly compressed colors on $\Omega$\\
\pc \>\textbf{if} $|\Omega|\leq N_0$ \textbf{then return} the answer by direct enumeration\\
\pc \>Set $N\gets|\Omega|$, $\lambda\gets N^\alpha$, and $h\gets2$\\
\pc \>Initialize each original orbit's hierarchy and interval bounds\\
\pc \>\textbf{return} $\Continue(\mathcal S)$ for the resulting state $\mathcal S$\\[3pt]
\pc \textbf{procedure} $\Continue(\mathcal S)$\\
\pc \>\textbf{if} $|\Omega|\leq N^\gamma$ \textbf{then return} $\Solve(\Omega,G,x,y)$\\
\pc \>\textbf{if} $G$ is intransitive \textbf{then return} the inherited orbit-chain answer\\
\pc \>\textbf{if} the active column floor has been entered \textbf{then}\\
\pc \>\>Clear the row record and floor mode; set $h\gets2$ against the incoming descent\\
\pc \>Expose the next primitive quotient above any retained column partition\\
\pc \>\textbf{if} enumeration or a product refinement applies \textbf{then}\\
\pc \>\>Apply that reduction and solve its children with $\Continue$\\
\pc \>\>\textbf{return} their exact combined answer\\
\pc \>Normalize the single alternating factor, with at most two cosets\\
\pc \>\textbf{if} the normalized group is intransitive \textbf{then}\\
\pc \>\>\textbf{return} the combined inherited orbit-chain answers for these cosets\\
\pc \>Run $\proc{GiantStep}$ immediately in each coset, absorbing its representative\\
\pc \>\textbf{return} the exact combination of these answers\\[3pt]
\pc \textbf{procedure} $\proc{GiantStep}(\mathcal S,\Gamma)$\\
\pc \>Put $q\gets|\Gamma|$ for the supplied alternating label set\\
\pc \>\textbf{if} $q>\lambda$ \textbf{then}\\
\pc \>\>Recover the one-label or two-label map for the physical kernel orbits\\
\pc \>\>\textbf{if} this map is natural and (no rows are retained, or $q$ exceeds their degree)\\
\pc \>\>\>\textbf{then return} $\proc{RowStep}(\mathcal S,\Gamma)$\\
\pc \>Freeze $\mathcal S_0\gets\mathcal S$; compute local certificates and ordered comparisons\\
\pc \>$\mathcal A\gets$ an empty list of exact answers\\
\pc \>\textbf{for each} aggregation branch from these certificates \textbf{do}\\
\pc \>\>\textbf{if} it supplies large known symmetry \textbf{then}\\
\pc \>\>\>Append its exact answer from inherited half-domain calls to $\mathcal A$\\
\pc \>\>\textbf{else if} it supplies a balanced full-$\Gamma$ overcoset $E$ \textbf{then}\\
\pc \>\>\>Append $\proc{LiftAndContinue}(\mathcal S_0,E)$ to $\mathcal A$\\
\pc \>\>\textbf{else}\\
\pc \>\>\>Extend $\mathcal A$ by the list $\proc{AuxiliaryReduction}(\mathcal S_0,C_X,C_Y)$\\
\pc \>\textbf{return} $\proc{CombineExact}(\mathcal A)$
\end{tabbing}
\end{pseudocode}

The orbit-chain operation processes a largest original orbit first
and carries its exact answer into the remaining queries. The primitive
step always uses an actual quotient of the current action. An
enumerative reduction includes the bounded alternating-degree cases;
a product reduction refines the stored intervals. At a column floor,
the active domain has entered a proper column. The algorithm drops
the row record and the fixed-floor dispatcher mode in place, then uses
the general dispatcher on the same inherited hierarchy. The old column
partition remains a tower level, but no longer acts as a dispatcher floor.
Section~\ref{sec:analysis}
charges the corresponding domain or hierarchy decrease. These
structural steps are the dispatcher operations of
Section~\ref{sec:frozen-band}.

The single-factor normalization has index at most two above the bounded
exceptional degrees. Its preimage is normal. If that preimage is
intransitive, its equal physical orbits have size at most half the
parent domain, and the orbit-chain answers already have a physical
charge. Otherwise its immediate $\proc{GiantStep}$ is part of the
same operation: no independent same-domain recursive call is charged
for that normalization. A second normalization of the same unchanged
constituent is unnecessary, since its entire image is already
$\Alt(\Gamma)$. The constant coset factor is bundled with the next
charged giant operation or orbit descent in the analysis.

Child states use the following carrier policy. An independent certificate
or rectangle child that may lose the carrier clears its row record and
floor mode, and sets $h=2$ against its already established physical or
symmetry decrease.
A genuine hierarchy refinement may likewise fund a reset if it discards
the record and floor mode. Entry into one proper retained column clears
both and resets in place, against the preceding factor-two descent.
The first largest orbit
under a full quotient preimage retains both the carrier and $h$; later
conditional queries lie in other original orbits of size at most half
the parent domain and may clear and reset there. A cap contraction or
a coordinate change alone does not reset $h$. These updates belong to
the incoming child operation, before its potential is evaluated.

For the remaining branches, Algorithm~\ref{alg:local-certificate}
and its lockstep comparisons supply the input to aggregation. The
large-symmetry branch uses Lemma~\ref{lem:large-symmetry-constant-cert}.
For $q>\lambda$, Corollary~\ref{cor:physical-support-thresholds}
computes the action on the actual kernel orbits and its map from
$\Omega$. The row test uses this physical action: a natural
constituent somewhere in the hierarchy alone would not suffice.
When the physical action is natural and no sufficiently large row
action is retained, we put it in row form before aggregation. This
changes coordinates on the same state and keeps the current phase.

All matrix or configuration subroutines finish and return their
complete cosets before the parent continues. Their internal
branching is therefore an additive cost, not an additional
multiplier on the later continuation.

\subsection{Why the cases cover every large action}

For $q>\lambda$ and sufficiently large $N$, we have
$\binom q3>N$. Also $q$ exceeds the logarithmic transitive-kernel
threshold. The support classification therefore gives precisely
the one-label and two-label cases. An action with transitive
kernel on the whole active domain is excluded at this scale.

In the natural case, the row partition is already an interval
of the retained hierarchy. Indeed, a selected Johnson
constituent has $\binom qa\geq q$ physical blocks, all fixed
by the giant kernel. The support classification gives exactly
$q$ kernel orbits, so there are exactly $q$ physical blocks
and $a=1$. Normalizing the rows does not insert an uncharged
new level.

The row size is at most $N^{1-\alpha}$. Since
$1-\alpha<2\alpha<\gamma$, the row terminal makes only fresh
calls below the phase threshold. In an intransitive row
variation, the number of columns is also at most
$N^{1-\alpha}$. Any later column giant of degree
$q>\lambda$ must be natural, since even its unordered-pair
action would require $\binom q2>N^{1-\alpha}$ columns.

\subsection{The row and column step}

Algorithm~\ref{alg:row-step} is used when a natural action is too
wide for a fresh call on its complete pair structure. The normal
form makes even row permutations independent of the within-row
coordinate. If the remaining local variation is transitive, the
row terminal combines exact answers on individual rows. Otherwise
its orbits give a column partition that every allowed relabeling
preserves, and the algorithm continues above that partition.

\begin{pseudocode}{Normalize rows and retain columns}{alg:row-step}
\textbf{Input:} An inherited state whose physical kernel orbits have
a natural action on $q>\lambda$ labels; no retained rows have degree
at least $q$.\\
\textbf{Output:} The complete isomorphism coset of the same state.
\begin{tabbing}
\hspace{1.6em}\=\hspace{1.4em}\=\hspace{1.4em}\=\kill
\pc \textbf{procedure} $\proc{RowStep}(\mathcal S,\Gamma)$\\
\pc \>Set $h\gets h-1$ for this initial construction or enlargement\\
\pc \>Construct the independent-row normal form and its within-row variation $L$\\
\pc \>\textbf{if} $L$ is transitive on a row \textbf{then}\\
\pc \>\>Compute all needed row types and local isomorphism cosets using $\Solve$\\
\pc \>\>\textbf{return} the full row-terminal answer from these tables and the normal form\\
\pc \>Let $D_1,\ldots,D_t$ be the within-row orbits of $L$\\
\pc \>Make column $j$ the union of the copies of $D_j$ in all rows\\
\pc \>Refactor the old hierarchy by splitting and reordering its interval factors\\
\pc \>Inherit every interval bound, clamped by its new interval ratio\\
\pc \>Retain the degree-$q$ row action and the resulting column partition\\
\pc \>\textbf{return} $\Continue(\mathcal S)$ in the same phase
\end{tabbing}
\end{pseudocode}

The row-terminal operation in this algorithm is
Theorem~\ref{thm:transitive-gat-terminal}. Its input tables contain
complete local cosets, not only row-type names. The operation imposes
one common residual choice across the rows, together with the required
parity, and lifts the surviving representative and all stabilizer
generators. After the row tables are computed, this takes
$\exp(O((\log N)^2))$ additional work and makes no further calls
on a larger domain. Each row call has size at most $N^{1-\alpha}$,
below the fresh-phase threshold. The column refactoring is the
explicit operation in Lemma~\ref{lem:normal-transpose}; it preserves
the old interval bounds and origin labels.

\subsection{Why a row action is enlarged at most once}

Suppose the retained row degree is $m>\lambda$ and a later
natural column action has degree $q>m$. The old row factor
and the new column factor are distinct actual intervals of
the same hierarchy. To see why this persists through inherited calls,
write the carrier quotient as $G/M=\Alt(m)\times Q$. A full restriction
above the column floor has quotient $\Alt(m)\times Q'$. Its first
largest orbit contains all rows over a column orbit, and its faithful
image has quotient $\Alt(m)\times(Q'/Z)$ for an appropriate normal
$Z\triangleleft Q'$, as proved in
Section~\ref{subsec:carrier-inheritance}. Thus the full row factor
survives any sequence of these restrictions.

The retained natural ratio $m$ lies below the column floor. The newly
selected natural ratio $q$ lies above it, because its kernel contains
the full column kernel. Their whole constituent actions are
$\Alt(m)$ and $\Alt(q)$, so their valid clamped caps are exactly
$m$ and $q$. They cannot be two descriptions of the same interval.
All remaining interval ratios have
product at most
\begin{equation}
 \frac{N}{mq}<N^{1-2\alpha}<N^\alpha=\lambda.
 \label{eq:promotion-product}
\end{equation}
The whole current action consequently has alternating width
at most $q$: the two exposed factors have bounds $m,q$, and
all other factors are smaller than $\lambda$.

After constructing independent rows of degree $q$, every later
large quotient in a subgroup or quotient of this action has
degree at most $q$. It can therefore be handled by a diagonal
rectangle. Refactoring rows and columns splits and reorders
existing interval factors; it does not reset their bounds.
Lemma~\ref{lem:one-carrier-promotion} gives the precise
inherited-state argument.

There are two occasions on which the row action may be lost.
A certificate or rectangle child may restrict to a subgroup
that no longer contains it; that child has already paid for a
domain decrease or a decrease in $\sigma$. Also, entering one
proper column gives a domain decrease of at least a factor of
two. At that point we continue directly with the general
dispatcher on the inherited state. We do not start a fresh
phase merely because the row description has changed.

\subsection{Correctness}

The following proof separates coverage of isomorphisms from
the running-time accounting. Termination and the quantitative
bound are established in Section~\ref{sec:analysis}.

\begin{lemma}[Exact recursive output]
\label{lem:main-correctness}
Assuming exact answers to the recursive calls, every invocation
of $\Continue$ returns exactly the allowed relabelings taking
its first coloring to its second.
\end{lemma}

\begin{proof}
The orbit chain rule intersects the constraints one orbit at
a time, always retaining the full feasible coset. Enumerating
an actual quotient partitions the allowed relabelings into
exhaustive cosets; product refinement only changes the
description of the same action.

The local tests are equivariant by
Lemma~\ref{lem:main-local-test}. Consequently every parent
isomorphism is covered by an aggregation branch. The
large-symmetry branch returns the full answer using known
automorphisms and smaller kernel-orbit calls. A partition,
configuration, or rectangle branch returns an overcoset of
the parent isomorphisms in that branch. Its lift can contain
extra relabelings, but the continuation tests the original
colors, so none can survive without being an actual
isomorphism.

For each auxiliary call, projection of the full answer is
exact. In the pair case this follows from the equal-fiber
construction. In the row case it follows from the diagonal
rectangle lemma. Parity is included in both the group and
the coset representative. The row terminal also returns a
full coset; it verifies the common residual choice across
all row types.

Changing coordinates or retaining a column partition loses
no relabeling. Restrictions above the column partition are
lifted by full preimages. Finally, combining an exhaustive
family of exact branch answers returns exactly their union,
which is the isomorphism coset of the current instance.
\end{proof}

An auxiliary subgroup used to realize a rectangle is never
substituted for the parent group. Similarly, the original
giant-parent action is retained while constructing an
aggregation branch. Intermediate anchor choices are branch
conditions, and the final overcoset is lifted to the original
parent action. This avoids an unjustified general group
intersection and is needed for the coverage argument above.

\section{Running time}
\label{sec:analysis}\label{sec:runtime}

The main issue is a call that keeps almost all current
positions. Three types of such calls occur, and they are
paid for differently. A local-test call consumes the
symmetry measure. A genuine partition refinement consumes
hierarchy rank. A restriction from an auxiliary graph
consumes the alternating-width bound of an existing
interval. We combine these with domain size in one
potential that remains attached to the recursive instance.

\subsection{The potential}

We analyze transitive inherited states and their ordered orbit forests.
Expand an intransitive conditional query into its transitive orbit calls,
including any further splitting caused by earlier conditional answers.
These domains are disjoint subsets of the original query domain. Until
that expansion, the original transitive orbit's resources are upper-bound
credits; we do not evaluate $\sigma$ on an intransitive action. Every
eventual child is an actual subgroup-orbit restriction of that original
action and inherits its tower.

For a transitive state on $s$ positions, put $x=\log(s+2)$ and retain
\begin{align}
 \rho&=\sum_i\bigl(2\lfloor\log_2b_i\rfloor-1\bigr),
 &\rho&\leq2\log_2s,\label{eq:main-rho}\\
 Q&=\sum_i(\log b_i)\log\max\{c_i,2\},
 &Q&\leq(\log s)^2.\label{eq:main-Q}
\end{align}
Only nontrivial interval ratios occur in the first sum.
Splitting a ratio into at least two nontrivial factors
decreases $\rho$. Restricting to an inherited orbit cannot
increase it. Splitting or restricting intervals also cannot
increase $Q$, because the logarithms of the interval ratios
add and every child inherits its old width bound.

The counter $h$ follows the policy in Section~\ref{sec:graph-algorithm}.
It starts at two, decreases at an initial row construction and its
possible single enlargement, and increases by at most two at a charged
reset. The one-promotion argument ensures that it never becomes negative.
We use
\begin{equation}
 \Phi=A\bigl(x^2+\rho x+\sigma x+Q\bigr)+Bh.
 \label{eq:main-potential}
\end{equation}
The weighted majorant is
\begin{equation}
 W(\mathcal S)=(s+2)^D\exp\bigl(\Phi(\mathcal S)\bigr).
 \label{eq:main-weight}
\end{equation}
Here $A,B,D$ are fixed constants chosen in
Section~\ref{sec:parameters}. At a phase root,
$\log W=O((\log N)^2)$. The counter pays bounded parity choices in
row constructions that have no immediate domain decrease; the
polynomial prefactor pays additive work on unbalanced orbit chains.

\subsection{Charging the recursive calls}

Fix an absolute exponent $a\geq1$ bounding the polynomial numbers of
local and auxiliary calls, the structural work, and the exponents in
small-quotient enumeration. Work within a phase may be polynomial in
the cover degree $N$, rather than the active degree $s$. At every
internal node, however,
\begin{equation}
 s>N^\gamma\quad\Longrightarrow\quad
 N^a\leq \exp(a_*x),\qquad a_*:=a/\gamma.
 \label{eq:main-phase-active}
\end{equation}
In contrast, a same-domain configuration or rectangle continuation has
at most $q^a$ choices, after increasing this fixed $a$. They come from
bounded tuples, anchors, and orbital choices on the $q$ labels.
Polynomial-$N$ construction of certificate tables is additive work;
it is not a multiplier on such a continuation. All bounded
single-factor normalization indices are included in these multiplicity
bounds, or in the constant $\mu_0$ for a row-construction event.

Take $1/2<\theta<1$ as in Section~\ref{sec:parameters}, and put
$c_\theta=\tfrac12\log(1/\theta)>0$. For $s\geq6$ we have
$s\geq2/\sqrt\theta$, whence
\begin{equation}
 x-\log(\theta s+2)\geq c_\theta,
 \qquad x^2-\log^2(\theta s+2)\geq c_\theta x.
 \label{eq:main-domain-payment}
\end{equation}
This pays affected local-test calls, rectangle calls, large-symmetry
calls, and entries into proper columns. An extra unaffected local-test
call decreases $\sigma$ by $\kcert$, giving a decrease at least
$\kcert x$ in $\sigma x$, even if its domain is unchanged.

A small primitive quotient of degree $b$ is an actual quotient of the
whole acting group. Its kernel orbits have size at most $s/b$, and
its index is at most $\exp(a(\log b)^2)$. Uniformly for $s\geq b\geq2$,
\begin{equation}
 \frac{s+2}{s/b+2}\geq\frac{b+2}{3}\geq b^{1/3},
 \qquad
 x^2-\log^2(s/b+2)\geq\tfrac13x\log b.
 \label{eq:main-enumeration-payment}
\end{equation}
The first ratio is minimized at $s=b$, and the second inequality in
that chain is the arithmetic--geometric mean inequality.

For a product exponent $r\geq2$, the actual coordinate refinement
decreases $\rho$ by at least $r-1$. Its outer index is at most
$2^r r!$, by \cite[Theorem~1.1]{Maroti}. Since $r\leq\log_2s$,
\[
 \log(2^r r!)\leq r\log(2r)\leq4(r-1)x.
\]
This index is paid by $\rho x$. For $r=1$, the at-most-two
normalization cosets are processed with their next giant step or
their already smaller normal orbits, as
specified in Algorithm~\ref{alg:main-recursion}; no separate potential
decrease is asserted for that normalization alone.
An ordinary interval refinement with $r$ new factors and no product
index satisfies the same estimate with that index equal to one.

A configuration or rectangle answer changes the selected
interval bound from at least $q$ to at most $\beta q$.
If that interval has ratio $b$, its contribution to $Q$
drops by at least
\begin{equation}
 (\log b)\log(1/\beta)\geq c_\beta\log q,
 \qquad c_\beta:=\log(1/\beta)>0.
 \label{eq:main-cap-payment}
\end{equation}
The old cap is at least $q$ because the selected alternating section
belongs to that whole constituent; the new cap bounds its entire
restricted action. We choose the bounded cutoff so that $\beta q\geq2$
whenever this charge is used. This pays the actual $q^{O(1)}$
continuation multiplicity.
It is not necessary to claim a new block level here:
a natural action can restrict to a primitive Johnson
action on the same domain, leaving $\rho$ unchanged.

\begin{center}
\small
\begin{tabular}{@{}p{0.30\linewidth}p{0.29\linewidth}p{0.31\linewidth}@{}}
\toprule
Recursive operation & Multiplicity bound & Potential decrease \\
\midrule
Affected test or rectangle & $N^a$ & $Ac_\theta x-2B$ \\
Visible unaffected test & $N^a$ & $A\kcert x-2B$ \\
Product refinement, $r\geq2$ & $N^a2^r r!$ & $A(r-1)x-2B$ \\
Primitive enumeration & $N^a\exp(a(\log b)^2)$ & $\tfrac A3x\log b-2B$ \\
Auxiliary-graph continuation & $q^a$ & $Ac_\beta\log q-2B$ \\
Row initialization or enlargement & $\mu_0$ & $B$ \\
\bottomrule
\end{tabular}
\end{center}

The table allows the worst possible increase $2B$ from an admissible
counter reset, even in rows where no reset is needed. An entry counts
branches; within each branch an intransitive restriction is expanded
into a disjoint orbit forest. The polynomial factors in the product
and enumeration rows are harmless upper bounds. The row-construction
constant $\mu_0$ includes its bundled parity factor, and this event
decreases $h$ without resetting it.

The rectangle query and the original-domain continuation are separate
children. The query is paid by its physical decrease and the
continuation by~\eqref{eq:main-cap-payment}. The complete auxiliary
coset is assembled before any parent continuation is made. Hence the
internal leaves of an auxiliary query do not multiply the number of
parent continuations.

\subsection{Unequal orbits and changes of coordinates}

The largest-orbit-first rule is also needed structurally.
Before any coloring constraint is imposed in that chain,
the first orbit receives the unmodified row action.
Its faithful image retains the row factor, as proved in
Section~\ref{subsec:carrier-inheritance}. Later conditional
queries lie in the other original orbits, each of size
at most $s/2$, and may discard the row action with this
already paid decrease.

The orbit chain rule gives an additive sum over disjoint domains.
If their sizes are $s_1,\ldots,s_t$ with $t\geq2$ and
$\sum_i s_i\leq s$, then for an integer $D\geq3$,
\begin{equation}
 \sum_i(s_i+2)^D\leq(s+1)^D+3^D,
 \qquad
 (s+2)^D-(s+1)^D-3^D\geq\tfrac D2s^{D-1}
 \quad(s\geq6).
 \label{eq:main-orbit-slack}
\end{equation}
For a fixed $t$, convexity maximizes the sum at
$(s-t+1,1,\ldots,1)$. Its maximum decreases with $t$, since
$4^D\geq2\cdot3^D$, proving the first inequality. For the second,
use $(s+2)^D-(s+1)^D\geq Ds^{D-1}$ and
$3^D\leq(D/2)s^{D-1}$. Pruned branches only increase this slack.

Each child potential is at most the original orbit's credit. This is
immediate for the first largest child; for a later child,
\eqref{eq:main-domain-payment} also pays any increase of $h$.
Consequently~\eqref{eq:main-orbit-slack}, multiplied by the parent
exponential credit, absorbs additive polynomial work even along
successive one-point losses. We take $D>a/\gamma+2$; then at every
internal phase size the displayed slack exceeds $N^a$.

Row and column refactoring splits and permutes old
interval factors. It preserves their origin labels and
does not increase $\rho,Q,$ or $\sigma$.
Equation~\eqref{eq:promotion-product} permits at most
one enlargement while a row action is retained. The
bounded counter $h$ is reset only after a charged
domain decrease, symmetry decrease, or actual primitive
refinement. The slack in that charge pays the constant
reset. Entering a column is a change of dispatcher on
the inherited state, not an additional same-domain
recursive call.

\subsection{The phase recurrence}

We can now complete the running-time analysis. Every fresh leaf has at most
$N^\gamma$ positions.
Ordinary inherited calls stop at that threshold.
An auxiliary configuration contributes at most
$N^{2\alpha}$ positions. A transitive-variation row
terminal calls only on rows of size at most
$N^{1-\alpha}$. Both exponents are smaller than
$\gamma$ by~\eqref{eq:scale-constraints}. The terminal's
additional work is $\exp(O((\log N)^2))$.

Let $F$ be the monotone worst-case cost for compressed fresh inputs,
put $M_N=\lceil N^\gamma\rceil$, and define
\begin{equation}
 E_N=\exp\bigl(c_T(\log(N+2))^2\bigr),\qquad
 K_N=E_N\bigl(1+F(M_N)\bigr),
 \label{eq:main-phase-toll}
\end{equation}
where $c_T$ bounds the row terminal's extra work. After increasing
$a$, all structural work and all fresh configuration or row calls at
one node have total cost at most $N^aK_N$. These fresh calls are
terminal oracle work for the current phase; their newly initialized
potentials are not inherited by this phase. An ordinary threshold
leaf has cost at most $K_N$ and keeps its old potential only as credit
on its incoming edge.

\begin{lemma}[Weighted phase induction]
\label{lem:main-phase-bound}
With the fixed constants chosen in Section~\ref{sec:parameters}, a
transitive inherited state $\mathcal S$ costs at most
$K_NW(\mathcal S)$, including its fresh oracle calls. The same bound
applies to a conditional orbit forest using the original transitive
orbit's inherited credit. In particular, a fresh phase costs
$\exp(O((\log N)^2))(1+F(M_N))$.
\end{lemma}

\begin{proof}
First assume that fresh calls of degree at most $M_N<N$ terminate.
Every charged transition in the table decreases $\Phi$ by a fixed
positive amount after the counter updates. A pure orbit transition
does not increase $\Phi$ and strictly decreases the positive integer
domain size. Normalization is bundled with its next giant operation,
and coordinate changes or floor-entry changes are performed in place.
Thus an infinite inherited path would either have infinitely many
uniform positive potential decreases, contradicting $\Phi\geq0$, or
eventually consist only of strict domain decreases. Each node has
finitely many children, so the phase terminates. Induction on the
fresh degree, starting at $N_0$, proves termination of all calls.

We may therefore induct over the finite phase computation. At a
threshold leaf, $W\geq1$ and the claimed bound follows from
\eqref{eq:main-phase-toll}. At a charged node, group the inherited
children into at most $J$ classes from the table, where $J$ is an
absolute constant. Within one branch, the prefactors of its disjoint
orbit children sum to at most $(s+2)^D$. For a class of multiplicity
$\mathcal M$ and minimum potential decrease $\Delta\Phi$, its total
child weight is consequently at most
\[
 \mathcal M\exp(-\Delta\Phi)W(\mathcal S).
\]
The inequalities in Section~\ref{sec:parameters} make this at most
$W(\mathcal S)/(8J)$ for every class, including any bundled parity
factor. In particular,
\begin{equation}
 \sum_j W(\mathcal S_j)\leq\tfrac12W(\mathcal S),
 \qquad W(\mathcal S)\geq2N^a.
 \label{eq:main-charged-induction}
\end{equation}
The second inequality follows from $s>N^\gamma$ and the choice of
$D$, after enlarging the absolute base size. The induction hypotheses
and the additive node toll now give
\[
 T_N(\mathcal S)
 \leq K_N\left(N^a+\sum_jW(\mathcal S_j)\right)
 \leq K_NW(\mathcal S).
\]
Here $T_N$ includes the costs of the fresh calls through the factor
$K_N$. The auxiliary children and parent continuations appear in the
same sum; their recursive costs are not multiplied.

At a pure orbit node, use the original transitive orbit's resource
credit until the conditional action has been expanded into its ordered
orbit forest. Every child potential is at most that credit, and
\eqref{eq:main-orbit-slack} supplies
\[
 W(\mathcal S)-\sum_jW(\mathcal S_j)\geq N^a.
\]
This closes the same induction without demanding a fixed-fraction
decrease on its largest child. The polynomial work of expanding the
forest is included in $N^a$; it has at most $s$ nonempty leaf domains.

At a phase root, $\rho=O(\log N)$,
$\sigma=O(\log(N+2))$, $Q\leq(\log N)^2$, and $h\leq2$.
Thus $W\leq\exp(C_0(\log(N+2))^2)$ for a fixed $C_0$.
For a fresh intransitive input, initialize each original orbit before
conditional restrictions and sum its root weight. The disjoint-domain
prefactors give the same bound, with polynomial initialization work
absorbed in $C_0$. Multiplication by $E_N$ adds $c_T$ to the constant
in the quadratic logarithmic exponent and proves the phase estimate.
\end{proof}

Consequently,
\begin{equation}
 F(N)\leq \exp(C(\log(N+2))^2)
       \bigl(1+F(\lceil N^\gamma\rceil)\bigr)
 \label{eq:main-recurrence}
\end{equation}
above an absolute base size. Fix $\widehat\gamma$ with
$\gamma<\widehat\gamma<1$ and enlarge that base size so that
$\log(\lceil N^\gamma\rceil+2)\leq
\widehat\gamma\log(N+2)$ whenever the recurrence is used.
The logarithmic contributions then sum to at most
\[
 C(\log(N+2))^2\sum_{j\geq0}\widehat\gamma^{2j}
 =\frac{C}{1-\widehat\gamma^2}(\log(N+2))^2.
\]
The additive constants and polynomial input costs fit
within the same bound. Together with
Lemma~\ref{lem:main-correctness}, this proves the following
bound for the general procedure. Its constants are fixed in
Section~\ref{sec:parameters}.

\begin{proposition}[The general procedure]
\label{prop:general-procedure}
For an arbitrary group given by generators, the procedure
$\Solve$ of Algorithm~\ref{alg:main-recursion} solves string
isomorphism on $N$ positions in time
$\exp(O((\log(N+2))^2))$, times a polynomial in the explicit
input description. It returns either the empty set or one
isomorphism and generators for the entire color-preserving
subgroup.
\end{proposition}

Section~\ref{sec:degree} uses
Proposition~\ref{prop:general-procedure} only on instances with at
most $d^2$ positions, where $d$ is the degree parameter of
Theorem~\ref{thm:main-string}.

\subsection{Choice of parameters}
\label{sec:parameters}

We collect the choices here so that numerical cutoffs
do not obscure the algorithm. The proof only needs
fixed constants satisfying the following conditions.

\begin{enumerate}[label=(\roman*)]
\item The test size $\kcert$ is above the absolute
threshold in the hereditary-budget proof. The bounded
ideal cutoff satisfies $q_0\geq4\kcert+2$ and exceeds the bounded
exceptional degrees and support thresholds. Enlarge it as needed for
the coherent-configuration cap bounds and so that $\beta q_0\geq2$.
\item The scale exponents satisfy
$1/3<\alpha<1/2$ and $2\alpha<\gamma<1$.
These imply $1-\alpha<\gamma$ and
$1-2\alpha<\alpha$.
\item The coherent-domain fraction satisfies $0<\delta\leq1/4$,
the dominant-class threshold satisfies $2/3<\eta<1$,
and
\[
 \max\{1/2,1-\delta,1-\eta\delta,\eta\}<\beta<1.
\]
Choose $\theta<1$ above $1/2$ and
$2\eta(1-\eta)$, and enlarge $q_0$ so that
$2\eta(1-\eta)q/(q-1)\leq\theta$ for every $q\geq q_0$.
Put $c_\theta=\tfrac12\log(1/\theta)$ and
$c_\beta=\log(1/\beta)$.
\item Fix $a\geq1$ large enough for all multiplicities and node tolls
specified above, and put $a_*=a/\gamma$. Also require
$q_0!\leq\exp(a(\log2)^2)$. A primitive quotient has degree $b\geq2$,
so this makes the enumeration test $\exp(a(\log b)^2)$ accept every
single alternating factor of degree $m\leq q_0$, whose order is at
most $m!\leq q_0!$. Hence every normalized
single factor, and every giant step, has degree $q>q_0$. Let $J$ be a fixed upper
bound on the number of child classes at a node; the six rows of the
table and their bundled constant factors permit, for example, $J=8$.
Let $\mu_0$ bound the parity multiplicity of a row construction, including
its possible preceding single-factor normalization. Choose
\[
 B\geq\log(8J\mu_0),\qquad
 x_0:=\log8,\qquad H_*:=2B+\log(8J).
\]
Then choose one fixed $A$ satisfying
\begin{align}
 Ac_\theta&\geq a_*+H_*/x_0,
 &A\kcert&\geq a_*+H_*/x_0,\label{eq:main-constant-domain}\\
 Ac_\beta&\geq a+H_*/\log q_0,\label{eq:main-constant-cap}\\
 A/3&\geq a+\frac{a_*}{\log2}
                   +\frac{H_*}{x_0\log2},\label{eq:main-constant-enum}\\
 A&\geq4+a_*+H_*/x_0.\label{eq:main-constant-product}
\end{align}
Finally take an integer $D\geq3$ with $D>a/\gamma+2$.
\end{enumerate}

These inequalities pay the complete sums of child costs. At an internal
node $x\geq x_0$, the physical and symmetry rows of the table have
logarithmic multiplicity at most $a_*x$, and
\eqref{eq:main-constant-domain} leaves a margin at least
$H_*-2B=\log(8J)$. The cap row has logarithmic multiplicity at most
$a\log q$ and the same margin by~\eqref{eq:main-constant-cap}.
For enumeration use $\log2\leq\log b\leq x$:
\[
 a(\log b)^2+a_*x+H_*
 \leq\left(a+\frac{a_*}{\log2}
                 +\frac{H_*}{x_0\log2}\right)x\log b.
\]
Thus~\eqref{eq:main-constant-enum} gives that margin after subtracting
the reset cost $2B$. For a product exponent $r\geq2$, use
\[
 \log(2^r r!)+a_*x+H_*
 \leq(4+a_*+H_*/x_0)(r-1)x.
\]
Equation~\eqref{eq:main-constant-product} pays this row as well.
The last row satisfies $\mu_0e^{-B}\leq1/(8J)$. Therefore each class
contributes at most $1/(8J)$ of the parent weight. All choices are
absolute constants, independent of the number of inherited restrictions.

For example, the following values satisfy the numerical
constraints:
\[
 \kcert=64,\qquad \delta=\frac14,\qquad
 \eta=\frac34,\qquad \beta=\frac9{10},\qquad
 \theta=\frac34,\qquad
 \alpha=\frac25,\qquad\gamma=\frac9{10}.
\]
Indeed $2\alpha=4/5<\gamma$, $1-\alpha=3/5<\gamma$,
$1-2\alpha=1/5<\alpha$, and
$1-\eta\delta=13/16<\beta$.
Also $2\eta(1-\eta)=3/8<\theta$.
Choose an absolute base size $N_0$ large enough that,
for every $N\geq N_0$,
\[
 \binom{\lfloor N^\alpha\rfloor}{3}>N,
 \qquad
 \binom{\lfloor N^\alpha\rfloor}{2}>N^{1-\alpha},
 \qquad
 N^\alpha>q_0,
\]
and that the row threshold admits an odd integer
$k>\max\{63,12\log_2 N\}$ with $2k-1\leq N^\alpha$.
Also require $N^\gamma\geq6$, $\lceil N^\gamma\rceil<N$, and
$N^{\gamma D-a}\geq2$ above the base size. These ensure the lower
size assumptions and the second inequality
in~\eqref{eq:main-charged-induction}. The orbit slack already exceeds
$N^a$ because $\gamma(D-1)>a$.
These inequalities hold eventually because
$3\alpha>1$ and a positive power of $N$ dominates
$\log N$. Below $N_0$ use direct enumeration.

\section{Groups with restricted composition factors}
\label{sec:degree}

We now prove Theorem~\ref{thm:main-string}. The plan is the one in the
last part of Section~\ref{sec:overview}. For a group in $\Gd$, two
facts replace the term $x^2$ of the
potential~\eqref{eq:main-potential} by $x\log(d+2)$. Every alternating
action that the recursion meets has at most $d$ labels
(Sections~\ref{sec:degree-caps} and~\ref{sec:degree-primitive}). And
the calls made by the local tests split into $q^{O(1)}$ families with
disjoint domains (Section~\ref{sec:degree-windows}).
Sections~\ref{sec:degree-algorithm}--\ref{sec:degree-proof} give the
recursion and its analysis.

The recursion of this section keeps the inherited hierarchy and caps
of Section~\ref{sec:model}. It does not use rectangles, rows, or
phases. Its only fresh calls are on instances with at most $d^2$
positions. Proposition~\ref{prop:general-procedure}
solves each of them in time $\exp(O((\log(d+2))^2))$, times a
polynomial in the input description.

Throughout this section $d\geq6$ and $G\in\Gd$. This is no loss.
The class $\widehat\Gamma_{d}$ is contained in
$\widehat\Gamma_{d'}$ for $d\leq d'$, and replacing $d$ by
$\max\{d,6\}$ changes $\log(d+2)$ by at most a factor of three. Put
\begin{equation*}
 \ell=\log(d+2)\geq\log8 .
\end{equation*}
The group of every state of the recursion is a subgroup of the input
group. Hence it lies in $\Gd$, by \cite[Lemma~2.4]{GNS18}, which
states that $\Gd$ is closed under subgroups and homomorphic images.

\subsection{Alternating width and caps}
\label{sec:degree-caps}

The caps of Section~\ref{sec:model} bound the alternating width of
each interval constituent. For groups in $\Gd$ this width is at
most $d$, so every cap can start at most $d$.

\begin{lemma}[Width in $\Gd$]
\label{lem:degree-width}
Every group in $\Gd$ has alternating width at most $d$.
\end{lemma}

\begin{proof}
Let $H\in\Gd$, and suppose $K\leq H$ and $L\triangleleft K$ satisfy
$K/L\cong\Alt(t)$ with $t\geq5$. By \cite[Lemma~2.4]{GNS18}, the
group $K/L$ lies in $\Gd$. It is simple, so it is its own only
composition factor, and it embeds in $\Sym(d)$. Hence
$t!/2\leq d!$. If $t\geq d+1$, then $t!/2\geq(d+1)!/2>d!$. Thus
$t\leq d$.
\end{proof}

At the root of the recursion we give interval $i$ the cap
$\min\{d,b_i\}$, in place of the cap $b_i$ of
Section~\ref{sec:model}. This cap is valid. The constituent of an
interval is the action of a block stabilizer on the smaller blocks
inside one block. It is a section of $G$, so each of its sections is
a section of $G$, and Lemma~\ref{lem:degree-width} bounds their
alternating width by $d$. The bound $b_i$ holds because the
constituent acts on $b_i$ points. Later caps are obtained by the
inheritance rules of Section~\ref{sec:model}, by the coordinate rule
below, and by cap contractions. None of these rules increases a cap,
so every cap stays at most $d$. At a root on $s$ positions the ratios
satisfy $b_1\cdots b_t=s$. Since $\log\max\{c_i,2\}\leq\log d$, the
cap sum then satisfies
\begin{equation}
 Q=\sum_i(\log b_i)\log\max\{c_i,2\}\leq(\log s)\log d .
 \label{eq:degree-Q}
\end{equation}
This replaces the bound $(\log s)^2$ of~\eqref{eq:main-Q}. It is the
first place where the degree enters the potential.

\subsection{Primitive quotients}
\label{sec:degree-primitive}

The general procedure enumerates small primitive quotients of order
$b^{O(\log b)}$, where $b$ is the degree of the quotient. For groups
in $\Gd$, Grohe, Neuen, and Schweitzer proved a sharper dichotomy: a
primitive group either has order $b^{O(\log d)}$, or it is a product
of Johnson actions on at most $d$ labels~\cite[Theorem~3.13]{GNS18}.
We use their result in the following form. Its index bound comes from
their proof, and its lower bound on $q$ lets the giant steps use the
cutoff $q_0$ of Section~\ref{sec:parameters}.

\begin{lemma}[Primitive groups in $\Gd$]
\label{lem:degree-primitive}
There are absolute constants $c_1,c_2$ with the following property.
Let $P\in\Gd$ be a primitive group of degree $b$ with
$|P|>b^{c_1\log d+c_2}$. Then $\Soc(P)\cong\Alt(q)^r$ with
$q_0<q\leq d$, and there is an integer $h$ with
\begin{equation*}
 1\leq h\leq q/2,\qquad q>4\log_2\binom qh,\qquad b=\binom qh^{\!r},
\end{equation*}
such that the following hold.
\begin{enumerate}[label=(\roman*)]
\item The points of $P$ can be identified with $r$-tuples of
$h$-subsets of a set of $q$ labels, so that $P$ permutes the $r$
coordinates and each coordinate factor of $\Soc(P)$ acts as $\Alt(q)$
on the $h$-subsets in its coordinate.
\item Let $K\leq\Sym(r)$ be the action of $P$ on the $r$
coordinates. Then $[P:\Soc(P)]\leq2^r|K|\leq2^rd^{r-1}$. In
particular, the index is at most two when $r=1$.
\item Given generators of $P$, one can compute in polynomial time the
socle, the coordinates, and for each coordinate a bijection to the
$h$-subsets of $q$ labels that carries $\Alt(q)$ to its natural
action.
\end{enumerate}
\end{lemma}

\begin{proof}
We first fix the constants. Logarithms in~\cite{GNS18} are binary,
and ours are natural. The first option of
\cite[Theorem~3.13]{GNS18} is the bound
$|P|\leq b^{c_1'\log_2d+c_2'}$ for absolute constants $c_1',c_2'$.
Let $c_1$ be at least $c_1'/\log2$ and at least $1/\log5$. Let $c_2$
be at least $c_2'$ and at least $\log(2\cdot q_0!)/\log5$.

Then $c_1\log d+c_2\geq c_1'\log_2d+c_2'$. So $|P|$ exceeds the bound
in the first option of \cite[Theorem~3.13]{GNS18}, and the second
option holds. Its proof shows
more: $P$ is a primitive group of Type~II or Type~IV in the
classification used there, and the small-order alternatives of
\cite[Lemmas~3.7 and~3.9]{GNS18} fall under the first option. In the
first case,
\cite[Lemma~3.7]{GNS18} therefore shows that $\Soc(P)$ is
permutationally
equivalent to $\Alt(q)$ on $h$-subsets, with $q\leq d$ and
$h\leq q/2$, and that $[P:\Soc(P)]\leq2$. Put $r=1$. In the second
case, \cite[Lemma~3.9]{GNS18} shows that $P\leq P_0\wr K_0$ in the
product action, where $P_0$ is a group of the first case and
$K_0\leq\Sym(r)$ is transitive. The socle $T$ of $P_0$ satisfies
$\Soc(P)=T^r$. Let $K$ be the image of $P$ in $\Sym(r)$. The kernel
of this action is $P\cap P_0^r$, so
\begin{equation*}
 [P:\Soc(P)]=[P:P\cap P_0^r]\,[P\cap P_0^r:T^r]
 \leq|K|\,[P_0:T]^r\leq2^r|K| .
\end{equation*}
This is the computation in the proof of \cite[Lemma~3.9]{GNS18}. The
group $K$ is an image of $P$, so it lies in $\Gd$ by
\cite[Lemma~2.4]{GNS18}. As $d\geq6$, \cite[Lemma~2.2]{GNS18} gives
$|K|\leq d^{r-1}$. This proves (i) and (ii). The proof of
\cite[Theorem~3.13]{GNS18} also shows that $q\leq4\log_2\binom qh$
leads to its first option, so here $q>4\log_2\binom qh$.

Next suppose $q\leq q_0$. The socle factor $\Alt(q)$ is nonabelian
simple, so $q\geq5$ and $b\geq q^r\geq5^r$. By (ii),
\begin{equation*}
 \log|P|\leq\log\bigl(2^rd^{r-1}(q_0!)^r\bigr)
 \leq r\bigl(\log d+\log(2\cdot q_0!)\bigr)
 \leq\frac{\log b}{\log5}\bigl(\log d+\log(2\cdot q_0!)\bigr).
\end{equation*}
By the choice of $c_1$ and $c_2$, this gives
$|P|\leq b^{c_1\log d+c_2}$, contrary to the hypothesis. Hence
$q>q_0$.

For (iii), the order $|P|$ is computable in polynomial time, and
\cite[Theorem~3.13]{GNS18} includes a polynomial-time algorithm that
computes the socle and its coordinate partitions in the second option.
For each coordinate, the constituent of the socle is permutationally
equivalent to $\Alt(q)$ on $h$-subsets with $q>4\log_2\binom qh$.
This is the hypothesis of \cite[Lemma~4.4]{GNS18}, which computes the
required bijection in polynomial time.
\end{proof}

\begin{remark}[Recognition]
\label{rem:degree-recognition}
For Johnson recognition, the general procedure uses a fallback that
enumerates a constituent of order $b^{O(\log b)}$
(Section~\ref{sec:frozen-band}). The degree recursion cannot afford
this. Its potential pays only $b^{O(\log d)}$ for an enumeration. It
uses \cite[Lemma~4.4]{GNS18} instead. The algorithm first compares
$|P|$ with $b^{c_1\log d+c_2}$, and it calls recognition only when
the order is larger. By Lemma~\ref{lem:degree-primitive}, the
hypothesis of \cite[Lemma~4.4]{GNS18} then holds.
\end{remark}

\subsection{Local tests have disjoint windows}
\label{sec:degree-windows}

Lemma~\ref{lem:main-local-test} bounds the number of local calls by a
polynomial in $s$, since a test may make one call at each of up to $s$
window updates. Multiplying the largest call cost by this number would
lose the degree dependence. The updates check disjoint new parts of
the domain, and this gives a stronger statement.

\begin{lemma}[Disjoint windows]
\label{lem:degree-windows}
Let $\mathcal S$ be a transitive state on $s$ positions with group
$G$ and a supplied alternating action on $q$ labels. The recursive
calls made by the endpoint procedures and ordered comparisons of
Section~\ref{sec:graph-tests} split into at most $(\kcert!+4)q^{2\kcert}$
families such that the calls in one family have pairwise disjoint
domains inside $\Omega$. Let such a call have group $H$ and domain
$\Omega'$. Then every point $\alpha\in\Omega'$ satisfies
\begin{equation*}
 |\alpha^H|\leq s/\kcert
 \qquad\text{or}\qquad
 \sigma(H,\Omega',\alpha)\leq\sigma(G,\Omega)-\kcert .
\end{equation*}
\end{lemma}

\begin{proof}
We first treat one endpoint, for a string $x$ and a test set $T$.
Its stage $i$ has the window $W_i$, the increment $V_i$, the group
$A_i$, and the kernel $N_i=\ker(\phi_T|_{A_i})$ of
Section~\ref{sec:certificates}. The update at stage $i$ enumerates the
$|\Alt(\kcert)|$ cosets of $N_i$ in $A_i$. For each coset it makes one
orbit-chain call on $V_i$ under $N_i$. Label each coset by its image
in $\Alt(T)$. The increments are pairwise disjoint, since $V_i$ lies
outside $W_i$ and $W_{i+1}=W_i\cup V_i$. So for one label, the calls
of all stages have disjoint domains. One endpoint therefore gives at
most $|\Alt(\kcert)|=\kcert!/2$ families, however many stages it has.

For an ordered comparison, the partial comparison coset at stage $i$
is a coset of $N_i$. The next call is string isomorphism on the new
window under $N_i$ (Section~\ref{sec:certificates}). So one comparison
makes one call per stage, and these calls form one family.

There are at most $2q^\kcert$ endpoints, for the two strings and all
test sets. There are at most $4q^{2\kcert}$ ordered comparisons,
including self-comparisons. This gives at most
$\kcert!\,q^\kcert+4q^{2\kcert}\leq(\kcert!+4)q^{2\kcert}$ families.

The two bounds on points come from Section~\ref{sec:certificates}.
Let $\Delta$ be the $A_i$-orbit of a point $\alpha\in V_i$. If
$\Delta$ is affected, then $|\alpha^{N_i}|\leq|\Delta|/\kcert\leq
s/\kcert$. If $\Delta$ is an extra orbit of step~4, then
$\sigma(N_i|_\Delta,\Delta)\leq\sigma(G,\Omega)-\kcert$. In both
cases the call has group $N_i$ and $\alpha^{N_i}\subseteq\Delta$. By
the last assertion of Lemma~\ref{lem:budget-inheritance}, the budget
$\sigma(N_i,V_i,\alpha)$ is the budget of the faithful image of $N_i$
on $\alpha^{N_i}$. In the second case this orbit is all of $\Delta$,
since $N_i$ is transitive on $\Delta$. So both bounds hold for the
call, as stated.
\end{proof}

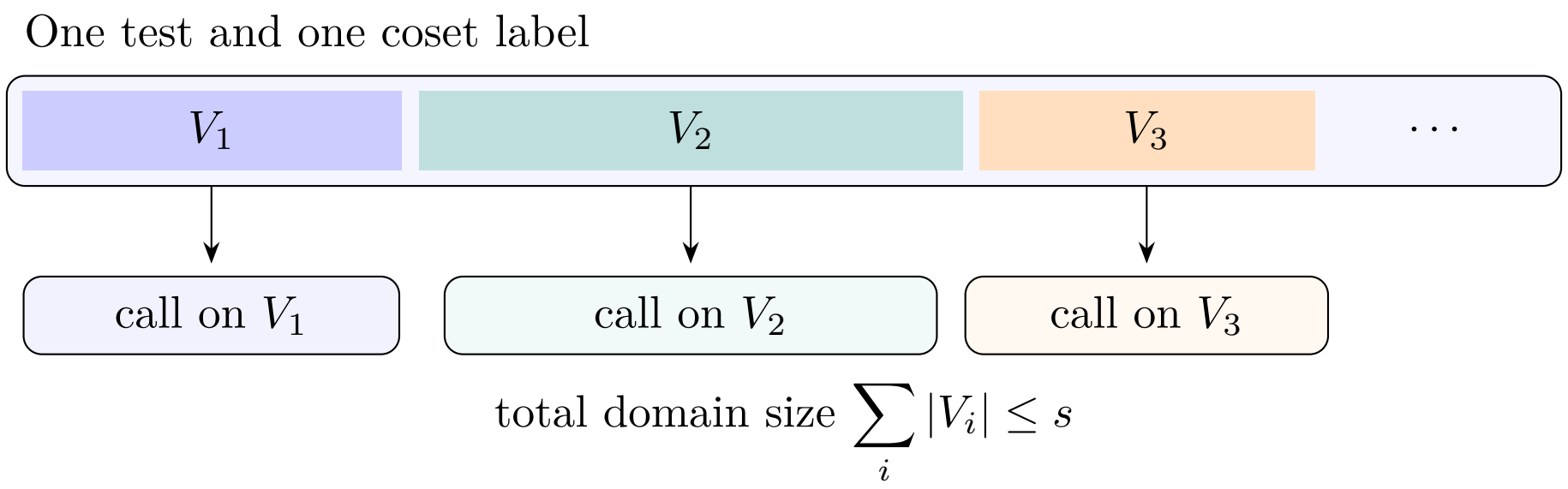
\begin{figure}[t]
\centering
\begin{tikzpicture}[font=\small,>=Stealth]
 \draw[rounded corners,fill=blue!4] (0,0) rectangle (12,0.85);
 \node[anchor=west] at (0,1.2) {One test and one coset label};
 \fill[blue!20] (0.12,0.12) rectangle (3.05,0.73);
 \fill[teal!25] (3.18,0.12) rectangle (7.38,0.73);
 \fill[orange!25] (7.51,0.12) rectangle (10.1,0.73);
 \node at (1.58,0.43) {$V_1$};
 \node at (5.28,0.43) {$V_2$};
 \node at (8.8,0.43) {$V_3$};
 \node at (11.05,0.43) {$\cdots$};
 \draw[->] (1.58,0) -- (1.58,-0.6);
 \draw[->] (5.28,0) -- (5.28,-0.6);
 \draw[->] (8.8,0) -- (8.8,-0.6);
 \node[draw,rounded corners,fill=blue!5,minimum width=2.9cm,minimum height=0.6cm] at (1.58,-1)
 {call on $V_1$};
 \node[draw,rounded corners,fill=teal!5,minimum width=3.8cm,minimum height=0.6cm] at (5.28,-1)
 {call on $V_2$};
 \node[draw,rounded corners,fill=orange!5,minimum width=2.8cm,minimum height=0.6cm] at (8.8,-1)
 {call on $V_3$};
 \node at (6,-1.9) {total domain size $\displaystyle\sum_i|V_i|\leq s$};
\end{tikzpicture}
\caption{The calls of one endpoint with one coset label have disjoint
domains, so together they weigh at most as much as one call on the
whole domain. There are only $q^{O(1)}$ such families over all tests
and comparisons.}
\label{fig:degree-windows}
\end{figure}

Figure~\ref{fig:degree-windows} shows the families. The calls are
made one after another, and their groups depend on earlier answers.
This does not matter for the count: only their domains enter the
weight below.

\subsection{The degree recursion}
\label{sec:degree-algorithm}

A state of the degree recursion has the form of a state of
Section~\ref{sec:graph-algorithm}, without rows, columns, the counter
$h$, or a phase size. It holds the active domain $\Omega$, the group,
the two colorings, and the retained hierarchy with its caps. We keep
the covering action of the input group, as inherited calls do in
Section~\ref{sec:graph-algorithm}. Passing to the faithful image on
$\Omega$ would change none of the quantities used below. The
hierarchy consists of partitions of $\Omega$. A section of an image
of a group is a section of the group, so the caps stay valid. The
budget $\sigma$ is defined through the faithful image, by
Lemma~\ref{lem:budget-inheritance}.

An intransitive state is created from a transitive state by a coset
enumeration, a normalization, a local test, the large-symmetry
branch, or a parent continuation. It keeps the partitions of that
transitive state, with their stored ratios and caps. This includes an
inserted refinement and a cap lowered by the continuation. When the
orbit chain passes to one orbit, the child intersects the stored
partitions with that orbit and clamps each cap by its new ratio, as in
Section~\ref{sec:model}. No ratio increases in this step. At the root,
the state stores one hierarchy for each original orbit, and each
orbit-chain child uses the hierarchy of the original orbit that
contains it.

Algorithm~\ref{alg:degree-recursion} gives the recursion. The
constants $c_1,c_2$ are those of Lemma~\ref{lem:degree-primitive},
and $N_0$ is the base size of Section~\ref{sec:parameters}. The
primitive quotient is selected as in the general procedure: it is a
quotient of the whole current group, compatible with the hierarchy
(Section~\ref{sec:frozen-band}).

\begin{pseudocode}{The degree recursion}{alg:degree-recursion}
\textbf{Input:} Two colorings and their allowed relabelings
$G\in\Gd$, with $d\geq6$ fixed for the whole computation; inherited
calls receive a state.\\
\textbf{Output:} The complete permitted isomorphism coset.
\begin{tabbing}
\hspace{1.6em}\=\hspace{1.4em}\=\hspace{1.4em}\=\hspace{1.4em}\=\kill
\pc \textbf{procedure} $\proc{DegreeIsomorphism}(\Omega,G,x,y)$\\
\pc \>\textbf{if} $d\geq|\Omega|$ \textbf{then return} $\Solve(\Omega,G,x,y)$\\
\pc \>Compress the colors of $x,y$ jointly\\
\pc \>Initialize each original orbit's hierarchy, with caps $\min\{d,b_i\}$\\
\pc \>\textbf{return} $\proc{DegreeContinue}(\mathcal S)$ for the resulting state $\mathcal S$\\[3pt]
\pc \textbf{procedure} $\proc{DegreeContinue}(\mathcal S)$\\
\pc \>\textbf{if} $|\Omega|\leq N_0$ \textbf{then return} the answer by direct enumeration\\
\pc \>\textbf{if} $G$ is intransitive \textbf{then return} the inherited orbit-chain answer\\
\pc \>Expose a primitive quotient $P$ of degree $b$, compatible with the hierarchy\\
\pc \>\textbf{if} $|P|\leq b^{c_1\log d+c_2}$ \textbf{then}\\
\pc \>\>Enumerate the cosets of its kernel and solve the children with $\proc{DegreeContinue}$\\
\pc \>\>\textbf{return} their exact combined answer\\
\pc \>Compute $\Soc(P)\cong\Alt(q)^r$ and its coordinates (Lemma~\ref{lem:degree-primitive})\\
\pc \>\textbf{if} $r\geq2$ \textbf{then}\\
\pc \>\>Enumerate the cosets of the preimage of $\Soc(P)$; insert the $r$ coordinate partitions\\
\pc \>\>Solve the children with $\proc{DegreeContinue}$; \textbf{return} their exact combined answer\\
\pc \>Normalize the single alternating factor, with at most two cosets\\
\pc \>\textbf{if} the normalized group is intransitive \textbf{then}\\
\pc \>\>\textbf{return} the combined inherited orbit-chain answers for these cosets\\
\pc \>Run $\proc{DegreeGiantStep}$ immediately in each coset, absorbing its representative\\
\pc \>\textbf{return} the exact combination of these answers
\end{tabbing}
\end{pseudocode}

The procedure $\proc{DegreeGiantStep}(\mathcal S,\Gamma)$ is
$\proc{GiantStep}$ of Algorithm~\ref{alg:main-recursion} with three
changes.
\begin{enumerate}
\item The test $q>\lambda$ and the row step that it guards are
omitted.
\item In Algorithm~\ref{alg:auxiliary}, the test $q\leq\lambda$ is
omitted and its first branch is always used. So every configuration
is solved by $\Solve$ on the colored-pair structures $C_X^2,C_Y^2$,
with at most $q^2$ positions, and extended as in
Lemma~\ref{lem:exact-configuration}. The rectangle branch is never
used.
\item Every inherited call uses $\proc{DegreeContinue}$ in place of
$\Continue$. These are the calls in $\proc{CheckNewWindow}$ and in
the ordered comparisons, the half-domain calls of the large-symmetry
branch, and the call in $\proc{LiftAndContinue}$.
\end{enumerate}
Thus the only fresh calls of the degree recursion are the
configuration calls of the second change and, at the root, the call
of $\Solve$ when $d\geq|\Omega|$.

The steps of $\proc{DegreeContinue}$ need three more comments. In an
enumeration, each child is the kernel of the action on the $b$ blocks,
with the target twisted by one coset representative. Its orbits have
at most $s/b$ points. In a product with $r\geq2$, the children use
the preimage of $\Soc(P)$. Their hierarchy replaces the selected
interval by the $r$ coordinate intervals, each of ratio $\binom qh$.
Each new interval receives the cap $\min\{c,q\}$, where $c$ is the
old cap of the selected interval. This cap is valid: the constituent
of a coordinate interval is $\Alt(q)$ on $h$-subsets, whose
alternating width is $q$, and it is a section of the old constituent.
The index of the enumeration is at most $2^rd^{r-1}$ by
Lemma~\ref{lem:degree-primitive}. Replacing the actual coordinate
group $K$ by all of $\Sym(r)$ would lose this bound, since $r!$ can
be much larger than $d^{r-1}$ when $r>d$.

For a single factor, Lemma~\ref{lem:degree-primitive} identifies the
blocks with the $h$-subsets of a set $\Gamma$ of $q$ labels. This
gives the epimorphism of the normalized group onto $\Alt(\Gamma)$ and
the Johnson block map used by $\proc{GiantStep}$. The normalized
group is normal in the current group. If it is intransitive, its
orbits have equal size, hence at most $s/2$ points. The label count
satisfies
\begin{equation*}
 q_0<q\leq d .
\end{equation*}
The lower bound is part of Lemma~\ref{lem:degree-primitive}. The
upper bound holds because $\Alt(q)$ is an image of a subgroup of $G$,
so Lemma~\ref{lem:degree-width} applies. In particular, every cutoff
that $\proc{GiantStep}$ needs holds, since these cutoffs were absorbed
into $q_0$ in Section~\ref{sec:parameters}.

\begin{lemma}[Exact output of the degree recursion]
\label{lem:degree-correctness}
Assuming exact answers to its recursive calls, every invocation of
$\proc{DegreeContinue}$ returns exactly the allowed relabelings taking
its first coloring to its second. The same holds for
$\proc{DegreeIsomorphism}$.
\end{lemma}

\begin{proof}
The orbit chain, the enumeration of an actual quotient, the product
refinement, and the single-factor normalization are steps of
$\Continue$. The product step enumerates exhaustive cosets of a normal
subgroup, and the coordinate partitions are invariant under that
subgroup, so no relabeling is lost. The proof of
Lemma~\ref{lem:main-correctness} shows that these steps are exact.

The procedure $\proc{DegreeGiantStep}$ uses only branches of
$\proc{GiantStep}$ without rows or rectangles: the local tests, the
large-symmetry branch, the balanced partitions, and the
small-configuration branch of Algorithm~\ref{alg:auxiliary}. The
proof of Lemma~\ref{lem:main-correctness} shows that these branches
are exact, and it does not use the size of $q$ relative to $\lambda$.
The configuration answers are exact by
Proposition~\ref{prop:general-procedure}. Finally,
$\proc{DegreeIsomorphism}$ either calls $\Solve$, which is exact by
the same proposition, or runs the orbit chain on the original orbits.
\end{proof}

\subsection{A potential defined at every point}
\label{sec:degree-potential}

In Section~\ref{sec:analysis} the potential was defined on transitive
states, and an intransitive query used the credit of its transitive
parent. Here one local query may contain both affected orbits and
orbits with a smaller budget, and it may cover most of the domain.
Neither the domain size nor the budget need drop for the query as a
whole. Every point of the query does make progress, by
Lemma~\ref{lem:degree-windows}. So we define the potential on every
state, as a maximum over points.

Let $\mathcal S$ be a state with domain $\Omega$ of size $s$ and group
$G$. For $\alpha\in\Omega$ put
\begin{equation*}
 \psi_\alpha(\mathcal S)=\log(|\alpha^G|+2)+\sigma(G,\Omega,\alpha),
 \qquad
 \psi(\mathcal S)=\max_{\alpha\in\Omega}\psi_\alpha(\mathcal S),
\end{equation*}
with $\sigma(G,\Omega,\alpha)$ from
Definition~\ref{def:hereditary-budget}. For a transitive state,
$\psi(\mathcal S)=x+\sigma(G,\Omega)$ with $x=\log(s+2)$. Let
$\rho(\mathcal S)$ and $Q(\mathcal S)$ be given
by~\eqref{eq:main-rho} and~\eqref{eq:main-Q}, using the stored ratios
and caps. In this section, $A$ denotes a new absolute constant, fixed
in Section~\ref{sec:degree-constants}, and the weight of a state is
\begin{equation}
 \Psi(\mathcal S)=A\bigl[(\psi(\mathcal S)+\rho(\mathcal S))\ell
                    +Q(\mathcal S)\bigr],
 \qquad
 W(\mathcal S)=(s+2)^3\exp\bigl(\Psi(\mathcal S)\bigr).
 \label{eq:degree-weight}
\end{equation}
The budget enters with the factor $\ell$, and the cap sum $Q$ enters
with no factor. By~\eqref{eq:degree-Q}, all terms are
$O(\log(s+2)\,\ell)$ at a root.

\begin{remark}[The budget is not bounded in terms of $d$]
\label{rem:degree-example}
The improvement does not come from replacing every $\log N$ by
$\log d$. The budget $\sigma$ can still be of order $\log N$ when $d$
is fixed. Fix $k\geq\kcert$, let $H=\F_2^k\rtimes\Alt(k)$ act on
$\F_2^k$ by translations and coordinate permutations, and let
$G=H^t$ act coordinatewise on $\Omega=(\F_2^k)^t$. Every composition
factor of $G$ is cyclic of order two or isomorphic to $\Alt(k)$, so
$G\in\widehat\Gamma_k$. Let $H_i$ be the subgroup in which the first
$i$ factors are replaced by their translation subgroups. Each step
$H_i\geq H_{i+1}$ has quotient $\Alt(k)$, and $H_{i+1}$ is transitive,
since it contains all translations. The action is faithful, so each
step is visible in the sense of~\eqref{eq:visible-mark}. Hence
\begin{equation*}
 \sigma(G,\Omega)\geq kt=\log_2|\Omega| .
\end{equation*}
So the potential must allow order $\log s$ units of budget. What
changes is the price of one unit. In Section~\ref{sec:analysis} it is
$O(\log N)$ in the exponent, because the local calls are counted one
by one. Lemma~\ref{lem:degree-windows} reduces it to $O(\log d)$.
\end{remark}

\begin{lemma}[Inherited potential]
\label{lem:degree-monotone}
Let $\mathcal S'$ be obtained from a state $\mathcal S$ by passing to
a subgroup $H\leq G$ and a union $\Omega'\subseteq\Omega$ of
$H$-orbits, with the stored hierarchy described in
Section~\ref{sec:degree-algorithm}, possibly refined. Then
$\psi(\mathcal S')\leq\psi(\mathcal S)$,
$\rho(\mathcal S')\leq\rho(\mathcal S)$, and
$Q(\mathcal S')\leq Q(\mathcal S)$. Suppose moreover that
$\mathcal S$ is transitive with $s\geq6$.
\begin{enumerate}[label=(\roman*)]
\item If every $\alpha\in\Omega'$ satisfies $|\alpha^H|\leq\theta s$
or $\sigma(H,\Omega',\alpha)\leq\sigma(G,\Omega)-\kcert$, then
$\psi(\mathcal S')\leq\psi(\mathcal S)-c_\theta$.
\item If every $H$-orbit in $\Omega'$ has at most $s/b$ points, where
$2\leq b\leq s$, then
$\psi(\mathcal S')\leq\psi(\mathcal S)-\tfrac13\log b$.
\end{enumerate}
\end{lemma}

\begin{proof}
Let $\alpha\in\Omega'$. Then $\alpha^H\subseteq\alpha^G$. By the last
assertion of Lemma~\ref{lem:budget-inheritance},
$\sigma(H,\Omega',\alpha)$ is the budget of the faithful image of $H$
on $\alpha^H$, so it equals $\sigma(H,\Omega,\alpha)$. A chain for $H$
at $\alpha$, prefixed by the unmarked step $G\geq H$, is a chain for
$G$ at $\alpha$. Definition~\ref{def:hereditary-budget} therefore
gives $\sigma(H,\Omega',\alpha)\leq\sigma(G,\Omega,\alpha)$. Hence
$\psi_\alpha(\mathcal S')\leq\psi_\alpha(\mathcal S)$. The stored
ratios do not increase when they are intersected with an orbit, and
the caps are clamped. A refinement does not increase $\rho$ or $Q$,
as shown after~\eqref{eq:main-Q}. This proves the first claims.

For~(i), let $\alpha\in\Omega'$. If $|\alpha^H|\leq\theta s$, then
\eqref{eq:main-domain-payment} gives
$\log(|\alpha^H|+2)\leq x-c_\theta$, while
$\sigma(H,\Omega',\alpha)\leq\sigma(G,\Omega)$. Otherwise
$\sigma(H,\Omega',\alpha)\leq\sigma(G,\Omega)-\kcert$ and
$\log(|\alpha^H|+2)\leq x$. Since $c_\theta<1<\kcert$, in both cases
$\psi_\alpha(\mathcal S')\leq x+\sigma(G,\Omega)-c_\theta
=\psi(\mathcal S)-c_\theta$. For~(ii),
\eqref{eq:main-enumeration-payment} gives
$(s+2)/(s/b+2)\geq b^{1/3}$, so
$\log(|\alpha^H|+2)\leq x-\tfrac13\log b$. The budget term does not
increase, as before.
\end{proof}

\subsection{Charging the recursive calls}
\label{sec:degree-charging}

We now bound the cost of the recursion by the weight. The structure
follows Section~\ref{sec:analysis}, with two differences. The local
calls are counted by families, not one by one. The cost of a giant
step is bounded by a quarter of $K$ times its weight, so that the at
most two giant steps after a normalization fit in the weight of their
parent.

Let $M$ be the length of the explicit input. Every state is
represented in the covering action of the input, so its structural
operations take time polynomial in $M$ (Section~\ref{sec:model}).
Fix an absolute exponent $a\geq1$ such that a giant step makes at
most $q^a$ families of local-test and large-symmetry calls, at most
$q^a$ parent continuations, and at most $q^a$ configuration calls.
Such an $a$ exists. Lemma~\ref{lem:degree-windows} bounds the
local-test families. There are polynomially many aggregation branches
in $q$ (Section~\ref{sec:graph-tests}). Each of them makes a constant
number of large-symmetry orbit chains
(Lemma~\ref{lem:large-symmetry-constant-cert}), or at most one
continuation, and at most one configuration call. Since $q>q_0\geq2$,
the constant factors are absorbed by enlarging $a$.

Fix absolute constants $a_0,a_1,c_4$ and put
\begin{equation}
 K=(M+2)^{a_0}\exp(c_4\ell^2),
 \label{eq:degree-K}
\end{equation}
so that the following hold. At a node of $\proc{DegreeContinue}$ that
enumerates $I$ cosets, the work outside recursive calls is at most
$(1+I)^{a_1}K/8$. Nodes without an enumeration have $I=0$, and a
single-factor normalization has $I\leq2$. A state with at most $N_0$
positions is solved within $K/8$. Finally, the work of
$\proc{DegreeGiantStep}$ outside its calls to
$\proc{DegreeContinue}$ is at most $K/8$, including its configuration
calls. For the last claim, a configuration call has at most
$q^2\leq d^2$ positions and an explicit description of size
polynomial in $M$. By Proposition~\ref{prop:general-procedure}, it
costs $\exp(O(\ell^2))$, times a polynomial in $M$. There are at most
$q^a\leq\exp(a\ell)$ such calls, and the rest of the giant step is
polynomial in $M$. A large enough $c_4$ therefore pays for them.

\begin{lemma}[Weighted induction for the degree recursion]
\label{lem:degree-induction}
With the constants of Section~\ref{sec:degree-constants}, the degree
recursion terminates. Every call of $\proc{DegreeContinue}$ on a
state $\mathcal S$, other than an intransitive root, costs at most
$KW(\mathcal S)$, and every
call of $\proc{DegreeGiantStep}$ on a state $\mathcal S$ costs at most
$KW(\mathcal S)/4$.
\end{lemma}

\begin{proof}
We first list the children of each kind of node and the weight they
lose. Lemma~\ref{lem:degree-monotone} shows that no child has a larger
$\Psi$ than its parent. Every node below has $s>N_0\geq6$, since
smaller states are solved directly. Within one family of children with
disjoint domains in $\Omega$, the first inequality
of~\eqref{eq:main-orbit-slack} with $D=3$ gives
\begin{equation}
 \sum_j(s_j+2)^3\leq(s+1)^3+27\leq(s+2)^3-1
 \label{eq:degree-slack}
\end{equation}
when there are at least two children. The second inequality uses the
second inequality of~\eqref{eq:main-orbit-slack}. For a single child
with $s_1<s$, the bound~\eqref{eq:degree-slack} holds directly, and
for a single child with $s_1=s$ the prefactor is unchanged.

\emph{Enumeration.} There are $I\leq b^{c_1\log d+c_2}$ children. Each
child has orbits of at most $s/b$ points, and $s\geq b$. By
Lemma~\ref{lem:degree-monotone}(ii), each child has
$\Psi\leq\Psi(\mathcal S)-\tfrac A3\ell\log b$. As $\log d\leq\ell$
and $1\leq\ell$, we have $\log I\leq(c_1+c_2)\ell\log b$, so the
children have total weight at most
\begin{equation*}
 \exp\bigl((c_1+c_2-A/3)\ell\log b\bigr)W(\mathcal S).
\end{equation*}

\emph{Product.} There are $I\leq2^rd^{r-1}$ children, with $r\geq2$.
Since $r\log2\leq(r-1)\log4\leq(r-1)\ell$, we have
$\log I\leq2(r-1)\ell$. Each child replaces one stored ratio $b$ by
$r$ ratios with product $b$. The remark after~\eqref{eq:rho} shows
that this decreases $\rho$ by at least $r-1$. The caps do not grow,
so $Q$ does not grow, and $\psi$ does not grow. The children have
total weight at most
\begin{equation*}
 \exp\bigl((2-A)(r-1)\ell\bigr)W(\mathcal S).
\end{equation*}

\emph{Single factor.} There are at most two cosets. If the normalized
group is intransitive, each child has orbits of at most
$s/2\leq\theta s$ points. Lemma~\ref{lem:degree-monotone}(i) shows that
the two children have total weight at most
$2\exp(-Ac_\theta\ell)W(\mathcal S)$. Otherwise each coset gives a
giant state with the same domain and weight at most $W(\mathcal S)$.

\emph{Giant step.} Each local-test call has
$\Psi\leq\Psi(\mathcal S)-Ac_\theta\ell$. This follows from
Lemma~\ref{lem:degree-windows} and Lemma~\ref{lem:degree-monotone}(i),
since $s/\kcert\leq\theta s$. The same decrease holds for each
large-symmetry chain, since its orbits have at most $s/2$ points. By
\eqref{eq:degree-slack}, the prefactors in one family sum to at most
$(s+2)^3$. A parent continuation keeps the domain, and its group is a
subgroup of $G$. It lowers the selected cap from at least $q$ to at
most $\beta q$, so~\eqref{eq:main-cap-payment} gives
$\Psi\leq\Psi(\mathcal S)-Ac_\beta\log q$. As $q\leq d$, we have
$q^a\leq e^{a\ell}$, and the children have total weight at most
\begin{equation*}
 \Bigl(\exp\bigl((a-Ac_\theta)\ell\bigr)
       +q^{a-Ac_\beta}\Bigr)W(\mathcal S).
\end{equation*}

\emph{Intransitive state.} The orbit chain makes transitive calls on
disjoint orbits, after any conditional splitting. There are at least
two of them, or one with a smaller domain. Their $\Psi$ is at most
$\Psi(\mathcal S)$, so~\eqref{eq:degree-slack} gives total weight at
most $W(\mathcal S)-\exp(\Psi(\mathcal S))\leq W(\mathcal S)-1$.

\emph{Termination.} Each child of an enumeration, a product, an
intransitive normalization, or a giant step has $\Psi$ smaller than
its parent by at least
$\varepsilon=\min\{\tfrac A3\ell\log2,\,A\ell,\,Ac_\theta\ell,\,
Ac_\beta\log q_0\}>0$. Each child of an orbit chain has no larger
$\Psi$ and a smaller domain. A giant state follows a normalization,
and all its children lose at least $\varepsilon$. So along any path of
calls, either infinitely many steps lose $\varepsilon$, which is
impossible since $\Psi\geq0$, or the path ends with orbit-chain steps
only. These strictly decrease the positive integer $s$, so such a tail
is finite. Every node has finitely many children, and the calls of
$\Solve$ terminate by Proposition~\ref{prop:general-procedure}. Hence
the recursion is finite.

\emph{Induction.} We prove the two bounds by induction on the height
of the finite recursion tree. A state with at most $N_0$ positions
costs at most $K/8\leq KW(\mathcal S)$. For a giant step,
Section~\ref{sec:degree-constants} makes the children weigh at most
$W(\mathcal S)/8$. With the induction hypothesis and $W\geq1$, the
cost is at most
\begin{equation*}
 \tfrac K8+\tfrac K8W(\mathcal S)\leq\tfrac K4W(\mathcal S).
\end{equation*}
At an enumeration or a product node, the children weigh at most
$W(\mathcal S)/2$, and
Section~\ref{sec:degree-constants} gives $(1+I)^{a_1}\leq W(\mathcal S)$.
So the cost is at most
$KW(\mathcal S)/8+KW(\mathcal S)/2\leq KW(\mathcal S)$. A single-factor
node costs at most $3^{a_1}K/8$ in its own work, plus either
$KW(\mathcal S)/2$ for intransitive children or $2\cdot
KW(\mathcal S)/4$ for two giant steps. Since $W(\mathcal S)\geq
3^{a_1}$, this is at most $KW(\mathcal S)$. An intransitive state
costs at most $K/8+K(W(\mathcal S)-1)\leq KW(\mathcal S)$.
\end{proof}

\subsection{Choice of the constants}
\label{sec:degree-constants}

Keep $\kcert,\theta,\beta,q_0,N_0$, and $c_\theta,c_\beta$ from
Section~\ref{sec:parameters}. Fix $a,a_0,a_1,c_4$ as in
Section~\ref{sec:degree-charging}, and choose one absolute $A$ with
\begin{equation*}
 A\geq3(a_1+1)(c_1+c_2+2),\qquad
 Ac_\theta\geq a+2,\qquad
 Ac_\beta\geq a+2.
\end{equation*}
We check the conditions used in the proof of
Lemma~\ref{lem:degree-induction}. Recall that $\ell\geq\log8>2$ and
$\log b\geq\log2$.

For a giant step, $\exp((a-Ac_\theta)\ell)\leq e^{-2\ell}\leq1/64$.
Also $q>q_0\geq4$, so $q^{a-Ac_\beta}\leq q^{-2}\leq1/16$. The
children therefore weigh at most $W(\mathcal S)/8$. For an
intransitive normalization, $2e^{-Ac_\theta\ell}\leq1/32$.

For an enumeration, $c_1+c_2-A/3\leq-2$, so the children weigh at
most $e^{-2\ell\log b}W(\mathcal S)\leq W(\mathcal S)/2$. Moreover
$\Psi(\mathcal S)\geq Ax\ell\geq A\ell\log b$, since $s\geq b$. As
$\log(1+I)\leq\log2+(c_1+c_2)\ell\log b\leq(c_1+c_2+1)\ell\log b$,
this gives $(1+I)^{a_1}\leq W(\mathcal S)$.

For a product, $A\geq4$, so the children weigh at most
$e^{-2\ell}W(\mathcal S)\leq W(\mathcal S)/2$. The selected ratio
$b=\binom qh^{r}\geq5^r$ contributes at least $4r-1\geq r-1$ to
$\rho(\mathcal S)$, so $\Psi(\mathcal S)\geq A(r-1)\ell$. As
$\log(1+I)\leq\log2+2(r-1)\ell\leq3(r-1)\ell$, this gives
$(1+I)^{a_1}\leq W(\mathcal S)$.

Finally, $\Psi(\mathcal S)\geq Ax\ell\geq A\log8$ at every node with
$s\geq6$, and this exceeds $a_1\log3$. So $W(\mathcal S)\geq3^{a_1}$
at a single-factor node.

\subsection{Proof of Theorem~\ref{thm:main-string}}
\label{sec:degree-proof}

Replace $d$ by $\max\{d,6\}$. If $d\geq N$,
Proposition~\ref{prop:general-procedure} solves the instance in time
$\exp(O((\log(N+2))^2))$, times a polynomial in $M$. This is within
the claimed bound, since $\log(N+2)\leq\ell$.

Assume $d<N$. By Lemma~\ref{lem:degree-induction} the recursion
terminates, and Lemma~\ref{lem:degree-correctness} shows that it
returns the complete isomorphism coset. For the time, consider an
original orbit with $s$ points and its initialized state. Its ratios
are exact, so~\eqref{eq:main-rho} gives $\rho\leq2\log_2s$, and
\eqref{eq:degree-Q} gives $Q\leq(\log s)\log d$.
Theorem~\ref{thm:hereditary-budget} gives
$\sigma\leq C\log(s+2)$ for an absolute $C$. Hence
\begin{equation*}
 \Psi\leq A\Bigl[\bigl(1+C+2/\log2\bigr)\log(s+2)\,\ell
          +\log(s+2)\,\ell\Bigr]
 \leq C'\log(N+2)\,\ell
\end{equation*}
for an absolute $C'$. If the input group is transitive, the root is
this initialized state, and Lemma~\ref{lem:degree-induction} bounds its
cost by $K(N+2)^3e^{C'\log(N+2)\ell}$. Otherwise the root runs the
orbit chain on the original orbits, with polynomial work of its own.
Each call of this chain is a restriction of the initialized state of
one original orbit, so its weight is at most
$(s_j+2)^3e^{C'\log(N+2)\ell}$ by Lemma~\ref{lem:degree-monotone},
where $s_j$ is the size of its domain. These domains are disjoint, so
by~\eqref{eq:degree-slack} and Lemma~\ref{lem:degree-induction} the
total cost is at most
\begin{equation*}
 \tfrac K8+K(N+2)^3\exp\bigl(C'\log(N+2)\,\ell\bigr)
\end{equation*}
in both cases.
Since $d<N$, we have $\ell\leq\log(N+2)$, so $K\leq(M+2)^{a_0}
\exp(c_4\log(N+2)\,\ell)$. The total is therefore
$\exp(O(\log(N+2)\log(d+2)))$, times a polynomial in $M$. The
replacement of $d$ by $\max\{d,6\}$ changes $\log(d+2)$ by at most a
factor of three. This proves Theorem~\ref{thm:main-string}.
\qed

\section{Graphs of maximum degree \texorpdfstring{$d$}{d}}
\label{sec:degree-graphs}

We now prove Theorem~\ref{thm:main-graph}. Luks reduced isomorphism
of graphs of maximum degree $d$ to string isomorphism for groups in
$\Gd$~\cite{Luks82}. Babai and Luks showed how to make the running
time of this reduction independent of $d$~\cite{BabaiLuks83}. Grohe,
Neuen, and Schweitzer state the result as a polynomial-time Turing
reduction for the decision problem~\cite[Theorem~8.1]{GNS18}.
Theorem~\ref{thm:main-graph} also asks for one isomorphism and for
generators of the automorphism group, so we give the reduction in a
form that returns them.

The idea is Luks's. Fix a root vertex in each graph, and grow a ball
around each root one layer at a time. Keep the set of all
isomorphisms between the two balls. To pass to the next layer, look
at the edges that leave the ball. An isomorphism of the balls extends
to the next layer when it can be made to respect how these edges meet
there. This condition compares two edges at a time, so one coloring of
the pairs of edges records it, and one string query returns all
extensions. Since every query returns all solutions, no separate
search step is needed.

Throughout this section, $X$ and $Y$ are graphs on $n\geq1$ vertices
each, of maximum degree at most $d$. Let $D$ be the larger of $2$ and
the maximum degree of $X$ and $Y$. Then $D\leq\max\{2,d\}$ and
$D\leq n+1$. We write maps on the left, and $h\circ a$ applies $a$
first.

\subsection{Balls around a root}

Fix a vertex $v$ of $X$ and a vertex $w$ of $Y$. For $r\geq0$, let
$L_r(X)$ be the set of vertices of $X$ at distance exactly $r$ from
$v$, and let $B_r(X)=L_0(X)\cup\cdots\cup L_r(X)$. Define $L_r(Y)$ and
$B_r(Y)$ in the same way from $w$. A shortest path from $v$ to a
vertex of $B_r(X)$ stays in $B_r(X)$. So the distance from $v$ to a
vertex of $B_r(X)$ is the same in $X$ and in the subgraph induced on
$B_r(X)$.

Let $I_r$ be the set of isomorphisms from the subgraph of $X$ induced
on $B_r(X)$ to the subgraph of $Y$ induced on $B_r(Y)$ that map $v$
to $w$. Let $A_r$ be the same set with $(Y,w)$ replaced by $(X,v)$.
It is a group of permutations of $B_r(X)$. We use three facts.
\begin{enumerate}[label=(\roman*)]
\item Every $h\in I_r$ keeps the distance from the root, by the
remark on shortest paths. So $h$ maps $L_j(X)$ onto $L_j(Y)$ for
every $j\leq r$, and the restriction of an element of $I_{r+1}$ to
$B_r(X)$ lies in $I_r$.
\item If $h\in I_r$, then $I_r=\{h\circ a:a\in A_r\}$. Indeed,
$h\circ a$ lies in $I_r$ for every $a\in A_r$, and
$h^{-1}\circ h'$ lies in $A_r$ for every $h'\in I_r$.
\item Distances in a graph on $n$ vertices are less than $n$, so
$B_{n-1}(X)$ is the connected component of $v$. Hence $I_{n-1}$ is
the set of isomorphisms from the component of $v$ to the component of
$w$ that map $v$ to $w$, and $A_{n-1}$ is the group of automorphisms
of the component of $v$ that fix $v$.
\end{enumerate}

\subsection{One layer as a string query}

Fix $r\geq0$ and an element $h\in I_r$. Fix also an order on the
vertices of each graph. A \emph{forward neighbor} of a vertex
$u\in L_r(X)$ is a neighbor of $u$ in $L_{r+1}(X)$. There are at most
$D$ of them.

A \emph{slot} is a pair $(u,i)$ with $u\in B_r(X)$ and
$1\leq i\leq D$. It \emph{lies over} $u$. Let $S_X$ be the set of
slots. The slot $(u,i)$ is \emph{live} if $u\in L_r(X)$ and $u$ has at
least $i$ forward neighbors. Its \emph{target} $t_X(u,i)$ is then the
$i$th forward neighbor of $u$ in the fixed order. Every vertex of
$L_{r+1}(X)$ has a neighbor in $L_r(X)$, so it is the target of at
least one live slot. Define $S_Y$, live slots, and $t_Y$ in the same
way for $Y$. The map $h$ carries slots to slots: put
$\hat h(u,i)=(h(u),i)$.

Live slots stand for the edges that leave the ball. We give every
vertex of the ball exactly $D$ slots, live or not, so that every
automorphism of the ball acts on the slots.

\paragraph{The coloring.}
Color a pair $(s,s')$ of slots of $S_X$ as follows. If $s$ or $s'$ is
not live, the color $c_X(s,s')$ is a fixed symbol~$*$. Otherwise it is
the pair of bits that records whether the targets $t_X(s)$ and
$t_X(s')$ are equal, and whether they are adjacent in $X$. Define
$c_Y$ on pairs of slots of $S_Y$ in the same way. There are five
colors.

\paragraph{The group.}
For a group $A$ of permutations of $B_r(X)$, let $W(A)$ be the set of
permutations $f$ of $S_X$ for which there is $a\in A$ such that $f$
maps the slots over $u$ to the slots over $a(u)$, for every
$u\in B_r(X)$. The element $a$ is determined by $f$, and we write
$a=\pi(f)$. The set $W(A)$ is a group, and
$\pi\colon W(A)\to A$ is a homomorphism. It is onto, because the
\emph{lift} $(u,i)\mapsto(a(u),i)$ of $a$ lies in $W(A)$. Its kernel
consists of the permutations that keep the set of slots over each
vertex. So the kernel is a direct product of $|B_r(X)|$ copies of
$\Sym(D)$.

\begin{lemma}
\label{lem:graphs-group}
If $A\in\widehat\Gamma_D$, then $W(A)\in\widehat\Gamma_D$. If a set
$T$ generates $A$, then $W(A)$ is generated by the lifts of the
elements of $T$ and the transpositions of two slots over the same
vertex.
\end{lemma}

\begin{proof}
The composition factors of $W(A)$ are those of the kernel of $\pi$
together with those of the quotient, which is isomorphic to $A$. The
composition factors of the kernel are composition factors of
$\Sym(D)$. Each of these is $\Alt(D)$ or cyclic of order two or
three, with order three only when $D\in\{3,4\}$, so it is isomorphic
to a subgroup of $\Sym(D)$. Hence $W(A)\in\widehat\Gamma_D$.

For the generators, let $f\in W(A)$ and $a=\pi(f)$. The group $A$ is
finite, so $a$ is a product of elements of $T$, and the lift of $a$
is the product of their lifts. Composing $f$ with the inverse of this
lift gives an element of the kernel. The kernel is generated by the
listed transpositions.
\end{proof}

\paragraph{The query.}
Let $W(A_r)$ act on pairs of slots by $f(s,s')=(f(s),f(s'))$. Let $Q$
be the set of all $f\in W(A_r)$ such that
\begin{equation*}
 c_Y\bigl(\hat h(f(s)),\hat h(f(s'))\bigr)=c_X(s,s')
 \qquad\text{for all } s,s'\in S_X .
\end{equation*}
So $Q$ is the set of elements of the group that carry the coloring
$c_X$ of $S_X\times S_X$ to the coloring
$(s,s')\mapsto c_Y(\hat h(s),\hat h(s'))$. Computing $Q$ is a string
isomorphism problem on $|S_X|^2\leq(nD)^2$ positions. The action on
pairs is faithful, so its group is isomorphic to $W(A_r)$.

\begin{lemma}[One layer]
\label{lem:graphs-step}
Let $h\in I_r$, and let $Q$ be as above.
\begin{enumerate}[label=(\alph*)]
\item Let $f\in Q$ and $a=\pi(f)$. There is a unique bijection
$b\colon L_{r+1}(X)\to L_{r+1}(Y)$ with
$b(t_X(s))=t_Y(\hat h(f(s)))$ for every live slot $s$ of $S_X$. The
map $\phi_f$ that equals $h\circ a$ on $B_r(X)$ and $b$ on
$L_{r+1}(X)$ lies in $I_{r+1}$.
\item Every element of $I_{r+1}$ equals $\phi_f$ for some $f\in Q$.
\item Suppose that $(Y,w)=(X,v)$, with the same vertex order, and
that $h$ is the identity. Then $Q$ is a subgroup of $W(A_r)$, and
$f\mapsto\phi_f$ is a homomorphism from $Q$ onto $A_{r+1}$.
\end{enumerate}
\end{lemma}

\begin{proof}
\emph{Part (a).} Put $g=\hat h\circ f$ and $\phi=h\circ a$. Then
$\phi\in I_r$ by fact~(ii), and $g$ is a bijection from $S_X$ to
$S_Y$ that maps the slots over $u$ to the slots over $\phi(u)$. A
pair $(s,s)$ has color $*$ exactly when $s$ is not live. Since $f\in
Q$, a slot $s$ is live if and only if $g(s)$ is live. So $g$ maps the
live slots of $S_X$ onto the live slots of $S_Y$.

Let $s,s'$ be live slots of $S_X$. The first bit of the color of
$(s,s')$ equals the first bit of the color of $(g(s),g(s'))$, so
$t_X(s)=t_X(s')$ if and only if $t_Y(g(s))=t_Y(g(s'))$. Every vertex
of $L_{r+1}(X)$ is the target of a live slot. So $b$ is defined on
$L_{r+1}(X)$, it does not depend on the choice of the slot, and it is
injective. It is onto $L_{r+1}(Y)$, because every vertex there is the
target of a live slot of $S_Y$, and that slot is $g(s)$ for a live
slot $s$ of $S_X$. This also shows that $b$ is unique.

By fact~(i), $\phi$ maps $L_j(X)$ onto $L_j(Y)$ for $j\leq r$. So
$\phi_f$ is a bijection from $B_{r+1}(X)$ to $B_{r+1}(Y)$ that maps
$v$ to $w$ and keeps the layers. It remains to compare edges. Let
$u,u'\in B_{r+1}(X)$.
\begin{itemize}
\item If both lie in $B_r(X)$, then $\phi\in I_r$ gives that $u,u'$
are adjacent if and only if $\phi(u),\phi(u')$ are.
\item If both lie in $L_{r+1}(X)$, choose live slots $s,s'$ with
targets $u,u'$. The second bit of the color of $(s,s')$ equals that of
$(g(s),g(s'))$. So $u,u'$ are adjacent if and only if $b(u),b(u')$
are.
\item Let $u\in L_r(X)$ and $u'\in L_{r+1}(X)$. If they are adjacent,
then $u'=t_X(s)$ for a live slot $s$ over $u$. The slot $g(s)$ is
live, lies over $\phi(u)$, and has target $b(u')$. So $\phi(u)$ and
$b(u')$ are adjacent. Conversely, let $\phi(u)$ and $b(u')$ be
adjacent. Then $b(u')$ is a forward neighbor of $\phi(u)$, so it is
the target of a live slot $s''$ over $\phi(u)$. The slot
$s=g^{-1}(s'')$ is live, and it lies over $u$ because $\phi$ is
injective. We get $b(t_X(s))=t_Y(s'')=b(u')$, so $t_X(s)=u'$ as $b$
is injective. Hence $u$ and $u'$ are adjacent.
\item Let $u\in B_{r-1}(X)$ and $u'\in L_{r+1}(X)$. The distances of
two adjacent vertices from the root differ by at most one. So $u,u'$
are not adjacent, and neither are their images, which lie in
$B_{r-1}(Y)$ and $L_{r+1}(Y)$.
\end{itemize}
Adjacency is symmetric, so these cases cover all pairs.

\emph{Part (b).} Let $\psi\in I_{r+1}$. By facts~(i) and~(ii), its
restriction to $B_r(X)$ is $h\circ a$ for some $a\in A_r$. Let
$u\in L_r(X)$. Since $\psi$ and its inverse keep layers and
adjacency, $\psi$ maps the forward neighbors of $u$ bijectively onto
the forward neighbors of $\psi(u)$. So both vertices have the same
number $m$ of forward neighbors, and there is a permutation $\tau_u$
of $\{1,\ldots,D\}$ that fixes every $i>m$ and such that $\psi$ maps
the $i$th forward neighbor of $u$ to the $\tau_u(i)$th forward
neighbor of $\psi(u)$ for $i\leq m$. Define $f(u,i)=(a(u),\tau_u(i))$
for $u\in L_r(X)$, and $f(u,i)=(a(u),i)$ for the other vertices of
$B_r(X)$. Then $f\in W(A_r)$ and $\pi(f)=a$. A slot $s$ is live if
and only if $\hat h(f(s))$ is live, and then
$t_Y(\hat h(f(s)))=\psi(t_X(s))$. Since $\psi$ is injective and keeps
adjacency on $L_{r+1}(X)$, both bits of every color are kept. So
$f\in Q$, and $\phi_f=\psi$.

\emph{Part (c).} Here $Q$ is the set of elements of $W(A_r)$ that fix
the coloring $c_X$, which is a group. Let $f,f'\in Q$. On $B_r(X)$,
the map $\phi_{f\circ f'}$ is $\pi(f\circ f')=\pi(f)\circ\pi(f')$.
Let $u\in L_{r+1}(X)$ be the target of a live slot $s$. Then $f'(s)$
is live with target $\phi_{f'}(u)$, so
$\phi_f(\phi_{f'}(u))=t_X(f(f'(s)))=\phi_{f\circ f'}(u)$. The map is
onto by part~(b).
\end{proof}

\begin{lemma}
\label{lem:graphs-gamma}
For every $r\geq0$, the group $A_r$ lies in $\widehat\Gamma_D$, and so
does $W(A_r)$.
\end{lemma}

\begin{proof}
The group $A_0$ is trivial. If $A_r\in\widehat\Gamma_D$, then
$W(A_r)\in\widehat\Gamma_D$ by Lemma~\ref{lem:graphs-group}.
Lemma~\ref{lem:graphs-step}(c) shows that $A_{r+1}$ is a homomorphic
image of a subgroup of $W(A_r)$. The class $\widehat\Gamma_D$ is
closed under subgroups and homomorphic
images~\cite[Lemma~2.4]{GNS18}.
\end{proof}

\subsection{Comparing two rooted graphs}

The \emph{rooted comparison} of $(X,v)$ and $(Y,w)$ runs as follows.
Start with $r=0$, the map $h$ that sends $v$ to $w$, and the empty
generating set of $A_0$. Repeat the following for $r=0,\ldots,n-2$.

Compute the set $Q$ of Lemma~\ref{lem:graphs-step} by one string
query. The group is $W(A_r)$, with the generators of
Lemma~\ref{lem:graphs-group}. By Theorem~\ref{thm:main-string}, the
answer is either empty, or it consists of one element $f_0\in Q$ and
generators of the group $K$ of elements of $W(A_r)$ that fix $c_X$.
If it is empty, stop and report that there is no isomorphism.
Otherwise replace $h$ by $\phi_{f_0}$. The group $K$ is the set $Q$
of Lemma~\ref{lem:graphs-step}(c). So the images of its generators
under the homomorphism of that part generate $A_{r+1}$. Reduce them to
at most $n^2$ generators by the standard algorithm~\cite{FHL80}.

\begin{lemma}
\label{lem:graphs-rooted}
If there is an isomorphism from the component of $v$ to the component
of $w$ that maps $v$ to $w$, the rooted comparison returns one,
together with generators of the group of automorphisms of the
component of $v$ that fix $v$. Otherwise it reports that there is
none. It makes at most $n-1$ string queries. Each query has at most
$(nD)^2$ positions, a group in $\widehat\Gamma_D$, and at most
$n^2+nD^2$ generators.
\end{lemma}

\begin{proof}
We show that at the start of round $r$ we have $h\in I_r$ and
generators of $A_r$. This holds for $r=0$. In round $r$,
Lemma~\ref{lem:graphs-step}(a) gives $\phi_{f_0}\in I_{r+1}$, and the
generators of $A_{r+1}$ were justified above. If $Q$ is empty, then
$I_{r+1}$ is empty by Lemma~\ref{lem:graphs-step}(b). Then $I_{n-1}$
is empty by fact~(i), and by fact~(iii) there is no isomorphism of
the required kind. If no round stops, the procedure ends with
$h\in I_{n-1}$ and generators of $A_{n-1}$, and fact~(iii) says that
these are the required objects. The group of each query lies in
$\widehat\Gamma_D$ by Lemma~\ref{lem:graphs-gamma}. Its generators are
at most $n^2$ lifts and fewer than $nD^2$ transpositions.
\end{proof}

\subsection{Proof of Theorem~\ref{thm:main-graph}}

Graphs need not be connected, so we combine rooted comparisons over
the components. Write $c(v)$ for the component of a vertex $v$.

\paragraph{Automorphisms.}
Let $\tau$ be an isomorphism from $c(v)$ to $c(w)$ with
$\tau(v)=w$, for vertices $v,w$ of $X$. Let $\sigma_\tau$ be the
permutation of the vertices of $X$ that equals $\tau$ on $c(v)$,
equals $\tau^{-1}$ on $c(w)$ when $c(w)\neq c(v)$, and is the identity
elsewhere. Every edge lies inside a component, so $\sigma_\tau$ is an
automorphism of $X$.

\begin{lemma}
\label{lem:graphs-aut}
Run the rooted comparison of $(X,v)$ and $(X,w)$ for all vertices
$v,w$ of $X$. The group $\Aut(X)$ is generated by the automorphisms
$\sigma_\tau$ for the returned isomorphisms $\tau$, together with the
returned automorphisms of components, extended by the identity.
\end{lemma}

\begin{proof}
Let $H$ be the group generated by these automorphisms, and let
$g\in\Aut(X)$. The map $g$ permutes the components. We show $g\in H$
by induction on the number of components $c$ with $g(c)\neq c$.

Suppose first that there is none. Then $g$ is the product, over the
components $c$, of the maps that equal $g$ on $c$ and the identity
elsewhere. Fix $c$ and a vertex $v\in c$, and put $w=g(v)$. The
restriction of $g$ is an isomorphism from $c$ to itself that maps $v$
to $w$. So the comparison of $(X,v)$ and $(X,w)$ returns some $\tau$,
and $\sigma_\tau$ equals $\tau$ on $c$ and the identity elsewhere. The
map $\tau^{-1}\circ g$ on $c$ is an automorphism of $c$ that fixes
$v$, so its extension lies in $H$ by
Lemma~\ref{lem:graphs-rooted}. Hence the factor of $g$ at $c$ lies in
$H$.

Now let $g(c)\neq c$ for a component $c$. Take $v\in c$ and $w=g(v)$.
The comparison returns an isomorphism $\tau$ from $c$ to $g(c)$, and
$\sigma_\tau$ swaps $c$ and $g(c)$ and keeps every other component.
Put $g'=\sigma_\tau^{-1}\circ g$. Then $g'(c)=c$. A component $c'$
with $g(c')=c'$ differs from $c$ and from $g(c)$, so $g'(c')=c'$.
Thus $g'$ moves fewer components than $g$, and $g'\in H$ by
induction. Hence $g=\sigma_\tau\circ g'\in H$.
\end{proof}

\paragraph{One isomorphism.}
Go through the components of $X$ in any order, and keep a list of
\emph{matched} components of $Y$. For a component $c$ of $X$, pick
$v\in c$ and run the rooted comparison of $(X,v)$ and $(Y,w)$ for
each vertex $w$ of $Y$ in an unmatched component, until one returns
an isomorphism $\tau$. Then record $\tau$ and mark $c(w)$ as matched.
If no $w$ works, report that $X$ and $Y$ are not isomorphic. If every
component of $X$ is treated, return the union of the recorded maps.

\begin{lemma}
\label{lem:graphs-iso}
If $X$ and $Y$ are isomorphic, this procedure returns an isomorphism.
Otherwise it reports that they are not isomorphic.
\end{lemma}

\begin{proof}
Suppose the procedure returns a map. The recorded maps are
isomorphisms from the components of $X$ onto distinct components of
$Y$. Their union is injective, and both graphs have $n$ vertices, so
it is a bijection and every component of $Y$ is matched. Every edge
lies inside a component, so the union is an isomorphism. In
particular, the procedure returns no map when $X$ and $Y$ are not
isomorphic.

Suppose now that $X$ and $Y$ are isomorphic. An isomorphism maps the
components of $X$ bijectively to isomorphic components of $Y$. So
every isomorphism type occurs equally often among the components of
$X$ and of $Y$. Let $c$ be the component being treated. Every matched
component of $Y$ of the type of $c$ was matched to an earlier
component of $X$ of this type, and $c$ is not one of these. So the
number of matched components of $Y$ of this type is less than the
number of components of $X$ of this type, which is the number of
components of $Y$ of this type. Hence some unmatched component $c'$
of $Y$ is isomorphic to $c$. An isomorphism from $c$ to $c'$ maps $v$ to
some $w\in c'$. By Lemma~\ref{lem:graphs-rooted}, the comparison of
$(X,v)$ and $(Y,w)$ returns an isomorphism. So the procedure does not
report failure at $c$.
\end{proof}

\begin{proof}[Proof of Theorem~\ref{thm:main-graph}]
For $n=0$ there is nothing to do. Let $n\geq1$. By
Lemmas~\ref{lem:graphs-aut} and~\ref{lem:graphs-iso}, the two
procedures decide isomorphism, return an isomorphism when one exists,
and return generators of $\Aut(X)$. Each of them runs at most $n^2$
rooted comparisons. By Lemma~\ref{lem:graphs-rooted}, this is fewer
than $2n^3$ string queries, each with $N\leq(nD)^2$ positions and a
group in $\widehat\Gamma_D$, given by polynomially many generators.
The rest of the work is polynomial in $n$.

We have $D\geq2$, so Theorem~\ref{thm:main-string} applies with $D$ in
place of $d$. Each query takes time
$n^{O(1)}\exp(O(\log(N+2)\log(D+2)))$. Since $D\leq n+1$, we get
$N+2\leq(n+2)^4$ and $\log(N+2)\leq4\log(n+2)$. Since
$D\leq\max\{2,d\}$, we get $D+2\leq(d+2)^2$ and
$\log(D+2)\leq2\log(d+2)$. A factor $n^{O(1)}$ is at most
$\exp(O(\log(n+2)\log(d+2)))$, because $\log(d+2)\geq\log2$. So the
total time is $\exp(O(\log(n+2)\log(d+2)))$, with constants that do
not depend on $d$.
\end{proof}

\medskip
\noindent\textbf{AI Use Statement.} This work resulted from extensive
interaction between the authors and OpenAI Astra, with the authors
providing suggestions and critically checking Astra's responses
through iterative exchanges. OpenAI Astra was also used in writing
and revising the manuscript through detailed prompts from the authors,
with Gemini Argon assisting in some parts. The authors directed the
process and evaluated the proposed ideas and arguments. Lean files
formalizing the main theorem are available on GitHub at
\url{https://github.com/mahdi-jfri/GraphIsomorphismLog2}.

\clearpage
\begingroup
\small\raggedright
\interlinepenalty=10000
\setlength{\bibsep}{4pt}
\bibliographystyle{plainnat}
\bibliography{graph_isomorphism_submission}

@article{Luks82,
  title={Isomorphism of graphs of bounded valence can be tested in polynomial time},
  author={Luks, Eugene M},
  journal={Journal of computer and system sciences},
  volume={25},
  number={1},
  pages={42--65},
  year={1982},
  publisher={Elsevier}
}

@inproceedings{Babai16,
  title={Graph isomorphism in quasipolynomial time},
  author={Babai, L{\'a}szl{\'o}},
  booktitle={Proceedings of the forty-eighth annual ACM symposium on Theory of Computing},
  pages={684--697},
  year={2016}
}

@misc{Babai18,
  author = {L{\'a}szl{\'o} Babai},
  title = {Graph Isomorphism in Quasipolynomial Time},
  year = {2018},
  note = {Author manuscript, version 2.5, November 2, 2018},
  url = {https://people.cs.uchicago.edu/~laci/quasi25.pdf}
}

@misc{BabaiFix17,
  author = {L{\'a}szl{\'o} Babai},
  title = {Graph Isomorphism: January 9, 2017 Update},
  year = {2017},
  note = {Quasipolynomial claim restored; with January 14 addendum linking the corrected recursive call},
  url = {https://people.cs.uchicago.edu/~laci/update.html}
}

@article{H,
  title={Graph isomorphisms in quasi-polynomial time},
  author={Helfgott, Harald Andr{\'e}s and Bajpai, Jitendra and Dona, Daniele},
  journal={arXiv preprint arXiv:1710.04574},
  year={2017}
}

@article{GNS18,
  title={A faster isomorphism test for graphs of small degree},
  author={Grohe, Martin and Neuen, Daniel and Schweitzer, Pascal},
  journal={SIAM Journal on Computing},
  volume={52},
  number={6},
  pages={FOCS18-1--FOCS18-36},
  year={2023},
  publisher={SIAM},
  note={Conference version in FOCS 2018. Theorem and lemma numbers refer to arXiv:1802.04659v3, November 30, 2019}
}

@InProceedings{Wiebking20,
  author =	{Wiebking, Daniel},
  title =	{{Graph Isomorphism in Quasipolynomial Time Parameterized by Treewidth}},
  booktitle =	{47th International Colloquium on Automata, Languages, and Programming (ICALP 2020)},
  pages =	{103:1--103:16},
  series =	{Leibniz International Proceedings in Informatics (LIPIcs)},
  ISBN =	{978-3-95977-138-2},
  ISSN =	{1868-8969},
  year =	{2020},
  volume =	{168},
  editor =	{Czumaj, Artur and Dawar, Anuj and Merelli, Emanuela},
  publisher =	{Schloss Dagstuhl -- Leibniz-Zentrum f{\"u}r Informatik},
  address =	{Dagstuhl, Germany},
  URL =		{https://drops.dagstuhl.de/entities/document/10.4230/LIPIcs.ICALP.2020.103},
  URN =		{urn:nbn:de:0030-drops-125106},
  doi =		{10.4230/LIPIcs.ICALP.2020.103}
}

@article{Neuen20,
  title={Hypergraph isomorphism for groups with restricted composition factors},
  author={Neuen, Daniel},
  journal={ACM Transactions on Algorithms (TALG)},
  volume={18},
  number={3},
  pages={1--50},
  year={2022},
  publisher={ACM, New York, NY}
}

@inproceedings{FHL80,
  title={Polynomial-time algorithms for permutation groups},
  author={Furst, Merrick and Hopcroft, John and Luks, Eugene},
  booktitle={21st Annual Symposium on Foundations of Computer Science (SFCS 1980)},
  pages={36--41},
  year={1980},
  organization={IEEE}
}

@article{Maroti,
  title={On the orders of primitive groups},
  author={Mar{\'o}ti, Attila},
  journal={Journal of Algebra},
  volume={258},
  number={2},
  pages={631--640},
  year={2002},
  publisher={Elsevier}
}

@article{GMP,
  title={Normalizers of primitive permutation groups},
  author={Guralnick, Robert M and Mar{\'o}ti, Attila and Pyber, L{\'a}szl{\'o}},
  journal={Advances in Mathematics},
  volume={310},
  pages={1017--1063},
  year={2017},
  publisher={Elsevier}
}

@article{LM,
  title={Polynomial-time normalizers},
  author={Luks, Eugene M and Miyazaki, Takunari},
  journal={Discrete Mathematics \& Theoretical Computer Science},
  volume={13},
  year={2011},
  publisher={Episciences. org}
}

@article{Du17,
  title={On graph isomorphism problem},
  author={Du, Wenxue},
  journal={arXiv preprint arXiv:1710.09526},
  year={2017}
}

@article{BA21,
  title={Asymmetric coloring of locally finite graphs and profinite permutation groups: Tucker's Conjecture confirmed},
  author={Babai, L{\'a}szl{\'o}},
  journal={Journal of Algebra},
  volume={607},
  pages={64--106},
  year={2022},
  publisher={Elsevier}
}

@article{Holt,
  title={Quotients of permutation groups by nonabelian minimal normal subgroups},
  author={Holt, Derek},
  journal={arXiv preprint arXiv:2405.19940},
  year={2024}
}

@inproceedings{BabaiCanon19,
  title={Canonical form for graphs in quasipolynomial time: preliminary report},
  author={Babai, L{\'a}szl{\'o}},
  booktitle={Proceedings of the 51st Annual ACM SIGACT Symposium on Theory of Computing},
  pages={1237--1246},
  year={2019}
}

@article{KannanBachem79,
  title={Polynomial algorithms for computing the Smith and Hermite normal forms of an integer matrix},
  author={Kannan, Ravindran and Bachem, Achim},
  journal={siam Journal on Computing},
  volume={8},
  number={4},
  pages={499--507},
  year={1979},
  publisher={SIAM}
}

@article{CorneilGotlieb70,
  title={An efficient algorithm for graph isomorphism},
  author={Corneil, Derek G and Gotlieb, Calvin C},
  journal={Journal of the ACM (JACM)},
  volume={17},
  number={1},
  pages={51--64},
  year={1970},
  publisher={ACM, New York, NY, USA}
}

@article{ReadCorneil77,
  title={The graph isomorphism disease},
  author={Read, Ronald C and Corneil, Derek G},
  journal={Journal of graph theory},
  volume={1},
  number={4},
  pages={339--363},
  year={1977},
  publisher={Wiley Online Library}
}

@inproceedings{HopcroftWong74,
  title={Linear time algorithm for isomorphism of planar graphs (preliminary report)},
  author={Hopcroft, John E and Wong, Jin-Kue},
  booktitle={Proceedings of the sixth annual ACM symposium on Theory of computing},
  pages={172--184},
  year={1974}
}

@inproceedings{BabaiLuks83,
  title={Canonical labeling of graphs},
  author={Babai, L{\'a}szl{\'o} and Luks, Eugene M},
  booktitle={Proceedings of the fifteenth annual ACM symposium on Theory of computing},
  pages={171--183},
  year={1983}
}

@techreport{Babai79,
  title={{Monte-Carlo} algorithms in graph isomorphism testing},
  author={Babai, L{\'a}szl{\'o}},
  institution={Universit{\'e} de Montr{\'e}al},
  number={DMS 79-10},
  year={1979},
  url={https://people.cs.uchicago.edu/~laci/lasvegas79.pdf}
}

@inproceedings{BKL83,
  title={Computational complexity and the classification of finite simple groups},
  author={Babai, L{\'a}szl{\'o} and Kantor, William M and Luks, Eugene M},
  booktitle={24th Annual Symposium on Foundations of Computer Science (Sfcs 1983)},
  pages={162--171},
  year={1983},
  organization={IEEE}
}

@article{WeisfeilerLeman68,
  title={A reduction of a graph to a canonical form and an algebra arising during this reduction},
  author={Leman, Andrei and Weisfeiler, Boris},
  journal={Nauchno-Technicheskaya Informatsiya},
  volume={2},
  number={9},
  pages={12--16},
  year={1968}
}

@article{CFI92,
  title={An optimal lower bound on the number of variables for graph identification},
  author={Cai, Jin-Yi and F{\"u}rer, Martin and Immerman, Neil},
  journal={Combinatorica},
  volume={12},
  number={4},
  pages={389--410},
  year={1992},
  publisher={Springer}
}

@inproceedings{NeuenSchweitzer18,
  title={An exponential lower bound for individualization-refinement algorithms for graph isomorphism},
  author={Neuen, Daniel and Schweitzer, Pascal},
  booktitle={Proceedings of the 50th Annual ACM SIGACT Symposium on Theory of Computing},
  pages={138--150},
  year={2018}
}

@incollection{Luks93,
  title={Permutation Groups and Polynomial-Time Computation},
  author={Luks, Eugene M.},
  booktitle={Groups and Computation},
  series={DIMACS Series in Discrete Mathematics and Theoretical Computer Science},
  volume={11},
  publisher={American Mathematical Society},
  pages={139--175},
  year={1993}
}

@inproceedings{StorjohannLabahn96,
  title={Asymptotically fast computation of Hermite normal forms of integer matrices},
  author={Storjohann, Arne and Labahn, George},
  booktitle={Proceedings of the 1996 international symposium on Symbolic and algebraic computation},
  pages={259--266},
  year={1996}
}

@article{Bochert1889,
  title={Ueber die {Z}ahl der verschiedenen {W}erthe, die eine {F}unction gegebener {B}uchstaben durch {V}ertauschung derselben erlangen kann},
  author={Bochert, Alfred},
  journal={Mathematische Annalen},
  volume={33},
  number={4},
  pages={584--590},
  year={1889},
  publisher={Springer}
}

@book{DixonMortimer1996,
  title={Permutation Groups},
  author={Dixon, John D. and Mortimer, Brian},
  series={Graduate Texts in Mathematics},
  volume={163},
  publisher={Springer-Verlag},
  address={New York},
  year={1996}
}

@article{Wielandt1960,
  title={{\"U}ber den {T}ransitivit{\"a}tsgrad von {P}ermutationsgruppen},
  author={Wielandt, Helmut},
  journal={Mathematische Zeitschrift},
  volume={74},
  number={1},
  pages={297--298},
  year={1960},
  publisher={Springer}
}

@book{Seress,
  title={Permutation Group Algorithms},
  author={Seress, {\'A}kos},
  series={Cambridge Tracts in Mathematics},
  volume={152},
  publisher={Cambridge University Press},
  address={Cambridge},
  year={2003}
}
\endgroup

\clearpage
\appendix
\section{Permutation-group ingredients}
\label{app:groups}

This appendix contains the group theory used by the
algorithm. The main text needs only the interfaces
stated there. We include the proofs because the
hereditary hypotheses are part of the running-time
claim: an abstractly isomorphic action need not carry
the same progress measures.

Throughout, groups are given by generators in their
displayed permutation actions. Homomorphisms are
specified on generators, with polynomial-time lifting
in their explicit images. Every isomorphism algorithm
returns its full coset. The standard basic operations
are those of Furst--Hopcroft--Luks and Schreier--Sims
as presented in~\cite{FHL80,H}. Invocations of stronger
normalizer or representation algorithms are identified
separately.
The degree $n$ in an appendix statement refers to that
statement's displayed permutation action; it is the
active-domain size denoted by $s$ in the main analysis.

\subsection{The local permutation families}
\label{sec:local-groups}

A local test permits relabeling outside its test set as well as inside it. The comparison problem must use the full ordered-tuple transporter. The following explicit construction fixes this convention.

Let \(G\leq\Sym(\Omega)\), \(|\Omega|=n\), and suppose an explicit
surjection \(\pi:G\to\Alt(\Gamma)\) is given, with \(|\Gamma|=m\).
For \(T\subseteq\Gamma\), \(|T|=k\), write
\begin{align*}
 Q_T&=\{g\in G:T^{\pi(g)}=T\},\\
 P_T&=\{g\in Q_T:\sgn(\restr{\pi(g)}T)=1\},\\
 J_T&=G_{(T)}=\{g\in G:\pi(g)\text{ fixes }T\text{ pointwise}\}.
\end{align*}
If \(k\geq3\) and \(m-k\geq2\), restriction gives a surjection
\(\phi_T:P_T\to\Alt(T)\), with kernel \(J_T\), and
\([Q_T:P_T]=2\).

\begin{proposition}
Generators for \(P_T,J_T,Q_T\) can be obtained in polynomial time in
\(n+m\), assuming a polynomial-size generating set and the explicit images
of its generators under \(\pi\).
\end{proposition}

\begin{proof}
Compute \(\ker\pi\) using the simultaneous action on \(\Omega\sqcup\Gamma\)
and a pointwise stabilizer. Lift explicit generators of
\(\Alt(T)\times\Alt(\Gamma\setminus T)\) by constructive membership in the
image. These lifts together with \(\ker\pi\) generate \(P_T\). Lifting only
the complementary alternating group gives \(J_T\). A paired transposition,
one in each part, supplies the other coset for \(Q_T\). All these operations
are basic Schreier--Sims operations~\cite[Section 2.1]{H}.
\end{proof}

\subsection{A hereditary budget for visible alternating quotients}
\label{sec:hereditary-budget}

This section replaces a certificate-size threshold by an inherited
analysis parameter. The parameter is defined using the actual
permutation action. Its proof uses the classification of finite simple
groups only through Schreier's conjecture, which states that the outer
automorphism group of every finite nonabelian simple group is solvable.
The algorithm does not have to compute this parameter.

\subsubsection{Definition, visibility, and inheritance}

Let \(G\leq\Sym(\Omega)\), and fix \(\alpha\in\Omega\).  We allow a
finite descending chain
\[
 G=H_0\geq H_1\geq\cdots\geq H_r,
 \qquad \Delta_i=\alpha^{H_i}.
\]
A selected step \(i\) is \emph{marked with weight \(k_i\)} if
\begin{equation}
 \label{eq:visible-mark}
 \begin{gathered}
 H_{i+1}\triangleleft H_i,\qquad
 H_i/H_{i+1}\cong\Alt(k_i),\qquad k_i\geq\kcert,\\
 H_{i+1}\text{ is transitive on }\Delta_i,
 \qquad (H_i)_{(\Delta_i)}\leq H_{i+1}.
 \end{gathered}
\end{equation}
The last condition says that the quotient is visible in the faithful
action on the active orbit.  Unmarked steps are arbitrary subgroup
restrictions.

\begin{definition}[Weighted hereditary budget]
\label{def:hereditary-budget}
Let \(\sigma(G,\Omega,\alpha)\) be the maximum of
\(\sum_{i\text{ marked}}k_i\) over these chains and choices of marks.
For a transitive action, conjugation shows that this value is independent
of \(\alpha\); write \(\sigma(G,\Omega)\).
\end{definition}

The maximum is finite, since every marked step is a proper subgroup
descent and the weights are bounded by the orders of the relevant
quotients.  The visibility condition is necessary.  For example, on a
disjoint union of one regular cyclic orbit and arbitrarily many natural
\(\Alt(\kcert)\)-orbits, one can delete the latter factors without changing
the action on the cyclic orbit.  Those deletions must not count as
progress on that active orbit.

The definition allows the restrictions made by the algorithm. A chain in a child action can be lifted and prefixed by the restriction that created the child; a marked kernel step then gives an actual decrease.

\begin{lemma}[Inheritance and a marked decrease]
\label{lem:budget-inheritance}
The budget does not increase under subgroup restriction followed by
passage to the faithful image on one orbit of a transitive action.
If \(N\triangleleft G\) is transitive on \(\Omega\) and
\(G/N\cong\Alt(k)\), \(k\geq\kcert\), then
\[
 \sigma(N,\Omega)\leq\sigma(G,\Omega)-k.
\]
More generally, \(\sigma(G,\Omega,\alpha)\) equals the budget of the
faithful image of \(G\) on \(\alpha^G\).
\end{lemma}

\begin{proof}
For the last assertion, map a chain to its action on \(\alpha^G\).
The kernel of this initial action, intersected with any later group,
fixes its later active orbit pointwise.  At every mark it is therefore
killed by the marked quotient.  Consequently all marks and their weights
survive.  In the reverse direction, lift every subgroup in a chain of
images to its full preimage; the marked quotients and transitivity are
unchanged, and visibility is automatic for the lifted action kernels.

For a subgroup \(H\leq G\) and an \(H\)-orbit \(D\), first conjugate
\(H\) by an element of the transitive group \(G\) so that \(D\)
contains the chosen base point.  Prefix the lifted chain on \(D\) by
the unmarked restriction to \(H\).  Repeating this argument permits
successive orbit choices: conjugation within the current group preserves
every earlier containing group as a group.  This establishes the stated
inheritance for the actual recursive orbit images.  Finally, prefix a
chain below \(N\) by the marked step \(G\geq N\).
\end{proof}

For an odd prime \(\ell\), write \(v_\ell\) for its order valuation.
For all \(k\geq\kcert\),
\[
 v_3(|\Alt(k)|)\geq\lfloor k/3\rfloor\geq k/4,
 \qquad
 v_5(|\Alt(k)|)\geq\lfloor k/5\rfloor\geq k/6.
\]
Indices telescope along the descending chain, so for either
\(\ell\in\{3,5\}\) we have the useful order bound
\begin{equation}
 \label{eq:budget-valuation}
 \sigma(G,\Omega,\alpha)\leq6v_\ell(|G|).
\end{equation}
It remains valid after any number of subgroup and orbit-image
restrictions.

\subsubsection{Localization into blocks and Cartesian coordinates}

We first record the elementary localization fact used throughout.

A nonabelian simple quotient cannot be simultaneously invisible in every component action. The commutator argument below is the reason that a single component can be charged for each marked step.

\begin{lemma}[Simple-quotient localization]
\label{lem:budget-simple-localization}
Let \(\phi:U\twoheadrightarrow S\) have nonabelian simple image, and
let \(K_1,\ldots,K_t\triangleleft U\) satisfy
\(\bigcap_jK_j\leq\ker\phi\).  Then \(\phi(K_j)=1\) for at least
one \(j\).  In particular, a visible simple quotient of a permutation
group factors through one of its orbit actions.
If \(U\triangleleft H\), the map \(\phi\) is the restriction of a
homomorphism on \(H\), and \(H\) is transitive on a domain \(D\)
with \(U_{(D)}\leq\ker\phi\), then
\(U_{(\Theta)}\leq\ker\phi\) for every \(U\)-orbit
\(\Theta\subseteq D\).
\end{lemma}

\begin{proof}
Each \(\phi(K_j)\) is either trivial or \(S\).  If they were all
\(S\), an iterated commutator of the \(K_j\) would lie in their
intersection and would still map onto \(S\), because \(S\) is perfect.
This is a contradiction.  For orbit actions take their pointwise
kernels.  In the last assertion their intersection is \(U_{(D)}\),
which is killed by hypothesis, so the first part kills one orbit kernel.
The \(U\)-orbits are conjugate under \(H\); their kernels are conjugate
and \(\ker\phi\) is invariant under that conjugation.  Thus every
orbit kernel is killed.
\end{proof}

For a partition into blocks, a visible quotient occurs either on the block labels or inside one block. The proof must make this choice along one descending chain, including when later active orbits are proper subsets.

\begin{lemma}[Hereditary block decomposition]
\label{lem:budget-blocks}
Suppose \(G\) is transitive and preserves a system \(\mathcal B\) of
\(b\) blocks of size \(a\).  Let \(C\in\mathcal B\) contain
\(\alpha\).  Then
\[
 \sigma(G,\Omega)
 \leq \sigma(G^{\mathcal B},\mathcal B)
       +\sigma(G_C^C,C).
\]
This inequality accounts for chains whose later active orbits are
proper subsets of \(\Omega\).
\end{lemma}

\begin{proof}
Consider a marked \(\phi:H\twoheadrightarrow S\), with kernel \(N\),
on \(\Delta=\alpha^H\).  Let \(\mathcal B_\Delta\) be the blocks
meeting \(\Delta\), and let \(U\) be the kernel of the \(H\)-action
on \(\mathcal B_\Delta\).  If \(\phi(U)=1\), the quotient is
visible on this active block orbit.  Its kernel remains transitive,
since \(N\) is transitive on \(\Delta\).  Charge this mark to the
top action.

Otherwise \(\phi(U)=S\).  Apply
Lemma~\ref{lem:budget-simple-localization} to the actions of \(U\) on
the nonempty slices \(\Delta\cap B\).  Their kernel intersection is
contained in \(H_{(\Delta)}\), which is killed.  The slices are
\(H\)-conjugate, so the kernel of the \(U\)-action on
\(\Delta\cap C\) is killed.  Let \(T\) be the pointwise kernel of
\(H_C\) on \(\Delta\cap C\).  Both \(U\) and \(T\) are normal
in \(H_C\), and
\[
 [U,T]\leq U\cap T\leq U_{(\Delta\cap C)}.
\]
Thus \(\phi(T)\) centralizes \(\phi(U)=S\) and is trivial.
The quotient is consequently visible on the local action of \(H_C\).
Moreover \(N_C\) is transitive on \(\Delta\cap C\): an element of
\(N\) sending one point of this slice to another must stabilize their
block.  Charge this mark, with its full weight, to that local action.

Along the original chain the top images and the images of the fixed
block stabilizers are descending chains.  Their active orbits are
\(\mathcal B_\Delta\) and \(\Delta\cap C\), respectively.  Every
charged step satisfies visibility and transitivity in its indicated
chain.  Summing the two budgets proves the claim.
\end{proof}

Cartesian coordinates introduce an additional coordinate-permutation action. We charge that action separately and then locate every remaining mark in one coordinate; no independence of the active coordinate projections is assumed.

\begin{lemma}[Cartesian decomposition with arbitrary active orbits]
\label{lem:budget-cartesian}
Let \(A\leq\Sym(D)\) and let
\(G\leq A\wr\Sym(s)\) act on \(D^s\) in product action.  Fix
\(v=(v_1,\ldots,v_s)\).  For each coordinate \(i\), let \(G_i\)
be its setwise stabilizer and let \(A_i\leq A\) be its component
image.  Then
\[
 \sigma(G,D^s,v)
 \leq 3(s-1)+\sum_{i=1}^s\sigma(A_i,D,v_i).
\]
No full-product transitivity is assumed at any stage of the chain.
\end{lemma}

\begin{proof}
At a marked step write \(\phi:H\twoheadrightarrow S\), with kernel
\(N\), and \(\Delta=v^H\).  Let \(K\) be the kernel of \(H\)
on the coordinate positions.  Transitivity of \(N\) gives
\(H=NH_v\), hence \(\phi(H_v)=S\).  Since
\(K_v=K\cap H_v\triangleleft H_v\), its image is trivial or \(S\).

If \(\phi(K_v)=1\), the quotient is visible in the coordinate
permutation images of \(H_v\) and \(N_v\).  Along the full chain
these images form one descending chain in \(\Sym(s)\).  All marks
of this kind have total weight at most
\[
 6v_3(s!)\leq3(s-1).
\]
This charge needs no transitivity on the coordinate positions.

Suppose instead that \(\phi(K_v)=S\).  In particular
\(\phi(K)=S\).  The map on \(K\) kills its pointwise kernel on
\(\Delta\).  By localization and conjugacy of the \(K\)-orbits,
it kills the kernel on the particular orbit \(\Theta=v^K\).
Let \(C_i\) be the projection of \(\Theta\) onto coordinate \(i\).
The kernel on \(\Theta\) is the intersection of the kernels on the
\(C_i\).  Localization therefore supplies a coordinate \(i\) with
\[
 \phi\bigl(\ker(K\longrightarrow\Sym(C_i))\bigr)=1.
\]
Write \(H_i\) for the coordinate stabilizer in \(H\), and put
\(D_i=v_i^{H_i}\), using its component action.  We have
\(K\leq H_i\) and \(C_i\subseteq D_i\).  If \(T_i\) is the
pointwise kernel of this \(H_i\)-action on \(D_i\), then
\[
 [K,T_i]\leq K\cap T_i
       \leq\ker(K\longrightarrow\Sym(C_i)).
\]
Since \(\phi(K)=S\), centerlessness gives \(\phi(T_i)=1\).
Thus the quotient is faithfully visible in this component action.
The group \(K_v\) fixes \(v_i\) and maps onto \(S\); the component
point stabilizer therefore maps onto \(S\).  Equivalently the induced
image of \(N\cap H_i\) is transitive on \(D_i\).  Charge the mark
to coordinate \(i\).

For each fixed coordinate, the coordinate-stabilizer images form a
descending chain on \(D\), with active orbits containing \(v_i\).
All charged steps are admissible marks in that chain.  Adding their
budgets and the coordinate-position charge proves the inequality.
\end{proof}

\subsubsection{Johnson actions: anchors and a refinement forest}

The remaining conversion concerns one action on subsets.  A subgroup
may have a correlated orbit of subsets, so transitivity on that orbit
does not imply transitivity on the underlying natural set.  The following
argument keeps the subset's individual points as anchors.

In a subset action, fixing a subset is weaker than fixing its individual points. We use those points as anchors. An entropy estimate bounds the total cost when their natural orbits split along the chain.

\begin{lemma}[Johnson anchor bound with an unchanged leading constant]
\label{lem:budget-johnson}
Let \(m\geq4\), \(2\leq t\leq m/2\), and
\(D=\binom{[m]}t\).  Put \(a_0=3\), choose \(C\geq{C_\star}\), and set
\(B=C+2a_0=C+6\).  Suppose
\[
 \sigma(\Sym(r),[r])\leq C\log r-B\quad(3\leq r\leq m),
 \qquad \sigma(\Sym(r),[r])=0\quad(r\leq2).
\]
Then
\[
 \sigma(\Sym(m),D)\leq C\log\binom mt-B.
\]
\end{lemma}

\begin{proof}
Fix the base subset \(T\in D\), and consider an arbitrary marked
\(\phi:H\twoheadrightarrow S\), with kernel \(N\), on
\(\Delta=T^H\).  Since \(N\) is transitive on \(\Delta\),
\(H=NH_T\) and \(\phi(H_T)=S\).  Let \(P=H_{(T)}\) be the
pointwise stabilizer of the \(t\) natural anchors.  Its image is trivial
or \(S\).

If \(\phi(P)=1\), the quotient is visible in the natural action of
\(H_T\) on the fixed set \(T\).  These images form one descending
chain in \(\Sym(t)\), so all such marks together have weight at most
\begin{equation}
 \label{eq:anchor-image-charge}
 6v_3(t!)\leq a_0(t-1).
\end{equation}

Suppose \(\phi(P)=S\).  Let
\(U=\bigcup_{h\in H}T^h\), the union of the natural \(H\)-orbits
meeting \(T\).  A permutation fixing \(U\) pointwise fixes every
subset in \(\Delta\).  Visibility therefore gives
\(\phi(H_{(U)})=1\).  Applying
Lemma~\ref{lem:budget-simple-localization} to the natural orbits in
\(U\) supplies an orbit \(E\) meeting \(T\) such that
\(\phi(H_{(E)})=1\).  Choose \(\alpha\in E\cap T\).
Since \(P\leq H_\alpha\) and \(\phi(P)=S\), the point stabilizer
maps onto \(S\).  Hence \(N^E\) is transitive on \(E\), and
\(H^E/N^E\cong S\).  This is a visible mark in the natural action
on \(E\).  Charge its weight to that anchored natural orbit.

We bound all charges of the latter kind simultaneously.  As the
groups in the chain decrease, their natural orbit partitions refine.
Track the orbits meeting \(T\), starting from the natural set
\([m]\).  Assign each tracked orbit of size at least three a path
token.  A token \(\nu\) started on an orbit of size \(w\) has a fixed
allowance \(b_\nu=C\log w-B\) and accumulated charge \(c_\nu=0\).
While it is active on an orbit \(E\) of the current group \(H\),
maintain the invariant
\[
 c_\nu+\sigma(H^E,E)\leq b_\nu.
\]
The natural-action hypothesis supplies this inequality when a token
is started.  In particular, the root token has allowance
\(C\log m-B\).

At a split, retain the children meeting \(T\) whose sizes are at
least three, and let a largest retained child continue the old token
with its existing charge and allowance.  Its actual remaining budget
cannot increase, by Lemma~\ref{lem:budget-inheritance}; it is not
replaced by a fresh upper bound.  Start a new token on each other
retained child of size \(w\), with allowance \(C\log w-B\).
Children of size at most two have zero budget and receive no token.
The same rule handles the initial restriction from \(\Sym(m)\) to
the first group of the chain.

If there is no retained child, terminate the old token at its last
node of size at least three and retain all its past charges.  Terminate
any still active tokens when the chain ends.  At a charged marked
step, the charged orbit does not split, and its actual remaining
budget falls by the weight of that mark.  Adding that weight to
\(c_\nu\) therefore preserves the invariant.  All other restrictions
preserve it by inheritance.  Since remaining budgets are nonnegative,
each terminated token has \(c_\nu\leq b_\nu\).  Consequently the total
weight of these charges is at most the sum of the root and fresh-child
allowances.

The retained orbits form a rooted refinement tree, with unary nodes
allowed.  Suppose it has \(r\) terminal paths.
If \(E\) is the last node of a terminal path, associate to it the
nonempty anchor set \(E\cap T\).  Terminal nodes are leaves of the
retained refinement tree, so these sets are pairwise disjoint, even
when a path terminated because all its children had size at most two.
Thus \(1\leq r\leq t\).  Each token continues along one path to
exactly one terminal node, so there are exactly \(r\) budget starts:
the root and all children which do not continue a largest-child path.
Let \(W\) be the sum of the logarithms of their sizes.  The charge is
at most \(CW-Br\).  We next prove
\begin{equation}
 \label{eq:anchor-entropy}
 W\leq r\log(m/r)+(r-1).
\end{equation}

More generally, consider any refinement tree with root size \(w\),
child sizes \(w_i\) satisfying \(\sum_iw_i\leq w\), and \(r\)
leaves.  Choose a largest child at every split and define \(W\) as
above.  For a leaf the claimed bound is equality.  Otherwise let child
\(i\) have \(r_i\) leaves, put \(r=\sum_i r_i\), and let
\(w_h=\max_iw_i\).  The exact recurrence is
\[
 W=\log(w/w_h)+\sum_iW_i.
\]
Write \(p_i=r_i/r\), \(x_i=w_i/w\), \(z=\sum_i x_i\leq1\),
and let \(d\) be the number of children.  For
\(F(w,r)=r\log(w/r)+(r-1)\),
\[
 F(w,r)-\sum_iF(w_i,r_i)
   =r\sum_i p_i\log(p_i/x_i)+(d-1)
   \geq r\log(1/z)+(d-1).
\]
The log-sum inequality gives the last step.  Since
\(\max_i x_i\geq z/d\), \(r\geq1\), and \(d-1\geq\log d\),
this difference is at least
\(\log(1/z)+\log d\geq\log(w/w_h)\).
Induction proves \eqref{eq:anchor-entropy}, including loss of untracked
mass and the retained terminal paths just described.

First assume \(t\leq m/4\).  Combining
\eqref{eq:anchor-image-charge} and \eqref{eq:anchor-entropy}, the total
weight is at most
\[
 C\{r\log(m/r)+r-1\}-Br+a_0(t-1).
\]
Let \(f(u)=u\log(m/u)\).  On \(1\leq u\leq t\leq m/4\),
\(f'(u)\geq\log4-1\).  Our constants satisfy
\(C(\log4-1)\geq a_0\), so the preceding expression is at most
\[
 Cf(t)-C-a_0r-a_0\leq Cf(t)-B.
\]
The elementary bound \(\binom mt\geq(m/t)^t\) gives the desired
conclusion.  If there were no charges of the second kind, the bound
\(a_0(t-1)\) alone is smaller than \(C\log\binom mt-B\).
All fresh budgets used above are nonnegative, since
\(C\log3-B=C(\log3-1)-6>0\).

If \(t>m/4\), use the inherited order valuation directly:
\[
 \sigma(\Sym(m),D)\leq6v_3(m!)\leq3m,
 \qquad
 \log\binom mt\geq t\log(m/t)>(m/4)\log2.
\]
For \(m\geq\kcert\), \(C\geq{C_\star}\) and \(B=C+6\) absorb \(3m\)
in \(C\log\binom mt-B\).  For \(m<\kcert\), no permitted marked
quotient can be a section of \(\Sym(m)\), by order, and the budget
is zero.  The right side is positive since \(\binom mt\geq6\).
\end{proof}

\subsubsection{The primitive cases with a logarithmic budget}

We first record how the O'Nan--Scott structure of a primitive group is
used. By the O'Nan--Scott theorem (see for example
\cite[Theorem~4.1A, p.~105]{DixonMortimer1996} and
\cite[Lemma~2.5]{Holt}), a finite primitive group $G\leq\Sym(\Omega)$
of degree $n$ falls into the affine case, the almost-simple case, the
product-action case, or the remaining nonabelian regular, diagonal, and
twisted wreath cases. A product action has degree $d^s$ with $d\geq2$
and $s\geq2$, and is contained in $\Sym(d)\wr\Sym(s)$ in product
action. In the remaining nonabelian regular, diagonal, and twisted
wreath cases, the solvable radical of $G$ is trivial and the socle
$E=\operatorname{soc}(G)\cong T^s$ is a direct product of $s\geq1$
copies of a finite nonabelian simple group $T$, with $n\geq|T|^{s/2}$
and hence
\begin{equation}
 \label{eq:budget-diagonal-degree}
 |E|\leq n^2.
\end{equation}
Product actions with diagonal components may instead be decomposed
directly by Lemma~\ref{lem:budget-cartesian}.

The full order of a primitive group can be too large for a logarithmic
bound. In the affine and non-Johnson alternating cases it is enough to
bound the exponent of one fixed odd prime in $|G|$, because each large
alternating quotient consumes linearly many copies of that prime. In
the regular, diagonal, and twisted wreath cases of
\eqref{eq:budget-diagonal-degree}, we instead combine the $3$-valuation
of the socle with Schreier's conjecture on the outer automorphism groups
of the simple factors.

\begin{lemma}[Primitive valuation cases]
\label{lem:budget-primitive-valuation}
There is an absolute constant $A_0>0$ with the following properties for
every primitive group $G\leq\Sym(\Omega)$ of degree $n$.
\begin{enumerate}
\item If $G$ is affine, or if $G$ is almost simple with alternating
socle and its action is neither natural nor a Johnson subset action,
then there is a prime $\ell\in\{3,5\}$ such that
$6v_\ell(|G|)\leq A_0\log n$.
\item If $G$ is in one of the cases of part~1, or if $G$ has trivial
solvable radical and its socle $E=\operatorname{soc}(G)$ satisfies
\eqref{eq:budget-diagonal-degree}, then
$\sigma(G,\Omega)\leq A_0\log n$.
\end{enumerate}
Bounded exceptional alternating degrees are included in this bound.
\end{lemma}

\begin{proof}
For $n\leq2$, the group order divides $2$, so $v_3(|G|)=0$ and
$\sigma(G,\Omega)=0$ by \eqref{eq:budget-valuation}. Assume now that
$n\geq3$.

We first treat the affine case. If $\ell$ is an odd prime not dividing
$q$, put $e=\operatorname{ord}_\ell(q)$ and $u=\lfloor d/e\rfloor$.
The order formula for $\GL_d(q)$ and lifting the exponent give
\begin{equation*}
 v_\ell(|\GL_d(q)|)
   =u\,v_\ell(q^e-1)+v_\ell(u!)
   \leq d\log_\ell q+\frac{d}{\ell-1}
   =O(d\log q).
\end{equation*}
The implied constant is uniform for $\ell\in\{3,5\}$. An affine
primitive group of degree $n=p^d$ embeds in $\operatorname{AGL}_d(p)$.
Choose $\ell\in\{3,5\}$ different from $p$. The translation subgroup
has order $p^d$ and therefore has zero $\ell$-valuation, so
$v_\ell(|G|)\leq v_\ell(|\GL_d(p)|)=O(d\log p)=O(\log n)$. By
\eqref{eq:budget-valuation}, this also gives
$\sigma(G,\Omega)=O(\log n)$.

Next suppose that $G$ has trivial solvable radical and its socle
$E=\operatorname{soc}(G)$ satisfies $|E|\leq n^2$ as in
\eqref{eq:budget-diagonal-degree}. Because $G$ has no nontrivial
solvable normal subgroup, the centralizer $C_G(E)$ is trivial, so
conjugation gives an injective homomorphism
$\psi\colon G\to\Aut(E)$. Moreover $E\cong T_1\times\cdots\times T_s$ is
the direct product of its minimal normal subgroups $T_1,\ldots,T_s$,
each of which is nonabelian simple. Every automorphism of $E$ permutes
the set $\{T_1,\ldots,T_s\}$. Let $K\triangleleft\Aut(E)$ be the kernel
of this permutation action on the $s$ factors. Then $\Aut(E)/K$ embeds
in $\Sym(s)$, while $K$ acts on each factor $T_i$ and therefore
$K/\operatorname{Inn}(E)$ embeds in
$\prod_{i=1}^s\operatorname{Out}(T_i)$. By Schreier's conjecture
\cite[p.~133]{DixonMortimer1996}, the outer automorphism
group $\operatorname{Out}(T_i)$ of each nonabelian simple factor is
solvable, so $K/\operatorname{Inn}(E)$ is solvable. Let
$N=\psi^{-1}(K)\triangleleft G$. Since $\psi(E)=\operatorname{Inn}(E)$,
we have $E\leq N$, the quotient $N/E$ is solvable, and $G/N$ embeds in
$\Sym(s)$.

Consider any marked step $\phi\colon H\twoheadrightarrow S$ with
$S\cong\Alt(k)$ and $k\geq\kcert$ along a chain in $G$. Because
$H\cap N$ is normal in $H$ and $S$ is simple, the image
$\phi(H\cap N)$ is either $S$ or $1$. If $\phi(H\cap N)=S$, then
$\phi(H\cap E)$ cannot be trivial, because $(H\cap N)/(H\cap E)$ embeds
in the solvable group $N/E$ and cannot map onto the nonabelian simple
group $S$. Thus $\phi(H\cap E)=S$, and this step reduces $v_3(|H\cap E|)$
by at least $v_3(|\Alt(k)|)\geq k/6$. If instead $\phi(H\cap N)=1$, then
$\phi$ factors through $H/(H\cap N)\cong HN/N\leq G/N$, and this step
reduces $v_3(|HN/N|)$ by at least $k/6$. Summing over the chain and
using $2v_3(s!)\leq s-1$ gives
\begin{equation*}
 \sigma(G,\Omega)
 \leq 6v_3(|E|)+6v_3(|G/N|)
 \leq 6\log_3|E|+6v_3(s!)
 \leq 12\log_3 n+3(s-1).
\end{equation*}
Since each factor $T_i$ has order at least $2$, we have
$2^s\leq|E|\leq n^2$ and hence $s\leq 2\log_2 n$. Therefore
$\sigma(G,\Omega)=O(\log n)$.

It remains to consider an almost-simple group $G$ with socle
$\Alt(m)$ for $m\geq7$. Its full group is $\Alt(m)$ or $\Sym(m)$. Let
$M$ be a point stabilizer in a faithful primitive action of degree $n$,
so $M$ is maximal in $G$. If $M$ is intransitive on the natural
$m$-set, choose an orbit or its complement of size $t\leq m/2$.
Maximality gives the full subset stabilizer, so the action is the
Johnson action on $t$-subsets (or the natural action when $t=1$). Equal
halves $t=m/2$ do not give a primitive action on single subsets, since
their subset stabilizer is properly contained in the stabilizer of the
partition into two complementary halves.

If $M$ is imprimitive on the natural set, write $m=ab$ for a preserved
system of $b$ blocks of size $a$, with $a,b\geq2$. Then
\begin{equation*}
 n\geq\frac{m!}{2(a!)^b b!}=\exp(\Omega(m))
\end{equation*}
uniformly in $a$ and $b$. Indeed, for $b\geq16$, elementary factorial
estimates give $\log n\geq(m-b)\log b-m-\log2\geq(\tfrac12\log16-1)m-\log2$.
For $2\leq b\leq15$, the number of equal-block partitions is at least
$\binom ma/b$ because every $a$-subset occurs as a block, so
$\log n\geq(m/b)\log b-\log b-\log2\geq(\log15/15)m-\log15-\log2$.

If $M$ is primitive on the natural $m$-set, it does not contain
$\Alt(m)$ because the degree-$n$ action is faithful and not natural. By
Bochert's theorem~\cite[p.~585]{Bochert1889} (see
also~\cite[Theorem~3.3B, p.~81]{DixonMortimer1996}), the index of a
primitive subgroup not containing $\Alt(m)$ in $\Sym(m)$ is at least
$\lfloor(m+1)/2\rfloor!$. Hence
\begin{equation*}
 n\geq\frac12\Bigl\lfloor\frac{m+1}{2}\Bigr\rfloor!=\exp(\Omega(m\log m)).
\end{equation*}
In both non-Johnson cases, $\log n\geq c_1 m$ uniformly after absorbing
bounded degrees. Therefore
\begin{equation*}
 6v_3(|G|)\leq 6v_3(m!)=O(m)=O(\log n),
\end{equation*}
and \eqref{eq:budget-valuation} gives $\sigma(G,\Omega)\leq 6v_3(|G|)=O(\log n)$.
The exceptional outer automorphisms at $m=6$ and all bounded alternating
degrees have bounded group orders, so they are absorbed into the
constant $A_0$.
\end{proof}

\subsubsection{The weighted logarithmic bound}

The preceding reductions fit an induction on the permutation degree and,
at each fixed degree, on the group order. An almost-simple group needs
no order-valuation bound of its own, because Schreier's conjecture
prevents any marked step from occurring at an almost-simple group.

\begin{theorem}[Hereditary visible-quotient budget]
\label{thm:hereditary-budget}
There are absolute constants $C\geq{C_\star}$ and $B=C+6$ such that
every faithful transitive permutation group $G\leq\Sym(\Omega)$
of degree $n$ satisfies
\begin{equation*}
 \sigma(G,\Omega)\leq
 \begin{cases}
  0,&n\leq2,\\
  C\log n-B,&n\geq3.
 \end{cases}
\end{equation*}
Thus an arbitrary chain of subgroup restrictions and faithful orbit
images, with visible transitive normal quotient steps
$\Alt(k_i)$, $k_i\geq\kcert$, has total marked weight
$\sum_i k_i=O(\log n)$. In particular, for any fixed
$k\geq\kcert$, it has at most $C\log n/k$ such $\Alt(k)$
steps.
\end{theorem}

\begin{proof}
Choose $C\geq{C_\star}$ large enough that the bound $A_0\log n$ from
Lemma~\ref{lem:budget-primitive-valuation} is at most
$C\log n-(C+6)$ for all $n\geq3$. This is possible because
$\log n-1\geq\log3-1>0$ for $n\geq3$. Fix $B=C+6$.

We argue by strong induction on the degree $n=|\Omega|$ and, for a
fixed degree $n$, by strong induction on the group order $|G|$. If
$n<\kcert$, no group of order dividing $n!$ has a section isomorphic to
$\Alt(k)$ with $k\geq\kcert$. Thus $\sigma(G,\Omega)=0$, while for
$n\geq3$ the right-hand side $C\log n-B$ is positive. Assume from now
on that $n\geq\kcert$.

If $G$ is transitive and imprimitive on $\Omega$, choose a nontrivial
block system with block size $a\geq2$ and $b\geq2$ blocks, so $n=ab$
and $a,b<n$. By Lemma~\ref{lem:budget-blocks} and the degree induction
hypothesis on the block action and on the faithful orbit image of the
block stabilizer, $\sigma(G,\Omega)$ is at most the sum of one
degree-$a$ bound and one degree-$b$ bound. If $a,b\geq3$, this sum is
\begin{equation*}
 (C\log a-B)+(C\log b-B)=C\log n-2B\leq C\log n-B.
\end{equation*}
If one of $a$ or $b$ is $2$, its budget is zero and the other term is at
most $C\log n-B$.

Suppose now that $G$ is primitive on $\Omega$. If $G$ is covered by
Lemma~\ref{lem:budget-primitive-valuation}, then
$\sigma(G,\Omega)\leq A_0\log n\leq C\log n-B$ by our choice of $C$.
If $G$ acts in product action of degree $n=d^s$ with $d\geq2$ and
$s\geq2$, pass each component action in
Lemma~\ref{lem:budget-cartesian} to its faithful image on the orbit of
the corresponding coordinate of the base tuple. Each such orbit has
size $r\leq d<n$, and Lemma~\ref{lem:budget-inheritance} preserves its
budget. By the degree induction hypothesis, each component budget is at
most $C\log d-B$ when $d\geq3$ (using the zero bound when $r\leq2$), so
Lemma~\ref{lem:budget-cartesian} gives
\begin{equation*}
 \sigma(G,\Omega)
 \leq 3(s-1)+s(C\log d-B)
 = C\log n-B-(s-1)(B-3)
 \leq C\log n-B.
\end{equation*}
When $d=2$, each component budget is zero and
$3(s-1)\leq Cs\log2-B$ holds for $s\geq2$ because $C\geq{C_\star}$.
If $G$ is a Johnson subset action of degree $n=\binom mt$ with
$2\leq t\leq m/2$, then $m<n$. The bounds for $\Sym(r)$ on $[r]$ for
all $r\leq m$ are available by the degree induction hypothesis, so
Lemma~\ref{lem:budget-inheritance} and Lemma~\ref{lem:budget-johnson}
give $\sigma(G,\Omega)\leq\sigma(\Sym(m),D)\leq C\log n-B$.

By the O'Nan--Scott classification, every remaining primitive group $G$
of degree $n$ is almost simple (including the natural giants $\Alt(n)$
and $\Sym(n)$ and all almost-simple groups with nonalternating socle).
Let $T=\operatorname{soc}(G)$ be its nonabelian simple socle. Then
$T\triangleleft G\leq\Aut(T)$, and the quotient
$G/T\leq\operatorname{Out}(T)$ is solvable by Schreier's conjecture
\cite[p.~133]{DixonMortimer1996}. We claim that no marked
step can occur at $G$ itself. Suppose for a contradiction that
$\phi\colon G\twoheadrightarrow S$ is a marked step with kernel $N$ and
$S\cong\Alt(k)$ for some $k\geq\kcert\geq5$. Because $T$ is a simple
normal subgroup of $G$, the intersection $T\cap N$ is either $1$ or
$T$. If $T\cap N=1$, then $[T,N]\leq T\cap N=1$, so $N$ centralizes $T$.
Since $C_G(T)=1$, this forces $N=1$, which contradicts the requirement
that $N$ is transitive on $n\geq\kcert\geq2$ points. If $T\cap N=T$,
then $T\leq N$, so $\phi$ factors through the solvable group $G/T$,
which cannot map onto the nonabelian simple group $S$.

Because no marked step occurs at $G$, any chain starting at $G$ with
positive marked weight must first restrict to a proper subgroup $H<G$.
By Lemma~\ref{lem:budget-inheritance}, passing from $H$ to its faithful
image $H^\Delta$ on the active orbit $\Delta=\alpha^H$ does not
increase the budget. If $|\Delta|<n$, the degree induction hypothesis
gives $\sigma(H^\Delta,\Delta)\leq C\log|\Delta|-B\leq C\log n-B$ (or
$0$ when $|\Delta|\leq2$). If $|\Delta|=n$, then $H^\Delta=H$ is a
transitive group of the same degree $n$ and strictly smaller order
$|H|<|G|$, so the order induction hypothesis at degree $n$ gives
$\sigma(H,\Omega)\leq C\log n-B$. Taking the maximum over all chains
completes the induction.
\end{proof}

\begin{remark}[Algorithmic use and boundary]
\label{rem:budget-use}
The budget is an analysis-only invariant.  A local-certificate query
already supplies the quotient map and tests whether its pointwise
active-orbit kernel is killed.  If that orbit's kernel is transitive,
the query consumes the known weight \(k\) by
Lemma~\ref{lem:budget-inheritance}; otherwise the usual affected-orbit
degree reduction is available.  No maximization over subgroup chains
is performed by the algorithm.  The theorem preserves the inherited
permutation action through subgroup and orbit images; it does not
assert invariance under arbitrary faithful re-representations of the
same abstract group.
\end{remark}

\paragraph{Constants for the hereditary bound.}
The arithmetic above is satisfied by $\kcert\geq64$ and $C_\star=128$.
After the primitive valuation constant $A_0$ is fixed, choose
$C\geq C_\star$ large enough that
$A_0\log n\leq C\log n-(C+6)$ for $n\geq3$, and put $B=C+6$.
Such a choice exists because $\log3-1>0$.
The other inequalities used in the anchor and product arguments are
$C(\log4-1)\geq3$, $C(\log3-1)>6$, and $B\geq3$;
the displayed choice of $C_\star$ satisfies them. Thus the same
constants work in every step of the induction.

\subsection{Constant-size certificates and exact comparisons}
\label{sec:certificates}

Use the notation of Section~\ref{sec:local-groups}, with a transitive
physical parent \(G\leq\Sym(\Omega)\).  The dispatcher processes
intransitive parents by inherited orbit chains before entering this
procedure.  Fix
\(k=\kcert\), with the constant chosen in Section~\ref{sec:parameters}. Fix an absolute ideal-degree cutoff
\(q_0\) sufficiently large relative to \(\kcert\), enlarged to include the bounded exceptions in the
structural cap lemmas below.  In particular, require
\(q_0\geq4\kcert+2\).  For \(m>q_0\), each test has at least two
complementary ideal labels, and every ideal set of size greater than
\(m/2\) has more than \(2\kcert\) labels.  These inequalities supply
the ordered-transporter parity convention and the Design Lemma's range.
When \(m>q_0\), all local tests have this
constant size. When \(m\leq q_0\), enumerate the alternating
quotient at its actual Cameron block level and recurse on its kernel
orbits.  This is the ordinary dispatcher reduction on a nontrivial
inherited block level; its absolute constant enumeration factor,
however large, is charged to the resulting physical-domain or
inherited-tower descent.  We do not charge an arbitrary small-degree
epimorphism with a transitive kernel to the budget defined below.
No unaffected-stabilizer
threshold depending on \(n\) is assumed.  The extra orbit rule below
is precisely what replaces such a threshold.

\subsubsection{Endpoint construction}

For a string \(x:\Omega\to\Sigma\) and a test set \(T\), maintain a window
\(W_i\) and its partial automorphism group
\[
 A_i=\Aut_{P_T}\bigl(x^{W_i}\bigr).
\]
Initially \(W_0=\varnothing\), \(A_0=P_T\). The invariant includes that
\(W_i\) is \(A_i\)-invariant. At every stage use the following rules.

\begin{enumerate}
\item If \(\phi_T(A_i)\ne\Alt(T)\), stop with a nonfull certificate.
\item Let \(E_i\) be the points outside \(W_i\) affected by \(A_i\): their
      stabilizers have proper images in \(\Alt(T)\). If \(E_i\ne\varnothing\),
      put \(V_i=E_i\).
\item Otherwise put \(U_i=\Omega\setminus W_i\). If
      \(\phi_T((A_i)_{(U_i)})=\Alt(T)\), return the full certificate
      \(K(T)=(A_i)_{(U_i)}\).
\item In the remaining case, put
      \[
       V_i=\bigcup\{\Delta:\Delta\text{ is an }A_i\text{-orbit in }U_i,
                 \ \phi_T((A_i)_{(\Delta)})=1\}.
      \]
\item Set \(W_{i+1}=W_i\cup V_i\) and compute
      \(A_{i+1}=\Aut_{A_i}^{V_i}(x)\).
\end{enumerate}

The extra union in step 4 is the modification. It is nonempty. Indeed, the
image of \((A_i)_{(U_i)}\), being normal in \(\Alt(T)\), is trivial in
that case. Thus \(\phi_T\) factors through the action on \(U_i\), and a
nonabelian simple quotient of a subdirect product factors through at least
one orbit action. Equivalently, at least one \(\Delta\) satisfies the stated
test. The union of all such orbits is equivariant and involves no choice.

For clarity, the groups are first defined by the displayed chain-rule
updates. For every finite prefix \(j\), they satisfy the exact identity
\[
 A_j=\Aut_{P_T}\bigl(x^{W_j}\bigr),
\]
where the complement receives a fresh marker. To prove the nontrivial
inclusion, let \(g\) preserve that marked string. Initially it normalizes
\(P_T\). Inductively, membership in \(A_i\) makes it preserve the next
equivariantly defined window increment \(V_i\). Since \(V_i\subseteq W_j\),
agreement with the marked string is agreement with \(x\) there; hence
\(g\in A_{i+1}\). This induction recovers every earlier window.
The reverse inclusion follows from the construction and invariance of
\(W_j\). Thus enlargement of a window does not silently allow mixing
of earlier pieces. The same prefix-recovery argument is used below for
transporters between two endpoints.

Every update adds a nonempty \(A_i\)-invariant set, so there are at most
\(n\) updates. The prefix-recovery argument proves the stated invariant for \(A_i\). At full
termination, \(K(T)\) preserves \(x\) on \(W_i\) and fixes its complement,
so \(K(T)\leq\Aut_G(x)\) and its image contains the required alternating
group. At nonfull termination, \(\Aut_{P_T}(x)\leq A_i\) has proper image.
The full setwise group \(\Aut_{Q_T}(x)\) therefore cannot contain
\(\Alt(T)\) in its induced action either.

\subsubsection{The two recursive call types}

At an update let \(N_i=\ker(\phi_T|_{A_i})\). Enumerate its
\(|\Alt(k)|\) cosets in \(A_i\), and apply the usual orbit chain rule only
on the newly added window \(V_i\). The old window has already been processed.

For an affected \(A_i\)-orbit \(\Delta\), the minimum-index property of
\(\Alt(k)\) gives
\[
 |a^{N_i}|=
 \frac{|\Delta|}{[\Alt(k):\phi_T((A_i)_a)]}\leq\frac{|\Delta|}{k}.
\]
Thus these recursive domains have length at most \(n/k\), exactly as in
Helfgott's exposition~\cite{H}.

For an extra orbit \(\Delta\) added in step 4, every point is unaffected,
so \(N_i\) is transitive on \(\Delta\). The test
\(\phi_T((A_i)_{(\Delta)})=1\) implies
\[
  (A_i|_\Delta)/(N_i|_\Delta)\cong\Alt(k).
\]
This is a visible marked step in the hereditary budget of
Section~\ref{sec:hereditary-budget}.  In particular,
\[
 \sigma(N_i|_\Delta,\Delta)
 \leq \sigma(A_i|_\Delta,\Delta)-\kcert
 \leq \sigma(G,\Omega)-\kcert.
\]
The second inequality uses the transitivity of the physical parent
and inherited passage to the \(A_i\)-orbit \(\Delta\).
Every recursive call obtained by subsequent subgroup and orbit
restrictions inherits this decrease.  The parent need not discover a
new block system, and the orbit may have as many as \(n-1\) points.
The pointwise-kernel test in step~4 is essential: an abstract alternating
quotient acting only outside \(\Delta\) would not justify the decrease.

Consequently the conservative recurrence contribution from all certificates
and their comparisons has the form
\begin{equation}
 n^{O(1)}\bigl(1+F(n/\kcert,\sigma)+F(n,\sigma-\kcert)\bigr),
 \label{eq:constant-recurrence}
\end{equation}
where the hidden polynomial exponent may depend on the fixed constant
\(\kcert\), and terms with impossible parameters are absent. The multiplicity absorbs the at most \(n\) updates, orbit calls,
\(k!\) cosets, and all test tuples and pairwise comparisons.

\subsubsection{Ordered comparisons and equivariance}

For an endpoint \((x,\boldsymbol t)\), let \(W(x,T)\) be its final window,
which depends on the underlying set \(T\), not its ordering. If \(\tau\in G\)
maps \(\boldsymbol t\) to \(\boldsymbol t'\), compare the endpoints using
\begin{equation}
 \Iso_{J_T\tau}\bigl(x^{W(x,T)},y^{W(y,T')}\bigr),
 \label{eq:comparison}
\end{equation}
where a new alphabet symbol marks the complement of each window.
The coset is the \emph{full} ordered-tuple transporter in \(G\). No support
restriction on \(T\cup T'\) is imposed.

Run the endpoint procedures in lockstep, keeping the marked-window
comparison coset \(I_i\): its elements send \(W_i\) to \(W_i'\)
and agree with the two strings on those windows. If nonempty, it is a coset of
\[
 \Aut_{J_T}\bigl(x^{W_i}\bigr)=A_i\cap J_T=N_i.
\]
Every representative conjugates \(A_i\) to the other endpoint group: the
even set-stabilizer \(P_T\) is conjugacy-equivariant, including for odd
bijections between the ordered test sets. Therefore it maps affected sets,
the extra orbit union, and stopping tests to the corresponding objects at
the other endpoint. The next comparison is string isomorphism on the new
window under \(N_i\), with the same two progress guarantees as above.

An empty partial comparison cannot revive at the final windows. Any map
satisfying \eqref{eq:comparison} maps the initial empty windows to one another.
Inductively it conjugates the endpoint groups, hence maps the next window;
its agreement on the final marked strings restricts to agreement there.
It must belong to every earlier comparison coset. This also proves that
incompatible stopping stages give an empty comparison.

The same induction, with the two endpoints equal, proves the endpoint
identity
\[
 \Aut_{P_T}\bigl(x^{W(x,T)}\bigr)=A_{\mathrm{final}}.
\]
Indeed, an automorphism of the final marked string recovers every earlier
window by the equivariant update rule, and consequently belongs to every
earlier partial automorphism group. The reverse inclusion follows from
invariance of the final window. Thus the masked-window interpretation used
in the color comparison is an equality, rather than an additional assumption.

Full transporter composition and inversion make nonemptiness in
\eqref{eq:comparison} a genuine equivalence relation. Its classes provide
the tuple colors needed for the Design Lemma.
Extend this coloring from injective tuples to all of $\Gamma^k$
by giving each noninjective tuple its equality-pattern color,
in a palette disjoint from the injective colors. Use the same
palette for both inputs. This extension is equivariant, preserves
the no-large-twin argument on injective tuples, and has polynomial
cost for fixed $k$. Canonical coherent refinement may then be
applied to the resulting full tuple structure.
 If all orderings of a nonfull
\(T\) had the same color, their transporters would realize all of
\(\Sym(T)\) on the final marked window. Its even part would force
\(\phi_T(A_i)=\Alt(T)\), a contradiction. Thus the no-large-twin argument
in Section 6.2 of Helfgott's exposition~\cite{H} survives.

If the exact subgroup output \(M(T)\leq\Sym(T)\) used there is desired,
perform the \(k!\) ordered self-comparisons and collect their feasible
permutation images. They form the image of the masked-window automorphism
group; its even part is proper. This extra factor is already allowed in
\eqref{eq:constant-recurrence}.

\subsubsection{Aggregation and orbital colors}
\label{subsec:aggregation-orbital-colors}

The aggregation in \cite[Section~6.2]{H} is used with one explicit
color convention.  The certificate-generated subgroup \(F\leq\Aut_G(x)\)
is canonical as an abstract subgroup, but an arbitrary numbering of its
orbitals need not be preserved by a parent isomorphism.  Such a numbering
is therefore not used as a fixed color alphabet.

In Case~2b of that aggregation, let $\Phi$ be the dominant orbit of
the ideal image of $F$, and let $d$ be its maximal degree of
transitivity. Since this action does not contain $\Alt(\Phi)$ and
Schreier's conjecture holds by the classification of finite simple
groups, Wielandt's theorem~\cite[Satz~1]{Wielandt1960} gives $d\leq7$.
Fix an ordered source tuple of $d-1\leq6$ distinct points and branch
over its at most $q^{d-1}\leq q^6$ possible target images. On
$C=\Phi\setminus\{x_1,\ldots,x_{d-1}\}$, the resulting subgroup
$U$ is transitive but not two-transitive. Choose one proper
non-diagonal source orbital $\mathcal O\subset C^2$.
Branch over its at most $|C|^2\leq q^2$ possible target orbitals.
For each branch, color the directed graph by membership in
$\mathcal O$, and apply the canonical coherent refinement from Helfgott's
exposition~\cite{H} to this binary coloring.

The resulting coherent configuration is homogeneous: \(U\) preserves
the binary graph and is transitive, and refinement preserves all these
automorphisms.  It is nontrivial because both \(\mathcal O\) and its
off-diagonal complement are nonempty, and refinement never merges their
colors.  Every genuine parent isomorphism conjugates the two anchored
subgroups and carries the chosen source orbital to one of the enumerated
target orbitals.  Consequently the branches cover every parent
isomorphism, while their further colors are canonically defined from the
binary graph.  The additional continuation factor is only
\(q^{O(1)}\), included in the \(q^{O(k)}\) Design factor or the
truncated Split-or-Johnson index.  No arbitrary permutation of orbital
color names is assumed to be known.

\paragraph{Balance parameters.}
Keep individualized labels as singleton colors. In Case~3, the
discarded support has size below $q/2$, and the Design Lemma's
dominant class contains more than half the remaining labels.
For this we apply the Design Lemma \cite[Proposition~5.1]{H} with
$\alpha=3/4$, not with $\alpha=1/2$ as in \cite[Section~5.1]{H}.
At $\alpha=1/2$ the stated lemma fails. Take four points, tuples of
length two, and the coloring of pairs by the orbitals of
$\Sym(\{1,2\})\times\Sym(\{3,4\})$. No twin class has more than
two points, and yet neither conclusion holds. For $1/2<\alpha<1$ the
proof in \cite{H} is correct under the hypothesis that every twin class
has fewer than $\alpha$ times as many points as the ground set. That
hypothesis holds here. By the no-large-twin argument, every twin class
among the remaining labels has fewer than $\kcert$ points. The remaining labels number more than
$q/2$, and $\kcert<\tfrac38q$ because $q>q_0\geq4\kcert+2$. An
$\alpha$-dominant class then has at least $3/4$ of the remaining
labels, which is more than half of them.
In Case~2b, the dominant orbit has size greater than $q/2$ and
only a bounded number of labels are anchored. Thus $|C|\geq\delta q$
for $0<\delta\leq1/4$, above a fixed absolute base degree.
Use the dominant-class threshold $\eta>2/3$ in the first
Split-or-Johnson step. Unless a balanced coloring has already
been found, $|W|>\eta|C|\geq\eta\delta q$.
The rectangle lemma below bounds the left twin classes by $|W|/2$.
Extending the left answer over the omitted labels therefore gives
width at most $\max\{1/2,1-\eta\delta\}q$.
Balanced colored partitions give width at most
$\max\{\eta,1-\delta\}q$. The choice of $\beta$ in
Section~\ref{sec:parameters} dominates all these bounds.

\subsubsection{The remaining aggregation branches at constant certificate size}
\label{subsec:aggregation-constant-threshold}

The aggregation in \cite[Section~6.2]{H} uses two properties of the
full certificates: the generated subgroup $F_x$ lies in
$\operatorname{Aut}_G(x)$, and the assignment $x\mapsto F_x$ is
equivariant under the parent isomorphisms. The modified endpoints
proved above retain both properties. The following verification shows
explicitly that the large-symmetry branch does not need the former
certificate threshold $k>2+\log_2 n$.

When the certificates already provide a large alternating family of actual automorphisms, those automorphisms can remove the corresponding search. The physical shrinkage comes from the supplied Johnson block map, not from the existence of an arbitrary alternating quotient.

\begin{lemma}[Large symmetry at an actual Johnson constituent]
\label{lem:large-symmetry-constant-cert}
Let $G\leq\operatorname{Sym}(\Omega)$ be transitive, $|\Omega|=n$,
and let $\pi:G\twoheadrightarrow\operatorname{Alt}(\Gamma)$,
$|\Gamma|=q$, be the supplied map of a selected Cameron constituent.
Assume its actual equivariant block map is
\[
 p:\Omega\longrightarrow\binom{\Gamma}{h},
 \qquad 1\leq h\leq q/2.
\]
Suppose the equivariantly assigned subgroup
$F_x\leq\operatorname{Aut}_G(x)$ has a dominant ideal orbit
$\Phi$, $|\Phi|>q/2$, and
$\pi(F_x)|_\Phi\geq\operatorname{Alt}(\Phi)$.
Then the corresponding aggregation branch of exact string isomorphism
reduces to a constant number of ordinary orbit-chain computations under
subgroups whose physical orbits have size at most $n/2$, together with
polynomial structural work. All recursive calls are inherited
subgroup-orbit calls, and the complete output coset is recovered.
\end{lemma}

\begin{proof}
The dominant orbit and the property in the hypothesis are equivariant.
For two strings, differing cases or dominant-orbit sizes certify
nonisomorphism. Otherwise choose an even ideal permutation aligning
their dominant orbits, lift it through $\pi$, and absorb it into the
target string. We may therefore assume that both dominant orbits
are the same set $\Phi$. Every remaining parent isomorphism lies in
\[
 H=\pi^{-1}\bigl((\operatorname{Sym}(\Phi)
              \times\operatorname{Sym}(C))
                 \cap\operatorname{Alt}(\Gamma)\bigr),
 \qquad C=\Gamma\setminus\Phi.
\]
Put
\[
 H_0=\pi^{-1}\bigl(\operatorname{Alt}(\Phi)
                    \times\operatorname{Alt}(C)\bigr),
 \qquad
 K=\pi^{-1}\bigl(1_\Phi\times\operatorname{Alt}(C)\bigr).
\]
Here $K\triangleleft H_0$ and
$H_0/K\cong\operatorname{Alt}(\Phi)$. The index $[H:H_0]$ is
at most two. When a second coset exists, an even paired transposition,
one transposition in each of $\Phi$ and $C$, supplies its
representative. Thus the parity issue requires at most two target
twists, exactly as in \cite[Section~4.2]{H}.

We first obtain actual automorphisms lifting the large alternating
factor. Let $L=\pi(F_x)\leq\operatorname{Sym}(\Phi)
\times\operatorname{Sym}(C)$, and let $E\leq L$ be the inverse
image of $\operatorname{Alt}(\Phi)$ under restriction to $\Phi$.
This restriction maps $E$ onto $\operatorname{Alt}(\Phi)$.
Its subgroup $E_{(C)}$ is normal in $E$, so its image on $\Phi$
is either trivial or all of $\operatorname{Alt}(\Phi)$.
The first possibility would make $\operatorname{Alt}(\Phi)$ a
quotient of a subgroup of $\operatorname{Sym}(C)$. This is
impossible, since $|C|<|\Phi|$ and
$|\operatorname{Alt}(\Phi)|>|C|!$ above the fixed base degree.
Consequently $L$ contains the pure supported
$\operatorname{Alt}(\Phi)\times1_C$.

Choose two standard generators of that pure alternating group.
Constructive membership in $\pi(F_x)$ lifts them to
$a_1,a_2\in F_x\cap H_0$, so both lifts preserve $x$.
They map onto generators of $H_0/K$. Hence
\[
 \operatorname{Aut}_{H_0}(x)
   =\langle\operatorname{Aut}_K(x),a_1,a_2\rangle,
\]
and, for any target string $z$,
\[
 \operatorname{Iso}_{H_0}(x,z)\ne\varnothing
 \quad\Longleftrightarrow\quad
 \operatorname{Iso}_{K}(x,z)\ne\varnothing.
\]
Indeed, any $H_0$-transporter can be multiplied on the source side
by an automorphism with the same image in $H_0/K$, leaving a
$K$-transporter. If $t\in\operatorname{Iso}_K(x,z)$ is found,
the full answer is
$\operatorname{Aut}_{H_0}(x)t$.

It remains to justify the physical degree decrease. Transitivity of
$G$ makes the fibers of $p$ equal, of size $n/\binom qh$.
Fix $\omega\in\Omega$ and write $A=p(\omega)$. If
$A\cap\Phi$ contains $a$, every block in the $K$-orbit of $A$
contains this fixed point $a$. The fraction of all Johnson blocks
containing $a$ is
\[
 \frac{\binom{q-1}{h-1}}{\binom qh}=\frac hq\leq\frac12.
\]
If $A\cap\Phi=\varnothing$, the orbit stays among the
$h$-subsets of $C$, whose fraction is at most
\[
 \frac{\binom{|C|}{h}}{\binom qh}\leq\frac{|C|}{q}<\frac12.
\]
Thus every physical $K$-orbit has size at most $n/2$.
The kernel need not be transitive in the fibers for this upper bound.
Subsequent conditional subgroups have orbits contained in the same
$K$-orbits, so the ordinary full-coset chain rule uses only inherited
domains of size at most $n/2$, with total domain size at most $n$
per chain.

Compute $\operatorname{Aut}_K(x)$ and the $K$-transporters for the
at-most-two parity twists by these orbit chains. Combine the successful
full cosets and undo the initial alignment. The quotient lifts
$a_1,a_2$, the complete $K$-stabilizer, and the successful
representatives recover the entire answer; no giant quotient is
enumerated. The number of orbit-chain families is an absolute constant.
\end{proof}

The actual Johnson block map in this lemma is essential. An arbitrary
small alternating epimorphism with a transitive physical kernel would
not imply the half-degree bound. The local maps on the fixed test sets
$T$ are not substituted for the Cameron map in this argument.

\paragraph{Case 1 and balanced colored partitions.}
In aggregation Case~1, let $S$ be the support of $\pi(F_x)$.
Then $|S|\geq q/2$ and every orbit has size at most $q/2$.
Color ideal points by their orbit lengths and, for a length
$\ell\geq2$, partition its color class into the actual
$\pi(F_x)$-orbits. The fixed-point class has size
$q-|S|\leq q/2$. Every other class is a union of $r$ blocks of
size $\ell$, with both $\ell\leq q/2$ and $r\leq q/2$.
Its full allowed group is a subgroup of
$\operatorname{Sym}(\ell)\wr\operatorname{Sym}(r)$.
Therefore the full colored-partition group has hereditary
alternating-section width at most $q/2$, by the extension and
subgroup properties of that width. The structure is equivariant,
so aligning the two colored partitions and lifting their full
allowed coset loses no parent isomorphism. Impose alternating parity
by the sign homomorphism, rejecting an empty parity coset. Its
preimage is taken inside the actual current parent group.

More generally, an admissible colored partition on
$C_0\subseteq\Gamma$, $|C_0|\geq\delta q$, with block sizes at most
$\eta|C_0|$, has block counts at most $|C_0|/2$: admissibility means
that all partition blocks have size at least two, apart from
individualized singleton color classes. Those singleton colors are
fixed and contribute no symmetry.
Products of the corresponding wreath groups have width at most
$\max\{\eta q,q/2\}$. Extending over the omitted points adds width
at most $q-|C_0|\leq(1-\delta)q$. Thus this branch has full ideal cap at
most $\max\{\eta,1-\delta\}q$, and in particular at most the uniform
$\beta q$ with $\beta$ as in Section~\ref{sec:parameters}.
The analogous balanced-coloring outcome of Design has all its color
classes bounded by a fixed fraction of $q$, and has the same cap
conclusion. No index of a colored-partition stabilizer is enumerated:
its full overcoset is lifted and the original string constraints are
then imposed there. Earlier inherited restrictions are retained in
that parent preimage.

These branches therefore have exactly the charges used below:
Case~2a gives inherited physical halving, while Case~1 and balanced
partition outcomes give a fixed-fraction ideal-cap contraction.
Together with the previously verified Case~2b orbital construction
and Case~3 ordered-tuple colors, this accounts for the entire source
aggregation with the fixed certificate size.

\subsection{Inherited towers, caps, and exact quotient interfaces}
\label{sec:frozen-band}

This section uses full isomorphism cosets at every interface.  An auxiliary
algorithm may perform a large internal search, but after its exact coset
has been assembled, the parent continues under one coset.  In particular,
its internal branching factor is an additive preprocessing cost; it is
not multiplied into an unrelated parent continuation.

\subsubsection{Inherited towers and the certificate budget}

Fix the absolute certificate size \(\kcert\) from Section~\ref{sec:parameters}.  The algorithm uses the
endpoint and ordered-comparison procedures of
Section~\ref{sec:certificates}, with their full-coset output.
Theorem~\ref{thm:hereditary-budget} supplies the analysis invariant
\(\sigma(G,\Omega)\): the maximum total alternating degree in a
hereditary chain of visible normal quotients with transitive kernels,
where every marked quotient is \(\Alt(t)\) with \(t\geq \kcert\).
It satisfies \(\sigma=O(\log(|\Omega|+2))\), is nonincreasing under
subgroup restrictions and faithful images of their inherited orbit
actions, and drops by at least
\(\kcert\) in each extra-unaffected-orbit certificate call.  The algorithm
does not need to evaluate this invariant.

For a transitive action retain a complete inherited block tower down
to singletons.  If its consecutive nontrivial ratios are
\(b_1,\ldots,b_s\), then \(\prod_i b_i=n\).  Write
\begin{equation}
 \rho=\sum_{i=1}^s\bigl(2\lfloor\log_2b_i\rfloor-1\bigr)
 \leq2\log_2 n.
 \label{eq:rho}
\end{equation}
Splitting one ratio into \(t\) nontrivial factors decreases \(\rho\)
by at least \(t-1\); restriction of the tower to a subgroup orbit
cannot increase it.  A local-certificate hidden quotient is charged
to \(\sigma\), without requiring a new block level.  The tower is
retained for ordinary primitive reductions, fixed floors, and the
normal-column transpose below.

\paragraph{Constructive recognition.}
Every selected Cameron constituent is accompanied by an explicit
natural or Johnson realization and a supplied quotient homomorphism.
The normal subgroup and product coordinates are obtained by the
constructive reduction stated in \cite[Theorem~3.1 and Section~2.8]{H}.
The index bounds used below come from Mar\'oti's form of Cameron's
theorem \cite[Theorem~1.1]{Maroti}, since \cite[Theorem~3.1]{H}
states only that the index of the normal subgroup is at most the
degree. For degree greater than $24$, Mar\'oti's theorem gives, in
the large case, a group that lies in $\Sym(m)\wr\Sym(s)$, acting in
the product action on $s$-tuples of $k$-subsets of $[m]$, and that
contains $\Alt(m)^s$. So its index over $\Alt(m)^s$ is at most
$|\Sym(m)\wr\Sym(s)|/|\Alt(m)^s|=2^s s!$.
For the Johnson recognition procedure in that source, the range not
covered by its simple polynomial construction is
$m\leq(k+1)^2-2$, for $k$-subsets of $m$ labels with $k\leq m/2$. In
this range $\log\binom mk\geq k\log(m/k)$ is at least a constant
times $\sqrt m\log m$, so $\log m!\leq m\log m=O((\log\binom mk)^2)$.
Hence the entire constituent has
order $b^{O(\log b)}$, where $b$ is the physical interval degree.
Enumerate that constituent and descend to its kernel orbits of size
at most $n/b$. The same $x^2$ charge used for Cameron-small groups
pays this exceptional recognition range. Thus no unimplemented abstract
identification oracle is assumed.

\subsubsection{Exact coherent-configuration replacement}

For a finite group \(A\), let \(\operatorname{aw}(A)\) be the largest \(t\geq5\) for
which a subgroup of \(A\) has a quotient isomorphic to \(\Alt(t)\),
with value zero if there is none.  This quantity is nonincreasing under
subgroups and quotients.  If \(N\triangleleft A\), then
\[
 \operatorname{aw}(A)\leq\max\{\operatorname{aw}(N),\operatorname{aw}(A/N)\}.
\]
Indeed, intersect an alternating section with \(N\); simplicity forces
its alternating quotient to occur either in that intersection or in
its image in \(A/N\).

A complete configuration isomorphism answer is itself useful even if no Johnson structure is extracted. Its automorphism group has no large natural alternating action on a dominant orbit, and this yields the hereditary width decrease needed by the parent.

\begin{lemma}[Exact-configuration oracle replacement]
\label{lem:exact-configuration}
Suppose the short-certificate aggregation and Design step produces a
nontrivial homogeneous coherent configuration on a set
\(C\subseteq\Gamma\), where \(|\Gamma|=q\) and \(|C|\geq\delta q\).
One exact string-isomorphism oracle call on at most \(q^2\) positions
returns an explicit overcoset of all parent isomorphisms whose group,
on the complete ideal set \(\Gamma\), has hereditary alternating-section
width at most \(\beta q\), for an absolute \(\beta<1\) and all
sufficiently large \(q\).  The parent then continues directly under
its lifted restriction on the same original domain.  No
Split-or-Johnson index or enumeration of the returned ideal group is
required.  Bounded \(q\) is handled by enumerating the original
alternating quotient and descending to its kernel orbits.
\end{lemma}

\begin{proof}
Encode the configuration as its color string on \(C^2\), with the
diagonal action of \(\Sym(C)\).  The oracle returns its entire
isomorphism coset and the full group \(K_C=\Aut(C)\).
Combine it with the Young groups on the outer color classes, align
those classes, fix all previously individualized ideal points, and
impose alternating parity by the sign homomorphism.  Denote the
resulting group on \(\Gamma\) by \(K\).  These operations need only
point stabilizers, direct products, and the kernel of a sign map.
They do not ask for arbitrary group intersection.  The corresponding
preimage under the supplied giant epimorphism is explicitly generated
by its kernel and lifts of generators.

We prove the cap bound, using a fixed \(\beta>\max\{1-\delta,1/2\}\), enlarged towards
one if needed.  The outer color classes together have size at most
\((1-\delta)q\).  If every \(K\)-orbit has size at most \(\beta q\), then
\(K\) embeds in a product of symmetric groups of those degrees, and
the assertion follows.  Otherwise its unique dominant orbit \(D\)
lies in \(C\), and \(|D|>\beta q>|C|/2\).  Let \(P=K^D\).
The kernel of this induced action has width at most
\(q-|D|<(1-\beta)q\).

If \(P\) is imprimitive, it embeds in
\(\Sym(u)\wr\Sym(v)\), where \(uv=|D|\) and
\(u,v\leq|D|/2\).  Hence its width is at most \(q/2\).
If \(P\) is primitive, it cannot contain \(\Alt(D)\): its
two-transitivity would make the configuration on \(D\) a clique,
contrary to \cite[Exercise~2.15]{H}.
Apply Cameron--Maroti only as a structural bound.  In the small case,
\(|P|=|D|^{O(\log|D|)}\), so the factorial order of an alternating
section gives
\[
 \operatorname{aw}(P)
 =O\!\left(\frac{(\log|D|)^2}{\log\log|D|}\right)
 <\beta q
\]
for large \(q\).  In the large case the normal product Johnson
subgroup is nonnatural; its alternating factors have degree
\(O(\sqrt{|D|})\).  Its quotient embeds in the outer wreath group
\(C_2\wr\Sym(s)\), up to bounded exceptional factors, where
\(s=O(\log|D|)\) is the number of product coordinates.
Extension closure gives this quotient alternating-section width at most
\(\max\{s,O(1)\}=O(\log|D|)\).  This argument does not assert that
the order \(2^s s!\) is bounded by \(|D|\).  Extension closure bounds the width
of \(P\), and then of \(K\), by \(\beta q\).

All parent isomorphisms lie in the returned overcoset.  Lift this one
coset and solve the original string constraints there.  The current
alternating interval receives cap \(\beta q\); its descendants
inherit that cap, while other intervals keep theirs.  The cap potential
below pays the ensuing same-domain continuation directly.  There is
no need to enumerate \(K^D\), to decompose original Johnson strata,
or to infer an actual block-tower decrease from this restriction.
The next ordinary dispatcher applies Cameron to the actual physical
permutation action and pays its usual degree or product-tower cost.
\end{proof}

\paragraph{Selection of the quotient and its cap.}
Ordinary Cameron dispatch uses the coarsest available primitive quotient
of the whole current group, compatible with the inherited tower and
above any retained floor. Its degree-$b$ constituent is therefore an
actual quotient of the acting group: enumerating it has its stated
quotient index and leaves kernel orbits of size at most $n/b$. No
small-index assertion is transferred from an arbitrary deeper local
action to the entire group.

After the constructive Cameron normalization, the entire selected
constituent is a product of alternating $q$-actions in explicit Johnson
representations. Refining its product coordinates gives each resulting
interval cap $q$, clamped by any already smaller valid cap. This is a
cap on the whole interval action, not an inference from the existence
of one alternating quotient. A later configuration or rectangle
reduction contracts only its selected interval and its descendants.
Other origin-labeled intervals retain their bounds.

During aggregation, keep the original ideal set $\Gamma$ and its
supplied epimorphism. A working coherent set $C\subseteq\Gamma$ or a
discarded support does not become a fresh ideal degree for the support
threshold or for the cap accounting. The returned overcoset is extended
to the original $\Gamma$ before lifting. Only a subsequent actual
Cameron dispatch can select a new ideal action.

More precisely, freeze the giant-parent group $G_0$ and the onto map
$\phi:G_0\to\Alt(\Gamma)$ throughout one aggregation. Compute the
endpoint tables and certificate-generated groups relative to $G_0$.
Support alignments, dominant-orbit alignments, anchor choices, and
orbital choices define branches and their ideal structures; they do
not replace the stored recursive state or earn an intermediate budget
decrease. Lift each final full-$\Gamma$ ideal overcoset into $G_0$.
This may admit permutations violating a branch's intermediate choices,
but imposing the original strings removes them, while the proved cap
on that full ideal overcoset remains valid. An auxiliary rectangle may
likewise use an inherited orbit of $G_0$ to cover its entire requested
ideal transporter. There is no arbitrary intersection with a separately
restricted branch image. All restrictions inherited before this
aggregation remain built into $G_0$. The large-symmetry Case~2a is
different: it returns the exact answer directly through its aligned
Young subgroup and half-degree children, so no such widening follows.

An invariant partition can be retained while the algorithm works only on the action above it. The bound below stops at a block of that partition or at a small auxiliary configuration, without restarting the hierarchy at each restriction.

\begin{corollary}[Fixed-floor reduction]
\label{cor:fixed-floor}
Suppose a given invariant partition has blocks of size at most \(B\)
and its induced block group has \(\operatorname{aw}\leq\lambda\).  Short-certificate
string isomorphism above this fixed floor reduces to
\[
 \exp\bigl(O(\log^2(n+2))\bigr)
\]
ordinary exact string-isomorphism calls of degree at most
\(\max\{B,\lambda^2\}\), and the same amount of non-oracle work.
\end{corollary}

\begin{proof}
Retain the floor under every subgroup restriction and intersect it with
active orbits.  Stop inside one floor block.  Every primitive quotient
selected above it is a section of the original block group, so its giant
degree is at most \(\lambda\).  Apply
Lemma~\ref{lem:exact-configuration} at each Design outcome.
Keep the complete inherited tower down to singletons, including any levels
already present below the floor; retain all subsequent ordinary refinements.
For each inherited interval retain its ratio \(b_i\) and a hereditary
cap \(c_i\leq b_i\), initialized by \(c_i=b_i\).  Put
\[
 Q=\sum_i(\log b_i)\log\max\{c_i,2\}\leq(\log n)^2.
\]
A refinement splits a ratio into factors with inherited caps, so the
additivity of \(\log b_i\) makes \(Q\) nonincreasing.  Orbit
restrictions also do not increase it.  A selected alternating-interval
cap decrease \(q\mapsto\beta q\) reduces \(Q\) by
\(\Omega(\log b_i)\geq\Omega(\log q)\); a Johnson cap decrease
is stronger.  This pays the actual same-size Design continuation index
\(q^{O(\kcert)}\).  Polynomial-\(n\) group computations and certificate
construction are additive work, not multipliers on that continuation.

With \(x=\log(n+2)\), the oracle-leaf count is bounded by
\[
 \exp\bigl(A(x^2+\rho x+\sigma x+Q)\bigr).
\]
Here \(A\) is one sufficiently large absolute constant, which may
absorb the fixed value \(\kcert\).  A unit actual tower decrease pays
\(O(x)\), a local \(n/\kcert\) call decreases \(x^2\) by
\(\Omega(x)\), and a balanced descent has the same effect.  The
extra-unaffected-orbit call in \eqref{eq:constant-recurrence} decreases
\(\sigma\) by at least \(\kcert\), even if its orbit has size \(n-1\),
so the term \(\sigma x\) pays its polynomial local multiplicity.
The term \(Q\) pays cap contractions even
when the literal block tower does not change.  This distinction is needed:
a natural \(\Alt(q)\) action can restrict to a primitive Johnson
\(\Alt(s)\) action on the same \(q=\binom{s}{2}\) points, leaving
\(\rho\) unchanged.
Cameron-small enumeration on \(b\) blocks costs \(O(\log^2 b+x)\)
and is paid by a decrease \(\Omega(x\log b)\).  For a large
primitive Cameron reduction with product exponent \(s\), the outer
index is at most \(2^s s!\), by \cite[Theorem~1.1]{Maroti}.  If \(s\geq2\), splitting the old
ratio \(b_i=e^s\) into its coordinate intervals decreases the literal
\(\rho\) by at least \(s-1\), paying the stated polynomial-degree
index.  If \(s=1\) and \(q\geq9\), the index is at most two:
the nonregular primitive socle is \(\Alt(q)\), its centralizer is
trivial, and \(\operatorname{Out}(\Alt(q))=C_2\).  This constant
is bundled with the ensuing giant operation.  Bounded exceptions are
enumerated.  This argument does not require the stored cap to equal
the interval degree before the Cameron step.  Intransitive recursion is a sum over disjoint orbits, rather than a
polynomial multiplier on a largest child.  Formally one may multiply the
displayed majorant by \((n+2)^D\), for one fixed sufficiently large
\(D\).  Child progress potentials do not increase, and their domain
sizes sum to at most \(n\); this weighted bound absorbs polynomial
work even along successive one-point losses, as in
\cite[Appendix~A]{H}.  The prefactor changes the initial logarithmic
majorant by only \(O(L)\).  Finally
\(\rho=O(\log n)\).  Each leaf is either inside a floor block or
is a configuration oracle on at most \(\lambda^2\) positions.
\end{proof}

\subsubsection{Normalization of wide natural giants}

The normal form permits rows and columns to exchange roles. For the running-time proof, it is not enough that the new description is correct: the old interval factors must only split and move, so that their old bounds remain valid.

\begin{lemma}[Retained normal-column transpose]
\label{lem:normal-transpose}
Suppose \(G\) is transitive on \(\Omega\) and its top action is
\(\Alt(m)\) on \(m\) blocks of size \(b\),
with a supplied clean row normal form as in
Theorem~\ref{thm:logarithmic-gat-normal-form}. In that form, if the local variation
group \(N\) is intransitive with \(t\) orbits of size \(u=b/t\),
one can replace the top row quotient of degree \(m\) by a column
quotient of degree \(t\leq b\).  The old \(\Alt(m)\) action remains
inside normal columns of degree \(mu=n/t\).  This is a polynomial
operation, preserves all inherited interval factors through splitting
and permutation, and does not increase \(\rho\).
\end{lemma}

\begin{proof}
Use the notation of Theorem~\ref{thm:transitive-gat-terminal}.
The \(\mathcal M\)-orbits are \(\{i\}\times\Delta_j\), where
\(\Delta_j\) ranges over the \(N\)-orbits.  The normal subgroup
\(J=\mathcal M R_0\) has orbits
\([m]\times\Delta_j\).  The diagonal quotient ensures that the local
actions on the different rows induce the same permutation of the
\(\Delta_j\), so these are invariant columns.

For compatibility, write a local block tower as a point-stabilizer
chain \(P=L_0<\cdots<L_s=H\), and let \(N\triangleleft H\).
Insert the chain through \(NP\) by lower intersections and upper
products.  Each old adjacent index splits as
\begin{equation}
 [L_i:L_{i-1}]
 = [L_i\cap N:L_{i-1}\cap N]\,[NL_i:NL_{i-1}].
 \label{eq:modular-index-split}
\end{equation}
For clarity, let \(B_i=L_i\cap NP=(L_i\cap N)P\). Then
\(B_i\cap L_{i-1}=B_{i-1}\), so the action on the right cosets
\(B_{i-1}\backslash B_i\) is exactly the orbit of the base coset
under \(B_i\) in the old action on
\(L_{i-1}\backslash L_i\). Its faithful image is therefore a
quotient of a subgroup of the old constituent. The upper action
on \(NL_{i-1}\backslash NL_i\) is equivalent to the action of
\(L_i\) on the cosets of \(L_i\cap NL_{i-1}\), with the
same image.
Indeed, \(N\) acts trivially on these cosets. For all \(x\in H\) and
\(y\in N\) we have \(NL_{i-1}xy=NL_{i-1}(xyx^{-1})x=NL_{i-1}x\),
because \(N\) is normal in \(H\). Every element of \(NL_i=L_iN\) is a
product \(ly\) with \(l\in L_i\) and \(y\in N\), so it permutes these
cosets as \(l\) does. Hence \(L_i\) is transitive on them, the
stabilizer of the base coset in \(L_i\) is \(L_i\cap NL_{i-1}\), and
\(L_i\) and \(NL_i\) induce the same permutation group.
Since
\[
 L_{i-1}\leq L_i\cap NL_{i-1}\leq L_i,
\]
it is a block quotient of the old constituent. This identifies
both new actions, not only their numerical indices; no normality
of \(L_{i-1}\) in \(L_i\) is needed. Both pieces inherit every
hereditary alternating-section bound on the old action.
Keep their origin labels. The inequality
\(f(a)+f(b)\leq f(ab)\), for
\(f(r)=2\lfloor\log_2r\rfloor-1\) and \(f(1)=0\), shows that
\(\rho\) does not increase.

The currently acted natural row-block level is included in the inherited
tower before this refactoring.  Since \(\mathcal M\) fixes every row,
that level survives the lower/upper construction.  Thus all intervening
cell-to-row factors are pure actions on the local orbit labels; none mixes
the row factor with a local factor. Explicitly, modulo \(\mathcal M\),
write the clean product as \(\Alt(m)\times Q\). The cell and row
stabilizers are \(\Alt(m-1)\times U\) and
\(\Alt(m-1)\times Q\), respectively. Each intermediate subgroup
contains \(\Alt(m-1)\times1\) and is therefore exactly
\(\Alt(m-1)\times V\) for some \(U\leq V\leq Q\).
Replace these groups by \(\Alt(m)\times V\), after the initial
step \(\Alt(m-1)\times U<\Alt(m)\times U\) of index \(m\).
This preserves all local indices. Thus the clean product interchanges the ratio
sequence \(r_1,\ldots,r_s,m\) with
\(m,r_1,\ldots,r_s\), without changing any factor.  The inherited
coarse column partition is retained thereafter.  While the column floor is retained, dispatch above it uses its quotient
action. The transition that first enters one column from an active union
of at least two columns decreases the physical degree by a factor of at
least two. At entry, switch in place to the general inherited dispatcher.
Do not count a further same-domain recursive call or reinitialize the
inherited tower, its caps, or the symmetry budget. The bounded row-event
counter may reset against that already paid physical decrease, as
specified in Section~\ref{sec:graph-algorithm}.
Conditional string constraints may destroy the carrier in later columns;
those queries remain inside the same proper floor domains and inherit
their actual subgroup-orbit actions and budgets.
\end{proof}

\subsection{Clean giant complements above a logarithmic fiber threshold}
\label{sec:logarithmic-gat}

A large alternating quotient cannot be hidden above a transitive kernel on a much smaller set. This threshold is used to localize permutations to selected rows.

\begin{lemma}[No large giant above a transitive kernel]
\label{lem:transitive-kernel-giant-threshold}
Suppose $N\triangleleft E\leq\operatorname{Sym}(B)$, the normal subgroup
$N$ is transitive on $B$, $|B|=f$, and $E/N\cong\operatorname{Alt}(k)$. Then
\begin{equation*}
 k\leq\max\{63,12\log_2 f\}.
\end{equation*}
\end{lemma}

\begin{proof}
If $k\leq 63$, the bound holds at once. Assume from now on that
$k\geq 64$, so $\operatorname{Alt}(k)$ is simple and nonabelian. Let
$\psi\colon E\twoheadrightarrow\operatorname{Alt}(k)$ be the quotient
homomorphism with kernel $N$. Since $E$ maps onto the nonabelian
group $\operatorname{Alt}(k)$, the set $B$ has size $f\geq 2$.

We first reduce by induction on $f$ to a primitive group
$G\leq\operatorname{Sym}(\Omega)$ on a finite set $\Omega$ of size
$2\leq|\Omega|\leq f$ and a surjective homomorphism
$\phi\colon G\twoheadrightarrow\operatorname{Alt}(k)$ whose kernel
$K=\ker\phi$ is transitive on $\Omega$. If $E$ is already primitive
on $B$, we take $\Omega=B$, $G=E$, and $\phi=\psi$. Otherwise, choose
a nontrivial block system $\mathcal{B}$ for $E$ on $B$, consisting of
$b\geq 2$ blocks of common size $a\geq 2$ with $ab=f$, and fix one
block $C\in\mathcal{B}$. Because $E\leq\operatorname{Sym}(B)$ is
faithful on $B$, the quotient
$\psi\colon E\twoheadrightarrow\operatorname{Alt}(k)$ with transitive
kernel $N$ is a single marked step on $B$ in the sense of
Section~\ref{sec:hereditary-budget}. The proof of
Lemma~\ref{lem:budget-blocks} shows that either $\psi$ factors
through the block action
$E^{\mathcal{B}}\leq\operatorname{Sym}(\mathcal{B})$ of degree $b<f$
with transitive kernel $N^{\mathcal{B}}$, or the restriction of
$\psi$ to the block stabilizer $E_C$ factors through the local action
$E_C^C\leq\operatorname{Sym}(C)$ of degree $a<f$ with transitive
kernel $N_C^C$. In either case, the induction hypothesis supplies the
desired primitive group $G\leq\operatorname{Sym}(\Omega)$ of degree
$n=|\Omega|\leq f$.

We now bound $k$ in terms of $n=|\Omega|$ according to whether the
primitive group $G$ has a nontrivial solvable normal subgroup.
Suppose first that $G$ has a nontrivial solvable normal subgroup. Any
minimal normal subgroup $V\triangleleft G$ contained in it is
abelian. Because $G$ is primitive on $\Omega$ and $V$ is a nontrivial
abelian normal subgroup, $V$ acts regularly on $\Omega$ and is
elementary abelian of order $n=p^d$ for a prime $p$ and an integer
$d\geq 1$. Identifying $\Omega$ with $V\cong\mathbb{F}_p^d$ embeds
$G$ into $\operatorname{AGL}_d(p)$ with $V$ as the translation
subgroup. Since $V$ is abelian and $\operatorname{Alt}(k)$ is
nonabelian simple, $\phi(V)=1$, so $\phi$ factors through
$G/V\leq\operatorname{GL}_d(p)$. Choose a prime $\ell\in\{3,5\}$
different from $p$. As in the proof of
Lemma~\ref{lem:budget-primitive-valuation}, lifting the exponent in
the order formula for $\operatorname{GL}_d(p)$ gives
$v_\ell(|G/V|)\leq v_\ell(|\operatorname{GL}_d(p)|)\leq 2d\log_2 p=2\log_2 n$.
Because $\operatorname{Alt}(k)$ is a quotient of $G/V$ and
$k\geq 64>\ell^2$, Legendre's formula gives
$k\leq 6v_\ell(|\operatorname{Alt}(k)|)\leq 6v_\ell(|G/V|)\leq 12\log_2 n\leq 12\log_2 f$.

Suppose from now on that $G$ has no nontrivial solvable normal
subgroup. Consider first the case where $G$ has a minimal normal
subgroup $M\triangleleft G$ that is not contained in $K$. Then
$M\cap K$ is normal in $G$ and properly contained in $M$, so
minimality of $M$ forces $M\cap K=1$. Because $M$ and $K$ are both
normal in $G$ with trivial intersection, $M$ centralizes $K$. Since
$K$ is transitive on $\Omega$, any subgroup of
$\operatorname{Sym}(\Omega)$ centralizing $K$ acts semiregularly on
$\Omega$, and therefore $|M|\leq n$. Moreover, $\phi(M)$ is a
nontrivial normal subgroup of $\operatorname{Alt}(k)$, so
$\phi(M)=\operatorname{Alt}(k)$ and the restriction
$\phi|_M\colon M\to\operatorname{Alt}(k)$ is an isomorphism. For
$k\geq 64$, this gives
$2^k<k!/2=|\operatorname{Alt}(k)|=|M|\leq n\leq f$ and hence
$k<\log_2 f$.

In the remaining case, every minimal normal subgroup of $G$ lies in
$K$, so the socle $S=\operatorname{soc}(G)$ satisfies $S\leq K$. As
in the proof of Lemma~\ref{lem:budget-primitive-valuation}, the
absence of nontrivial solvable normal subgroups gives $C_G(S)=1$, so
conjugation embeds $G$ into $\operatorname{Aut}(S)$, and
$S\cong T_1\times\cdots\times T_s$ is the direct product of its
minimal normal subgroups $T_1,\ldots,T_s$, each nonabelian simple.
Let $N_0\triangleleft G$ be the kernel of the conjugation action of
$G$ on the set $\{T_1,\ldots,T_s\}$. Then $S\leq N_0$, the quotient
$G/N_0$ embeds into $\operatorname{Sym}(s)$, and $N_0/S$ embeds into
$\prod_{i=1}^s\operatorname{Out}(T_i)$, which is solvable by
Schreier's conjecture~\cite[p.~133]{DixonMortimer1996}. Because
$S\leq K=\ker\phi$ and $N_0/S$ is solvable while
$\operatorname{Alt}(k)$ is nonabelian simple, we have $\phi(N_0)=1$.
Thus $\operatorname{Alt}(k)$ is a quotient of
$G/N_0\leq\operatorname{Sym}(s)$, which forces $k\leq s$ because
$k!/2>s!$ whenever $k>s\geq 1$.

Finally, we bound $s$ in terms of $n$. If $G$ has two distinct
minimal normal subgroups $M_1$ and $M_2$, each centralizes the other
and acts transitively on $\Omega$, so each acts regularly on $\Omega$
and $M_2$ is the full centralizer of $M_1$ in
$\operatorname{Sym}(\Omega)$. Thus $S=M_1M_2$ has order $n^2$, and
since each factor $T_i$ has order at least $2$, we get
$2^s\leq|S|=n^2$.

Suppose instead that $S$ is the unique minimal normal subgroup of $G$.
Then $G$ permutes $T_1,\ldots,T_s$ transitively by conjugation, since
the product of the factors in one orbit is a nontrivial normal
subgroup of $G$ contained in $S$. Fix $\omega\in\Omega$, let
$S_\omega$ be its stabilizer in $S$, and let
$\pi_i\colon S\to T_i$ be the coordinate projections. The nontrivial
normal subgroup $S$ of the primitive group $G$ is transitive, so
$G=SG_\omega$. Since $S$ normalizes each $T_i$, the stabilizer
$G_\omega$ alone permutes $T_1,\ldots,T_s$ transitively. Conjugation
by an element of $G_\omega$ maps $S_\omega$ onto itself and carries
each pair $(T_i,\pi_i)$ to some pair $(T_j,\pi_j)$, so all the
subgroups $\pi_i(S_\omega)\leq T_i$ have the same index $d$. If
$d\geq2$, then $S_\omega\leq\pi_1(S_\omega)\times\cdots\times
\pi_s(S_\omega)$ gives $n=[S:S_\omega]\geq d^s\geq2^s$. If $d=1$,
then $\pi_i(S_\omega)=T_i$ for every $i$. The subgroup
$S_\omega\cap T_i$ is normal in $S_\omega$, so it is normal in
$\pi_i(S_\omega)=T_i$. It is not all of $T_i$, because the set of
points fixed by the normal subgroup $T_i$ of the transitive group $S$
is $S$-invariant, so the nontrivial group $T_i$ fixes no point. Hence
$S_\omega\cap T_i=1$ for every $i$. The claim below then gives
$|S_\omega|^2\leq|S|$, so $|S|=n|S_\omega|\leq n|S|^{1/2}$ and
$2^s\leq|S|\leq n^2$.

In all cases $2^s\leq n^2$, so $s\leq2\log_2 n\leq2\log_2 f$, and
therefore $k\leq 2\log_2 f$.

It remains to prove the claim: \emph{if
$D\leq T_1\times\cdots\times T_s$ with nonabelian simple groups $T_i$,
$\pi_i(D)=T_i$ and $D\cap T_i=1$ for all $i$, then
$|D|^2\leq|T_1|\cdots|T_s|$.} We use induction on $s$. For $s=0$
the group $D$ is trivial. For $s=1$ the hypotheses cannot hold,
since they would give $D=T_1$ and $D\cap T_1=T_1\neq1$. Let
$s\geq2$ and put $K_i=D\cap\ker\pi_i$, so that $D/K_i\cong T_i$.
First, some $j\geq2$ satisfies $\pi_j(K_1)=1$. Otherwise, for every
$j\geq2$ the normal subgroup $\pi_j(K_1)$ of $T_j$ equals $T_j$, so
$K_1K_j=D$ and hence $\pi_1(K_j)=\pi_1(K_1K_j)=\pi_1(D)=T_1$. The
iterated commutator $[\cdots[[K_2,K_3],K_4],\ldots,K_s]$ then maps
onto $T_1$ under $\pi_1$, because $T_1$ is perfect. It lies in
$K_2\cap\cdots\cap K_s=D\cap T_1=1$, a contradiction. Fix such a
$j$. Since $\pi_j$ kills $K_1$, it induces a surjection from
$D/K_1\cong T_1$ onto $T_j$, which is an isomorphism because $T_1$ is
simple. So $|T_j|=|T_1|$. For $i\geq2$, the subgroup $\pi_i(K_1)$ is
normal in $T_i$, so it is $1$ or $T_i$. Let $J$ be the set of
$i\geq2$ with $\pi_i(K_1)=T_i$. Then $j\notin J$, and $K_1$ lies in
$\prod_{i\in J}T_i$, projects onto each of these factors, and meets
each of them trivially. By induction $|K_1|^2\leq\prod_{i\in J}|T_i|$,
and therefore
\begin{equation*}
 |D|^2=|T_1|^2|K_1|^2=|T_1||T_j||K_1|^2
 \leq|T_1||T_j|\prod_{i\in J}|T_i|\leq|T_1|\cdots|T_s|.
\end{equation*}
\end{proof}

If two permutations meet in exactly one row, their commutator cancels outside three rows. It also cancels the internal return maps, which is what makes the resulting row cycle independent of within-row coordinates.

\begin{lemma}[Commutator of two block-local lifts]
\label{lem:block-local-clean-commutator}
Suppose permutations $g,h$ have supports $P,Q$ with $P\cap Q=A$,
and $A^g\cap A=A^h\cap A=\varnothing$. Then, with right point
actions and $[g,h]=g^{-1}h^{-1}gh$,
\[
 [g,h]=\prod_{a\in A}(a,\ a^h,\ a^g).
\]
The displayed cycles are disjoint. In particular, if $A$ is one block
and the three displayed sets are three distinct blocks, the commutator
is a clean block $3$-cycle, regardless of the internal monodromy of
$g$ and $h$.
\end{lemma}

\begin{proof}
The sets $A,A^g,A^h$ are disjoint: the latter two lie in $P\setminus A$
and $Q\setminus A$, respectively. For $a\in A$, following the four
factors of $g^{-1}h^{-1}gh$ sends $a$ to $a^h$, sends $a^h$ to
$a^g$, and sends $a^g$ to $a$. Every other point lies either outside
both supports or in only one support with its intermediate inverse
image outside $A$; the two factors of the other permutation then fix
it and the remaining factors cancel. This proves the formula.
\end{proof}

One clean row cycle is not enough: arbitrary overlapping lifts need not satisfy the relations of an alternating group. We choose cycles with one common row and otherwise disjoint supports, so their coordinate identifications are compatible.

\begin{theorem}[Logarithmic-threshold clean complement]
\label{thm:logarithmic-clean-complement}
Let $G\leq\operatorname{Sym}(V)$ have a supplied block system of $m$
blocks, each of size $f$. Suppose the block action is
$\operatorname{Alt}(m)$ and its kernel $K$ is transitive on each block.
Choose an odd integer $k$ with
\begin{equation*}
 k>\max\{63,12\log_2 f\},\qquad 2k-1\leq m.
 \tag{LG1}
\end{equation*}
Then a clean complement $R\cong\operatorname{Alt}(m)$ to $K$ can be
constructed in polynomial time. More precisely, one can compute an
identification $V\cong B\times[m]$, $|B|=f$, under which $R$ acts
only on $[m]$ and $G=KR$, $K\cap R=1$.
Thus a threshold $m=O(\log(f+2))$ suffices under the transitive-block-
kernel hypothesis; no assumption $m>4f$ is used.
\end{theorem}

\begin{proof}
Write $B_i$ for the blocks and $\phi:G\to\operatorname{Alt}(m)$ for
the supplied action. We first prove a localization statement. For a
$k$-set $T\subseteq[m]$, put $L_T=\phi^{-1}(\operatorname{Alt}(T))$.
Here the ideal group $\operatorname{Alt}(T)$ fixes the complement of
$T$ pointwise.  For any $j\notin T$, the induced group $L_T[B_j]$ has transitive
normal subgroup $K[B_j]$. Its quotient is a homomorphic image of
$L_T/K\cong\operatorname{Alt}(k)$. By simplicity, this quotient is
trivial or isomorphic to $\operatorname{Alt}(k)$. The latter is
excluded by Lemma~\ref{lem:transitive-kernel-giant-threshold} and (LG1).
Consequently $L_T[B_j]=K[B_j]$.

It follows that the pointwise block stabilizer $(L_T)_{(B_j)}$
surjects onto $\operatorname{Alt}(T)$: match the action of any element
of $L_T$ on $B_j$ by an element of $K$ and cancel it. Each
$(L_T)_{(B_j)}$ is normal in $L_T$, since $B_j$ is $L_T$-invariant.
The intersection of finitely many normal subgroups surjecting onto
a perfect quotient still surjects onto that quotient: for two such
subgroups their commutator is contained in the intersection and has
full image, and induction handles the rest. Hence
\[
 \phi\bigl(G_{(V\setminus\bigcup_{i\in T}B_i)}\bigr)
       =\operatorname{Alt}(T).
 \tag{LG2}
\]
All groups in (LG2), and lifts of prescribed top permutations, are
computable by pointwise stabilizers and constructive membership.

Choose $k$-sets $T_1,T_2$ meeting in exactly one block index; (LG1)
ensures their existence. Use (LG2) to lift a $k$-cycle on $T_1$ to $g$
and a $k$-cycle on $T_2$ to $h$, fixing the complement pointwise.
The cycles are even because $k$ is odd. Each lift moves every point
of its supported blocks, so their supports meet in exactly one whole
block $A$. Lemma~\ref{lem:block-local-clean-commutator} produces a
clean block $3$-cycle $c=[g,h]$.

We now give a compatible completion; arbitrary overlapping conjugates
are not assumed to satisfy alternating-group relations. First suppose
$m=2r+1$ is odd. Choose a central block $B_0$ and partition the other
blocks into pairs $(B_{u_j},B_{v_j})$. Conjugate $c$ by suitable lifted
top permutations to obtain clean cycles with top
$(0,u_j,v_j)$. These supports share only $B_0$. Choose coordinates on
$B_0$, and transport them around each clean cycle to its two private
blocks. In this one coordinate system every chosen cycle is a pure
block permutation. The cycles $(0,u_j,v_j)$ generate
$\operatorname{Alt}(m)$: adjoining one such cycle to the alternating
group on the previously reached blocks enlarges that alternating group
by its two new points, as is verified by conjugating the central point
and multiplying the resulting $3$-cycles. Therefore their lifts
generate a clean $R\cong\operatorname{Alt}(m)$.

If $m$ is even, first perform the odd construction on $m-1$ blocks,
leaving one block $B_\ell$ out. This gives a clean
$R_0\cong\operatorname{Alt}(m-1)$ in fixed coordinates on those blocks.
Choose a pure $g\in R_0$ with top $(a,b,c)$. Choose a $k$-set $T$
containing $a,\ell$ and avoiding $b,c$; it exists by (LG1).
Using (LG2), choose $h$ with a $k$-cycle top on $T$ sending $a$ to
$\ell$. The supports of $g$ and $h$ meet in precisely $B_a$.
Their commutator is the clean cycle on
$B_a,B_\ell,B_b$ described in Lemma~\ref{lem:block-local-clean-commutator}.
Its correspondence from $B_b$ back to $B_a$ is exactly the inverse
of the already pure map $g|_{B_a}$. Define coordinates on the new
block $B_\ell$ by $h|_{B_a}$. The commutator is then pure in the
extended coordinates. Together with $R_0$ it generates a clean
$\operatorname{Alt}(m)$.

Every operation used above is polynomial in $mf$. A pure top group
has trivial intersection with the block kernel, and it maps onto the
full top group. Thus it is the claimed complement.
\end{proof}

Once row permutations are independent, symmetry forces most within-row variation to occur independently in the rows. The remaining coupling is an abelian zero-sum constraint and a common quotient action. This is the precise information used by the row terminal.

\begin{theorem}[Full normal form under the logarithmic threshold]
\label{thm:logarithmic-gat-normal-form}
Under Theorem~\ref{thm:logarithmic-clean-complement}, the entire
row normal form follows constructively. In its clean
coordinates, let $H=K[B]\leq\operatorname{Sym}(B)$. There exist
$N^*\triangleleft N\triangleleft H$ and
$M^*\triangleleft M\triangleleft G$, with $M\leq K$, such that
\[
 M^*=(N^*)^m,\qquad M=K\cap N^m,
\]
\[
 M/M^*=\{(a_1,\ldots,a_m)\in (N/N^*)^m:
                              \textstyle\sum_i a_i=0\},
\]
where $N/N^*$ is abelian, and $K/M$ is the diagonal $H/N$.
Moreover $MR\triangleleft G$ and $G/(MR)\cong H/N$.
The groups, quotient maps, and lifts are polynomial-time computable.
\end{theorem}

\begin{proof}
In the clean coordinates $K\leq H^m$ is subdirect and is normalized
by pure $\operatorname{Alt}(m)$. Define $N$ as the projection to one
coordinate of the kernel fixing another coordinate, and $N^*$ as the
projection of the subgroup supported on one coordinate. Alternating
coordinate permutations make these definitions independent of the
chosen coordinates. Both are normal in $H$.

For $n\in N$, choose $x=(1,n,x_3,\ldots,x_m)\in K$.
For the pure cycle $\tau=(1,2,3)$, direct multiplication gives
$[\tau,x]=(x_3^{-1},n,n^{-1}x_3,1,\ldots,1)$.
Conjugate this element by the even permutation $(1,4)(3,5)$ to obtain
$y$ with first three coordinates $(1,n,1)$. Then
$[y,\tau^{-1}]=(n,n^{-1},1,\ldots,1)$, so
\[
 (n,n^{-1},1,\ldots,1)\in K.
 \tag{LG3}
\]
The argument uses $m\geq5$, which is assured by (LG1).
Conjugating (LG3) gives such transfers between any two coordinates.
Commuting transfers supported on $(1,2)$ and $(1,3)$ shows
$[N,N]\leq N^*$. Thus $A=N/N^*$ is abelian, and
$(N^*)^m\leq K$ by the definition of $N^*$ and conjugation.

For $x\in K$, compare $x$ with a pure $3$-cycle conjugate. Their
quotient has at least one identity coordinate, so every coordinate
of that quotient lies in $N$ by the definition of $N$ and coordinate
symmetry. It follows that all coordinates of $x$ agree modulo $N$.
Therefore $K\to H/N$, given by one coordinate, is onto with kernel
$M=K\cap N^m$.

Modulo $M^*=(N^*)^m$, the transfers (LG3) generate the full zero-sum
subgroup of $A^m$. Conversely, if an element of $M/M^*$ had nonzero
total, multiplying by transfers would concentrate that total in one
coordinate. Its lift would give a single-coordinate element of $K$
with nontrivial image in $N/N^*$, contradicting the definition of
$N^*$. Hence $M/M^*$ is exactly the zero-sum subgroup.

All these groups are normal as asserted: $N,N^*$ are normal in $H$,
coordinate permutations preserve their powers, and $K$ is normal in
$G=KR$. The action of $R$ on the diagonal quotient $K/M$ is trivial,
so $G/M=(K/M)\times(RM/M)$ and $MR\triangleleft G$.

Computationally, $H,N,N^*$ are projections of pointwise block
stabilizers in $K$. The group $M$ is the preimage of $N$ under the
onto first-coordinate map $K\to H$, so generators follow by lifting
generators of $N$ and adjoining that map's kernel. The direct-product
group $M^*$ is explicit. Standard constructive quotient operations
then give all displayed maps and lifts. No enumeration of an
alternating quotient or a general coset-intersection oracle is used.
\end{proof}

\subsection{Exact quotient algorithms and a transitive-kernel terminal}
\label{sec:quotient-algorithms}

This section gives the exact quotient subroutine used by the natural-giant
case of the final recursion. Its input retains explicit subgroup cosets,
and every output is a full transporter coset.

\subsubsection{The implicit object model}

Let \(H\leq\Sym(B)\), \(|B|=b\), and let \(N\triangleleft H\). An
\emph{orbit color} specifies a subgroup \(L\leq H\) containing \(N\).
The objects of this color are right cosets \(Lh\), with the right
\(H\)-action. Different orbit colors remain distinct, even if their
subgroups happen to agree. The input consists of two colored multisets
\(X,Y\), each with at most \(M\) explicitly listed objects. Subgroups and
cosets are represented by generators and one representative. We seek
\[
 \Iso_H(X,Y)=\{h\in H:X^h=Y\}.
\]
Since \(N\) fixes every object, this action factors through \(H/N\).
Point stabilizers and point transporters are ordinary degree-\(b\) subgroup
and coset intersections. A general polynomial-time intersection algorithm
is not assumed.

A solvable quotient does not make its given permutation cover solvable. Even so, a solvable quotient always arises from a solvable subgroup of the cover, and we use this fact to bound the derived length of the quotient.

\begin{lemma}[Solvable supplement]
\label{lem:solvable-supplement}
Let $N \triangleleft H$ be finite groups. If $H/N$ is solvable, then there exists a solvable subgroup $S \leq H$ such that $SN = H$.
\end{lemma}

\begin{proof}
We argue by induction on $|N|$, following the Frattini reduction in the proof of \cite[Theorem~1.3]{LM}. If $N$ is nilpotent, then $N$ is solvable. Since $H/N$ is also solvable, the group $H$ itself is solvable, so we can take $S = H$.

Suppose now that $N$ is not nilpotent. A finite group is nilpotent if and only if every Sylow subgroup is normal. Thus $N$ has a Sylow subgroup $P$ that is not normal in $N$, so its normalizer $N_N(P)$ is a proper subgroup of $N$. For every $h \in H$, the conjugate $P^h$ is also a Sylow subgroup of $N$ because $N$ is normal in $H$. By Sylow's conjugacy theorem in $N$, there exists $n \in N$ such that $P^n = P^h$, which gives $hn^{-1} \in N_H(P)$. Writing $h = (hn^{-1})n$ shows that $H = N_H(P)N$.

Set $H_1 = N_H(P)$ and $N_1 = N_N(P) = H_1 \cap N$. The subgroup $N_1$ is normal in $H_1$. The inclusion of $H_1$ into $H$ induces an isomorphism from $H_1/N_1$ onto $H_1 N / N = H/N$, so $H_1/N_1$ is solvable. Because $N_1$ is a proper subgroup of $N$, we have $|N_1| < |N|$. The induction hypothesis applied to $N_1 \triangleleft H_1$ gives a solvable subgroup $S \leq H_1$ with $S N_1 = H_1$. Then $S \leq H$ and $SN = S N_1 N = H_1 N = H$.
\end{proof}

In Theorem~\ref{thm:solvable-multicoset} we use Lemma~\ref{lem:solvable-supplement} only to bound the derived length of $H/N$, and we never construct $S$. Let $S \leq H$ be a solvable subgroup with $SN = H$. The inclusion of $S$ into $H$ induces a surjective homomorphism from $S$ onto $SN/N = H/N$ with kernel $S \cap N$. Thus the derived length of $H/N$ is at most the derived length of the solvable permutation group $S \leq \Sym(B)$, which is at most $(5/2)\log_3 b$ by Dixon's theorem \cite[Section~12]{BA21}. Our algorithm works directly with the pair $(H, N)$.

\subsubsection{Solvable quotients: recursive profiles and affine gluing}

The coset spaces may be enormous, but only the explicitly listed objects must be matched. The recursion first matches profiles modulo the last abelian derived subgroup. An affine compatibility condition then forces the separately chosen translations to come from one common group element.

\begin{theorem}[Solvable implicit multicoset transport]
\label{thm:solvable-multicoset}
In the preceding model, if \(H/N\) is solvable, the full transporter is
computable deterministically in
\[
 \exp\bigl(O(\log(b+M+2)\log(b+2))\bigr)
\]
time. The implied constant is absolute. No string-isomorphism oracle
occurs in this bound.
\end{theorem}

\begin{proof}
We work directly with the pair $(H, N)$, maintaining the invariant that every listed object $Lh$ satisfies $N \leq L \leq H$ and $h \in H$. Induct on the derived length $\ell$ of the acting quotient $H/N$. If $\ell = 0$, then $H = N$, the action is trivial, and equality of the two multisets decides the problem. Otherwise let $A/N$ be the last nontrivial derived subgroup of $H/N$, found from the derived series of $H$ together with $N$ \cite[item~(3.12), p.~145]{Luks93}. The subgroup $A$ is normal in $H$, the quotient $A/N$ is abelian, and $H/A$ has derived length at most $\ell - 1$.

First partition $X$ into its nonempty intersections $S_1, \ldots, S_r$ with $A$-orbits, and likewise partition $Y$. For an object $x = Lh$ with $N \leq L \leq H$ and $h \in H$, its stabilizer $H_x = L^h$ is contained in $H$ and therefore normalizes $A$. Thus two objects $Lh$ and $L'k$ lie in the same $A$-orbit if and only if they have the same color and $h^{-1} k \in H_x A$, and normalized-product factorization \cite[Theorem~3.1(viii)(c)]{LM} computes a point transporter in $A$ in polynomial time. Likewise the point stabilizer $A_x = A \cap H_x$ is a normalized-subgroup intersection \cite[Theorem~3.1(viii)(b)]{LM}. For an abelian action on one orbit, the stabilizer of one point fixes that entire orbit. Consequently, transport of two submultisets inside one $A$-orbit is polynomial: fix one source point, try its at most $M$ target images, and test one representative for each image. This also computes $B_i = \Aut_A(S_i)$.

Next compute all pairwise subconfiguration transporters $\Iso_H(S_i, T_j)$. Fix $x = Lh \in S_i$ and try every $y = L'k \in T_j$. Because $L \leq H$ and $h, k \in H$, the objects $x$ and $y$ lie in the same $H$-orbit if and only if they have the same color, in which case $\tau = h^{-1} k \in H$ satisfies $x^\tau = y$. Align the target first: the recursive problem is $\Iso_{H_x}(S_i, T_j^{\tau^{-1}})$, and its answer is multiplied by $\tau$ on the right. Since $A$ is normal in $H$, both aligned subconfigurations lie in the $A$-orbit of $x$. The normal kernel of this call is $A_x = A \cap H_x$. This kernel fixes the whole orbit and
\begin{equation*}
 H_x/A_x \cong H_x A / A \leq H/A.
\end{equation*}
Thus every such call has quotient derived length at most $\ell - 1$, the same cover degree, and at most $M$ objects. To construct its input, rebase the explicitly occurring objects into their $H_x$-orbits. Every object $z$ in $S_i$ or $T_j^{\tau^{-1}}$ has the form $z = x^a$ for an element $a \in A$ found by normalized-product factorization in $H_x A$. The map $x^a \mapsto A_x a$ identifies the $A$-orbit of $x$ with the finite abelian group $A/A_x$, and under this identification an element $g \in H_x$ acts on $A/A_x$ by conjugation $A_x a \mapsto A_x a^g$. Because $A_x$ acts trivially on $A/A_x$, the linear action of $H_x$ on $A/A_x$ factors through the solvable quotient $H_x/A_x$. Using the abelian coordinates and the primary-layer vector-stabilizer procedure described below, we compute the stabilizer $H_x \cap H_z$ (which is the stabilizer of $A_x a$ in $H_x$ and contains $A_x$) and pairwise $H_x$-transporters between these objects in polynomial time. No orbit is expanded, and rebasing does not increase the number of listed occurrences. Combining the recursive nonempty cosets over the possible images of $x$ gives the entire pairwise transporter. Polynomially many recursive calls suffice.

Replace \(S_i\) by the profile \(P_i=A\cdot S_i\), its orbit as a
multiset under \(A\). Normality of \(A\) gives
\[
 H_{P_i}=A\Aut_H(S_i),\qquad
 \operatorname{Trans}_H(P_i,P_j)=A\Iso_H(S_i,S_j),
\]
with a consistent choice of multiplication side. The pairwise results
therefore give explicit coset representations of all profiles.
Recursively transport the multiset of profiles under \(H/A\).
This returns a coset \(B\tau\), with \(A\leq B\).
At this point different \(A\)-translations may align the different
subconfigurations. They must be replaced by one common translation.

For automorphisms, put
\[
 W=\bigoplus_{i=1}^r A/B_i,\qquad
 \delta:A\longrightarrow W,\quad
 \delta(a)=(aB_i)_i,\qquad D=\delta(A).
\]
These are finite abelian groups, with \(N\) factored out implicitly.
An element \(g\in B\) permutes the profiles by a permutation \(\pi_g\).
The profiles are distinct because their subconfigurations lie in
distinct \(A\)-orbits. There is a unique offset
\(u(g)\in W\), computed by abelian orbit transport, such that
\[
 S_i^g=S_{\pi_g(i)}^{\,u(g)_{\pi_g(i)}}.
\]
Also \(B_i^g=B_{\pi_g(i)}\), so conjugation and coordinate permutation
give a linear automorphism \(R_g\) of \(W\). On the independent
translations of the \(S_i\), the action is exactly
\[
 w\longmapsto R_g(w)+u(g).
\]
This is an affine group action; in particular its offsets satisfy the
cocycle identity. Moreover \(R_g\delta(a)=\delta(a^g)\), so \(D\) is
invariant. For \(a\in A\), \(R_a=1\) and \(u(a)=\delta(a)\).
The affine action on \(W/D\) therefore factors through \(B/A\).

A coset $gA$ contains an automorphism of $X$ if and only if $u(g) \in D$. Compute the stabilizer $C/A$ of zero in $W/D$. For each generator $c$ of its full preimage, solve $\delta(a) = -u(c)$. Then $ca$ is an actual automorphism. Each subgroup $B_i = \Aut_A(S_i)$ contains $N$, and $A/N$ is abelian, so every $B_i$ is normal in $A$ and $\Aut_A(X) = \bigcap_i B_i$ is computed by repeated normalized-subgroup intersection \cite[Theorem~3.1(viii)(b)]{LM}. The corrected generators $ca$, together with $\Aut_A(X) = \bigcap_i B_i$, generate all of $\Aut_H(X)$ because they map onto the exact liftable subgroup $C/A$ and contain its kernel. In particular, individually unliftable quotient generators are not simply discarded.

For transport from \(X\) to \(\!Y\), first use the profile transporter
\(\tau\). Write \(B_X\) for the source profile stabilizer and use
the target stabilizer \(B_Y=\tau^{-1}B_X\tau\).
Form \(W_Y,D_Y\) from the target subconfigurations. In these
coordinates \(X^\tau\) has an explicitly computed offset
\(w_0\in W_Y\). Find \(g\in B_Y\) transporting
\(w_0+D_Y\) to zero under the target affine action, and choose
\(a\in A\) to cancel the remaining common translation.
Then \(\tau g a\) transports \(X\) to \(Y\).
Together with the source automorphism group already constructed,
it gives the full transporter coset.

\paragraph{Constructing the abelian coordinates.}
The preceding operations use explicit presentations, not an oracle
for an abelian group of potentially exponential order.
Choose compressed generators \(a_1,\ldots,a_t\) of \(A\) modulo
\(N\), and put
\[
 U_i=\langle N,a_1,\ldots,a_i\rangle,\qquad
 e_i=|U_i|/|U_{i-1}|,\qquad U_0=N.
\]
Since \(A/N\) is abelian, each successive quotient is cyclic.
Constructive permutation-group membership expresses
\(a_i^{e_i}\) by a straight-line program in generators of
\(U_{i-1}\) \cite[Theorem~3.1(v) and Section~2.3]{LM}. Evaluate that program as an integer exponent vector,
assigning zero to generators of \(N\). This gives the relations
\[
 e_i[a_i]=\sum_{j<i}c_{ij}[a_j].
\]
They are a complete presentation: the presented abelian group
maps onto \(A/N\) and has order
\(\prod_i e_i=|A/N|\). The orders and the coefficients have
polynomial bit length. For the coefficients, each multiplication
in a polynomial-length straight-line program adds exponent vectors,
and inversion negates them.

Integer Hermite and Smith normal forms, including their
transformation matrices, have polynomial bit complexity
\cite{KannanBachem79}. Storjohann and Labahn \cite[Theorem~12]{StorjohannLabahn96}
give the row Hermite form of a matrix of full column rank with its
unimodular transformation matrix. These forms provide coordinates, relation tests,
solutions to homomorphism equations, and lifts. Append the
coordinates of generators of \(B_i\) to obtain \(A/B_i\);
use block-diagonal relations for \(W\), and append
\(\delta(a_j)\) to obtain \(W/D\). A compressed permutation
generating set has at most \(O(b^2)\) members, so these matrices
have dimension \(O(Mb^2)\) and polynomial-bit entries.
The invariant factors of the represented finite abelian groups
have no prime divisors above \(b\). Trial division by primes
at most \(b\) therefore suffices to obtain their primary
decompositions; no unrestricted integer-factorization algorithm
is assumed. The integer relation coefficients themselves need
not have this restriction on their prime divisors.

\paragraph{Solving the affine compatibility condition.}
Set \(V=W/D\), and split it into its characteristic primary
components \(V_p\). Use the filtrations
\[
 V_p\supseteq pV_p\supseteq p^2V_p\supseteq\cdots\supseteq0.
\]
After transport modulo one level, choose and apply one
representative of the resulting transporter coset. Work next
with the target stabilizer, conjugating the source stabilizer
when necessary. This subgroup preserves the target fiber.
If \(v\) is its chosen origin, the next action is
\[
 z\longmapsto R_g(z)+(v^g-v)
       \quad\text{in }p^jV_p/p^{j+1}V_p.
\]
The offset obeys the cocycle identity. Homogenization by one coordinate therefore gives an explicit linear representation over $\F_p$. Because the affine action on $V$ factors through the solvable quotient $B/A$, the linear image of the permutation group in this representation is an extension of a quotient of $B/A$ by the abelian translation kernel and is therefore solvable. In Luks and Miyazaki's vector-stabilizer algorithm \cite[Theorem~1.4 and Section~5.2]{LM}, solvability is used only in the linear image (via its derived series and preimages in the permutation cover), so the algorithm applies whenever the linear image is solvable, without requiring the permutation cover itself to be solvable. For a vector transporter, we use the two-copy wreath extension on the direct sum of two copies of the vector space. Its linear image lies in the wreath product of the solvable linear image with $C_2$ and is again solvable, and its stabilizer of $(v, w)$ contains a copy-swapping element if and only if $v$ and $w$ lie in one orbit. The cover degree increases to $2b$ only for this nonrecursive vector calculation.

The bit bounds are uniform throughout:
\[
 \log_2|V|\leq M\log_2(b!),\qquad
 \sum_{p,j}\dim_{\F_p}(p^jV_p/p^{j+1}V_p)
       \leq\log_2|V|.
\]
Every characteristic satisfies \(p\leq b\), and all fields
have extension degree one. This matters because the vector
algorithm is polynomial in \(p\), rather than only in
\(\log p\). Retain the full affine transformations on \(V\)
in the finite-abelian coordinates, not only their reductions on
the current vector-space layer. Generators of each restricted
subgroup are expressed by straight-line programs in the preceding
generators \cite[Theorem~3.1(v)]{LM}. Evaluate these programs in the full affine action,
then recenter at the next fiber and reduce to obtain its matrices.
Thus the entire affine calculation is polynomial in \(b+M\),
including lifting. No faithful degree-\(b\) representation of
the quotient itself is required.

\paragraph{The recursion bound.}
Compute the pairwise comparisons among subconfigurations in both
inputs. There are at most \(4M^2\) ordered pairs, each with at
most \(M\) candidate images of its chosen point, followed by
one profile call. All these calls retain the same cover degree,
at most \(M\) listed objects, and quotient derived length at
most \(\ell-1\). Consequently, for absolute \(C\) and a fixed
polynomial \(P\),
\[
 T_\ell(b,M)\leq C(M+1)^3T_{\ell-1}(b,M)+P(b,M).
\]
The polynomial is independent of $\ell$, and generating sets and coset descriptions are compressed after every operation. Finally, by Lemma~\ref{lem:solvable-supplement} and Dixon's bound $\operatorname{dl}(S) \leq (5/2)\log_3 b$ \cite[Section~12]{BA21}, the acting quotient $H/N$ has derived length $O(\log b)$. This proves the stated bound.
\end{proof}

\subsubsection{A radical-free order bound and its constructive use}

For a finite group \(Q\), write \(a(Q)\) and \(b(Q)\) for the products
of the orders of its abelian and nonabelian composition factors.
These symbols are distinct from the cover degree \(b\).

For a quotient $Q = H/N$ above a transitive kernel, the product $b(Q)$ of its nonabelian composition-factor orders is controlled. We want to find a subgroup $R'$ between $N$ and $H$ such that $R'/N$ is solvable and the index $[H:R']$ is at most $b(Q)^2$. After we pass to the radical-free quotient $G = H/R$ by the radical preimage $R$, its socle $E$ has order at most $b(Q)$. The outer automorphisms of the simple factors of $E$ are solvable, so they fall into the solvable radical of $G/E$, and only the permutation action on the simple factors remains above that radical. Combining $R$ with a solvable supplement of the socle preimage inside the outer-radical preimage gives the desired subgroup $R'$.

\begin{lemma}[Solvable-quotient index bound]
\label{lem:radical-free-order}
Let $N \triangleleft H$ be finite groups and put $Q = H/N$. Let $R \triangleleft H$ be the full preimage of $\operatorname{Rad}(H/N)$, let $E_H \triangleleft H$ be the full preimage of $\operatorname{soc}(H/R)$, and let $K^*_H \triangleleft H$ be the full preimage of $\operatorname{Rad}(H/E_H)$. There exists a solvable subgroup $S_1 \le K^*_H$ with $S_1 E_H = K^*_H$. For every such subgroup $S_1$, the product $R' = S_1 R$ is a subgroup with $N \triangleleft R' \le H$ such that $R'/N$ is solvable and
\begin{equation*}
  [H : R'] \le b(Q)^2.
\end{equation*}
\end{lemma}

\begin{proof}
Since $E_H$ is normal in $H$, it is normal in $K^*_H$. By definition of $K^*_H$, the quotient $K^*_H/E_H = \operatorname{Rad}(H/E_H)$ is solvable. Lemma~\ref{lem:solvable-supplement} applied to $E_H \triangleleft K^*_H$ therefore gives a solvable subgroup $S_1 \le K^*_H$ with $S_1 E_H = K^*_H$.

Fix any solvable subgroup $S_1 \le K^*_H$ with $S_1 E_H = K^*_H$, and put $R' = S_1 R$. Because $R$ is normal in $H$ and $S_1 \le K^*_H \le H$, the product $R'$ is a subgroup of $K^*_H$ with $R \triangleleft R'$. Since $N \le R \le R' \le H$ and $N$ is normal in $H$, the subgroup $N$ is also normal in $R'$. By the second isomorphism theorem, $R'/R = S_1 R / R$ is isomorphic to $S_1 / (S_1 \cap R)$, which is a quotient of the solvable group $S_1$ and is therefore solvable. Because $R/N = \operatorname{Rad}(H/N)$ is also solvable, the extension $R'/N$ of $R'/R$ by $R/N$ is solvable.

It remains to bound the index $[H : R'] = [H : K^*_H][K^*_H : R']$. We bound the two factors by $b(Q)$ in turn. Since $R \le E_H$ and $S_1 E_H = K^*_H$, we have $R' E_H = S_1 R E_H = S_1 E_H = K^*_H$. By the second isomorphism theorem, the index $[K^*_H : R'] = [R' E_H : R']$ equals $[E_H : E_H \cap R']$. Because $R \le E_H \cap R'$, this index is at most $[E_H : R] = |\operatorname{soc}(H/R)|$. Put $G = H/R$ and $E = \operatorname{soc}(G)$. Since $R/N = \operatorname{Rad}(Q)$ is solvable, its composition factors are abelian, so $b(G) = b(Q)$. Every composition factor of $E$ is a composition factor of $G$. Moreover, the quotient $G = Q/\operatorname{Rad}(Q)$ has trivial solvable radical, so every minimal normal subgroup of $G$ is nonabelian and all composition factors of $E$ are nonabelian. Thus $|E| = b(E) \le b(G) = b(Q)$, which gives
\begin{equation*}
  [K^*_H : R'] \le |E| \le b(Q).
\end{equation*}

To bound $[H : K^*_H]$, recall that the socle $E$ of the radical-free group $G = H/R$ is a direct product of minimal normal subgroups of $G$, and each of these is a direct product of nonabelian simple groups. So $E = T_1 \times \dots \times T_t$ with nonabelian simple groups $T_1, \dots, T_t$, and these factors are exactly the minimal normal subgroups of $E$. If $G$ is trivial, then $K^*_H = H$ and $[H : K^*_H] = 1 \le b(Q)$. Assume now that $G$ is nontrivial, so $t \ge 1$. The normal subgroup $C_G(E)$ is trivial, because a nontrivial normal subgroup of $G$ in $C_G(E)$ would contain a minimal normal subgroup of $G$ lying in $E \cap C_G(E) = Z(E)$, whereas $E$ has trivial center. Conjugation on $E$ therefore embeds $G$ into $\Aut(E)$. Conjugation by $G$ permutes the $t$ simple factors $T_1, \dots, T_t$, giving a homomorphism $\varphi \colon G \to \Sym(t)$ with image $P = \varphi(G) \le \Sym(t)$ and kernel $K = \ker(\varphi) \triangleleft G$.

Every element of $K$ normalizes each factor $T_i$, and every inner automorphism of $E$ is induced by an element of $E \le K$. Taking the conjugation action on each factor modulo $\operatorname{Inn}(T_i)$ therefore gives a homomorphism from $K$ to $\prod_{i=1}^t \Out(T_i)$ whose kernel is $E \cdot C_K(E) = E$. Thus $K/E$ embeds into $\prod_{i=1}^t \Out(T_i)$. By Schreier's conjecture, which follows from the classification of finite simple groups~\cite[p.~133]{DixonMortimer1996}, the outer automorphism group $\Out(T_i)$ of each nonabelian simple group $T_i$ is solvable. A direct product of solvable groups is solvable, and a subgroup of a solvable group is solvable, so $K/E$ is a solvable normal subgroup of $G/E$. Hence $K/E$ lies in $\operatorname{Rad}(G/E) = K^*_H/E_H$. Since $G/E$ is isomorphic to $(H/R)/(E_H/R) \cong H/E_H$, the third isomorphism theorem gives
\begin{equation*}
  [H : K^*_H] = [G/E : \operatorname{Rad}(G/E)] \le [G/E : K/E] = [G : K] = |P|.
\end{equation*}

Finally, we bound $|P| = a(P)b(P)$. Since $P \le \Sym(t)$ has degree $t \ge 1$, Proposition~6.1 of~\cite{GMP} gives $a(P) \le 24^{(t-1)/3} \le 24^t$. Every nonabelian simple group $T_i$ has order at least $60 \ge 24$, so $a(P) \le 24^t \le \prod_{i=1}^t |T_i| = |E|$. Furthermore, $E$ is normal in $K$, so $b(E) \le b(K)$, and since all composition factors of $E$ are nonabelian we have $|E| = b(E) \le b(K)$. Multiplying by $b(P) = b(G/K)$ gives
\begin{equation*}
  |P| = a(P)b(P) \le |E|b(P) \le b(K)b(G/K) = b(G) = b(Q).
\end{equation*}
Therefore $[H : K^*_H] \le b(Q)$, and multiplying the bounds on $[H : K^*_H]$ and $[K^*_H : R']$ yields $[H : R'] \le b(Q)^2$.
\end{proof}

We next compute the full preimage in $H$ of $\operatorname{Rad}(H/N)$
without constructing a faithful permutation representation of $H/N$,
when $N$ is transitive on $B$. We first build a descending chain
$H=U_0\geq U_1\geq\cdots\geq U_L=N$ of normal subgroups of $H$
containing $N$ such that every nonabelian factor $U_i/U_{i+1}$ is a
chief factor of $H/N$. Abelian factors of the chain do not need to be
minimal, because only the nonabelian factors enter the centralizer
intersection below. Given the current term $U$ with $N<U\leq H$, we
construct the next term $V$ as follows. Form $D=[U,U]N$, which is
normal in $H$ because $U$ and $N$ are normal in $H$. If $D<U$, set
$V=D$, so that $U/V$ is abelian. Otherwise $U=[U,U]N$, so $U/N$ has no
nontrivial abelian quotient. To isolate a bounded quotient at the top
of $U$, start with $J=U$ and repeat while $NJ=U$: choose a largest
orbit of $J$ on $B$, find a block system of $J$ on that orbit on
which $J$ acts primitively (its blocks are maximal, or they are single
points if $J$ is already primitive on the orbit), and replace $J$ by
the kernel $J'$ of the action of $J$ on those blocks. Each step halves $|J|$ at least, so after at most $b^2$ steps
we reach the last step before $NJ'<U$. At that step $NJ=U$, the
subgroup $Y=NJ'$ is normal in $U$ with $N\leq Y<U$, and $U/Y$ is a
quotient of the primitive permutation group $J/J'$ of degree $d\leq b$.
Every quotient $\Gamma/\Lambda$ of a primitive group $\Gamma$ of
degree $d$ has order at most
$\max\{24!,\,d^{1+\log_2 d}\}\cdot b(\Gamma/\Lambda)$. Indeed, if
$d\leq24$ then $[\Gamma:\Lambda]\leq d!\leq24!$, and if
$|\Gamma|\leq d^{1+\log_2 d}$ then $[\Gamma:\Lambda]\leq|\Gamma|$ gives
the bound. Otherwise,
by \cite[Theorem~1.1]{Maroti}, $\Gamma$ lies in
$\Sym(m)\wr\Sym(r)$ in the product action on $r$-tuples of
$k$-subsets of $[m]$, and contains $S=\Alt(m)^r$. Here $m\geq5$. To
see this, suppose $m\leq4$ and put $c=\binom mk$, so that $d=c^r$.
Then $c\leq6$, so $d>24$ forces $r\geq2$. We have $c\geq m\geq2$
because $d>1$, and $m!\,r\leq m^{1+r}$ for $m\in\{2,3,4\}$ and
$r\geq2$. Hence
$|\Gamma|\leq(m!)^r r!\leq(m!\,r)^r\leq c^{(1+r)r}\leq
d^{1+\log_2 d}$, contrary to the case assumption. Then
$[\Gamma:S]\leq2^r r!\leq d^{1+\log_2 d}$, because $d\geq5^r$. A normal
subgroup of $S$ is a product of some of its simple factors, so
$S\Lambda/\Lambda\cong S/(S\cap\Lambda)$ is a product of copies of
$\Alt(m)$, and its order is at most $b(\Gamma/\Lambda)$. Hence
$[\Gamma:\Lambda]\leq[\Gamma:S]\,|S\Lambda/\Lambda|\leq
d^{1+\log_2 d}\,b(\Gamma/\Lambda)$. Applied to the quotient $U/Y$ of
$J/J'$, this gives $[U:Y]\leq\max\{24!,\,d^{1+\log_2 d}\}\cdot b(U/Y)$. Since $N$ is transitive on
$B$, Theorem~1.4 of \cite{GMP} gives
$b(U/Y)\leq b(H/N)\leq b^{\log_2 b}$, so $[U:Y]\leq b^{O(\log b)}$.
Enumerate a right coset transversal of $Y$ in $U$ in time polynomial in
$b$ and $[U:Y]$. Scan the transversal and, for each representative $r$,
replace $Y$ by the normal closure of $\langle Y,r\rangle$ in $U$
whenever that normal closure is a proper subgroup of $U$. The resulting
subgroup $M$ is a maximal proper normal subgroup of $U$ containing $N$,
so $U/M$ is simple, and it is nonabelian because $U=[U,U]N$. Next,
initialize $W=M$ and repeatedly intersect $W$ with $hWh^{-1}$ for each
generator $h$ of $H$ until a full pass over the generators leaves $W$
unchanged, which takes at most $b^2$ passes. The subgroup $W$ is an intersection of conjugates $M'=hMh^{-1}$ with
$h\in H$. Each $M'$ is normal in $U$ with $U/M'\cong U/M$ nonabelian
simple. When $W$ is intersected with one more such $M'$, the group
$WM'/M'$ is normal in the simple group $U/M'$, so either
$W\cap M'=W$ or $W/(W\cap M')\cong WM'/M'=U/M'$. Thus $U/W$ has a
normal series whose factors are all isomorphic to $U/M$, so all
composition factors of $U/W$ are nonabelian, and
$[U:W]=b(U/W)\leq b(H/N)\leq b^{\log_2 b}$. Enumerate a right coset
transversal of $W$ in $U$ and run the same greedy enlargement using
normal closures in $H$. This yields a maximal proper $H$-normal
subgroup $V$ of $U$ containing $W$, so $U/V$ is a nonabelian chief
factor of $H/N$. After at most $b^2$ steps, the chain reaches $N$.

Now let $U$ and $V$ be consecutive terms of this chain in $H$, with
$N\leq V<U\leq H$, such that the factor $U/V$ is nonabelian. By
construction, $U/V$ is a nonabelian chief factor of $H/N$. We need the
preimage in $H$ of
\begin{equation*}
 C_{H/V}(U/V),
\end{equation*}
which consists of the elements $h\in H$ such that
$huh^{-1}u^{-1}\in V$ for all $u\in U$. Since $H$ normalizes both $U$
and $V$, conjugation by elements of $H$ permutes the right cosets $Vu$
of $V$ in $U$, and $h(Vu)h^{-1}=V\,huh^{-1}$. So the coset $Vu$ is fixed
by $h$ exactly when $huh^{-1}u^{-1}\in V$, and the required preimage is
the kernel of this action. We enumerate right coset
representatives of $V$ in $U$ by closing $\{1\}$ under multiplication
by generators of $U$ and testing membership in $V$ for ratios, in time
polynomial in the cover degree $b$ and $[U:V]$. We then form the
permutation action of each generator of $H$ on these $[U:V]$ cosets and
compute the kernel of this action in $H$ in time polynomial in $b$ and
$[U:V]$. All these centralizer preimages are normal in $H$, so we
intersect them using normalized-subgroup intersections.

Their intersection is the radical preimage. Every solvable normal
subgroup of $H/N$ centralizes each nonabelian chief factor $U/V$,
because its normal intersection with $U/V$ is solvable and must
therefore be trivial. Conversely, a normal subgroup of $H/N$ that
centralizes every nonabelian factor of the chain meets each nonabelian
chief factor $U/V$ in a normal subgroup of $H/N$ centralized by $U/V$,
which is trivial because $U/V$ has trivial center. Restricting the
chain to that normal subgroup therefore leaves only abelian factors, so
the subgroup is solvable.

\paragraph{Computing $R'$.}
We now compute a subgroup $R'$ as in Lemma~\ref{lem:radical-free-order}, again when $N$ is transitive on $B$. Let $Q=H/N$, let $R$ be the radical preimage computed above, and let $E_H$ and $K^*_H$ be as in Lemma~\ref{lem:radical-free-order}. There are four steps. We compute $E_H$ from a chain, then $K^*_H$ by the routine above, and then a list of coset representatives of $R$ in $E_H$. Last, we run the descent from the proof of Lemma~\ref{lem:solvable-supplement} on $E_H\triangleleft K^*_H$, but modulo $R$, so that every search runs over this short list. The proof of Lemma~\ref{lem:radical-free-order} shows that $[E_H:R]=|\operatorname{soc}(H/R)|\leq b(Q)$, and also that $b(H/R)=b(Q)$. Intersections of subgroups normalized by a common group are computed as normalized-subgroup intersections \cite[Theorem~3.1(viii)(b)]{LM}.

First we compute $E_H$. The subgroup $R$ is normal in $H$ and contains $N$, so it is transitive on $B$. So the chain construction above, run with $R$ in place of $N$, gives a chain $H=W_0\geq W_1\geq\cdots\geq W_L=R$ of normal subgroups of $H$ in which every nonabelian factor $W_j/W_{j+1}$ is a chief factor of $H/R$. The group $H/R=Q/\operatorname{Rad}(Q)$ has no nontrivial solvable normal subgroup. We keep two normal subgroups $E$ and $C$ of $H$ that contain $R$, such that $E\leq E_H$ and $C/R$ is the centralizer of $E/R$ in $H/R$. We start with $E=R$ and $C=H$. While $C\neq R$, let $j$ be the largest index with $W_j\cap C\neq R$, and put $I=W_j\cap C$. Then $I$ is normal in $H$, and $I\cap W_{j+1}=W_{j+1}\cap C=R$ by the choice of $j$. The factor $W_j/W_{j+1}$ is not abelian. Otherwise $[I,I]\leq I\cap W_{j+1}=R$, and $I/R$ would be a nontrivial abelian normal subgroup of $H/R$. So, by the choice of the chain, $W_j/W_{j+1}$ is a chief factor of $H/R$. We claim that $I/R$ is a minimal normal subgroup of $H/R$. To see this, let $J$ be a normal subgroup of $H$ with $R<J\leq I$. Then $JW_{j+1}$ is normal in $H$ and lies between $W_{j+1}$ and $W_j$. It differs from $W_{j+1}$, because $J\cap W_{j+1}\leq I\cap W_{j+1}=R$. Hence $JW_{j+1}=W_j$, and Dedekind's modular law gives $I=I\cap JW_{j+1}=J(I\cap W_{j+1})=J$. Thus $I\leq E_H$. We replace $E$ by $EI$ and $C$ by $C\cap Z$, where $Z/R$ is the centralizer of $I/R$ in $H/R$. As for the centralizer preimages above, $Z$ is the kernel of the action of $H$ by conjugation on the right cosets of $R$ in $I$, and there are at most $[E_H:R]\leq b(Q)$ of them. The intersection of $I/R$ with its centralizer is the center of $I/R$. This center is an abelian normal subgroup of $H/R$, so it is trivial, and $I\cap Z=R$. Since $R<I\leq C$, the new group $C\cap Z$ is a proper subgroup of $C$, so the order of $C$ at least halves in each round. Hence the loop stops after at most $\log_2|H|\leq b^2$ rounds, with $C=R$. At that point $E=E_H$. Indeed, let $A/R$ be a minimal normal subgroup of $H/R$. If $A\cap E=R$, then $[A,E]\leq A\cap E=R$, so $A\leq C=R$, which is impossible. So $A\cap E$ is a normal subgroup of $H$ with $R<A\cap E\leq A$, and the minimality of $A/R$ gives $A\leq E$. Hence $E_H\leq E$, and $E\leq E_H$ holds throughout.

Second, $E_H$ is normal in $H$ and contains $N$, so it is transitive on $B$. Moreover $b(H/E_H)\leq b(H/R)=b(Q)$, because $H/E_H$ is a quotient of $H/R$. So the radical-preimage computation above, run on $(H,E_H)$ in place of $(H,N)$, gives $K^*_H$.

Third, we list right coset representatives of $R$ in $E_H$, by closing $\{1\}$ under multiplication by generators of $E_H$ and testing membership in $R$ for ratios. We call this list the transversal. It has $[E_H:R]\leq b(Q)$ elements.

Fourth, we run the descent. We keep a subgroup $S$ with $R\leq S\leq K^*_H$ and $SE_H=K^*_H$, starting with $S=K^*_H$. Put $F=S\cap E_H$. It contains $R$, and it is normal in $S$ because $E_H$ is normal in $H$. Every prime that divides $[F:R]$ is at most $b$, since $F\leq\Sym(B)$. For each prime $p\leq b$ we build the full preimage $P$ of a Sylow $p$-subgroup of $F/R$, as follows. Let $p^a$ be the largest power of $p$ that divides $[F:R]$. Start with $P=R$. While some element $e$ of the transversal satisfies $e\in F$, $e\notin P$, and $|\langle P,e\rangle|$ divides $p^a|R|$, replace $P$ by $\langle P,e\rangle$. Throughout, $R\leq P\leq F$ and $|P|$ divides $p^a|R|$, so $P/R$ is a $p$-subgroup of $F/R$. When this stops, $P/R$ is a Sylow $p$-subgroup of $F/R$. Indeed, suppose that $P/R$ is a $p$-subgroup of $F/R$ but not a Sylow subgroup. By Sylow's theorem it lies in a larger $p$-subgroup $P'/R$ of $F/R$. Pick $x\in P'$ with $x\notin P$, and let $e$ be the representative of $Rx$ in the transversal. Then $e\in Rx\subseteq P'\leq F$, and $e\notin P$ because $R\leq P$. So $\langle P,e\rangle\leq P'$, and $|\langle P,e\rangle|$ divides $|P'|$, which divides $p^a|R|$. Each replacement multiplies $|P|$ by at least $2$, so there are at most $\log_2|H|\leq b^2$ replacements. Then we test whether every generator of $F$ normalizes $P$.

If $P$ is normal in $F$ for every prime $p\leq b$, then every Sylow subgroup of $F/R$ is normal, so $F/R$ is nilpotent, and we stop. Otherwise fix a prime $p$ for which $P$ is not normal in $F$. Let $s$ be a generator of $S$. Then $sPs^{-1}$ is again the full preimage of a Sylow $p$-subgroup of $F/R$, because $s$ normalizes $F$ and $R$. By Sylow's conjugacy theorem in $F/R$, there is $m\in F$ with $m(sPs^{-1})m^{-1}=P$. The representative $e$ of $Rm$ in the transversal lies in $Rm\subseteq F$, and it satisfies the same equation, because $R\leq P$. We find such an element $e$ by testing every element of the transversal, and we replace $s$ by $es$. Let $S'$ be the subgroup generated by $R$ and these elements $es$. Then $R\leq S'\leq S$, because each $e$ lies in $F\leq S$. Also $S'E_H=K^*_H$. Indeed, $S'E_H\leq SE_H=K^*_H$, and each generator $s$ of $S$ equals $e^{-1}(es)$, which lies in $E_HS'=S'E_H$. Finally, every generator of $S'$ normalizes $P$, so $S'\cap E_H$ lies in the normalizer of $P$ in $F$, which is a proper subgroup of $F$. By Lagrange's theorem $|S'\cap E_H|\leq|S\cap E_H|/2$. We replace $S$ by $S'$ and repeat. So the descent stops after at most $\log_2|H|\leq b^2$ rounds.

At the end $S/R$ is solvable. Indeed, $(S\cap E_H)/R=F/R$ is nilpotent, hence solvable, and $S/(S\cap E_H)$ is isomorphic to $SE_H/E_H=K^*_H/E_H=\operatorname{Rad}(H/E_H)$, which is solvable. We put $R'=S$. This subgroup has the form required in Lemma~\ref{lem:radical-free-order}. Indeed, Lemma~\ref{lem:solvable-supplement} applied to $R\triangleleft S$ gives a solvable subgroup $S_1\leq S$ with $S_1R=S$. Then $S_1\leq K^*_H$, and $S_1E_H=S_1RE_H=SE_H=K^*_H$ because $R\leq E_H$. So Lemma~\ref{lem:radical-free-order} applies to $S_1$, and $R'=S_1R$ satisfies $N\triangleleft R'\leq H$, the quotient $R'/N$ is solvable, and $[H:R']\leq b(Q)^2$. We never construct $S_1$.

\paragraph{Rebasing into $R'$.}
The subgroup $R'$ need not be normal in $H$. To apply Theorem~\ref{thm:solvable-multicoset} to $(R',N)$, we rebase the listed objects into $R'$. That is, we write each one as a coset $L'u$ with $N\leq L'\leq R'$ and $u\in R'$, and we give each $R'$-orbit its own color. We do this with pairwise transporters. Let $x$ be a listed object of $X$ and $y$ a listed object of $Y$. If $x$ and $y$ have different colors, then no element of $H$ carries $x$ to $y$. If $x=Lh$ and $y=Lk$ have the same color, then $g=h^{-1}k$ carries $x$ to $y$, and $\Iso_H(x,y)=gK$, where $K=L^k$ is the stabilizer of $y$ in $H$. The transporter in $R'$ is $\Iso_{R'}(x,y)=R'\cap gK$. To compute it, let $K$ act on the right cosets of $R'$ in $H$ by right multiplication. We list elements $s\in K$, one for each coset $R's$ in the orbit of the coset $R'$, by closing $\{1\}$ under multiplication by generators of $K$ and testing membership in $R'$ for ratios. The list has at most $[H:R']$ elements, and it is a right transversal of $R'\cap K$ in $K$. If $u\in R'\cap gK$, say $u=gc$ with $c\in K$, then the coset $R'c^{-1}$ lies in the orbit. So $R'c^{-1}=R's$ for a listed $s$, which means $c^{-1}s^{-1}\in R'$, and hence $gs^{-1}=u\,c^{-1}s^{-1}\in R'$. Conversely, if $gs^{-1}\in R'$ for a listed $s$, then $gs^{-1}\in R'\cap gK$, because $s^{-1}\in K$. So $R'\cap gK$ is nonempty exactly when $gs^{-1}\in R'$ for some listed $s$. In that case put $u=gs^{-1}$. Then $gK=uK$, and since $u\in R'$ we get $R'\cap gK=u(R'\cap K)$. By Schreier's lemma \cite[Lemma~4.2.1]{Seress}, the group $R'\cap K$ is generated by the elements $sc(s')^{-1}$, where $s$ is listed, $c$ is a generator of $K$, and $s'$ is the listed element with $R'sc=R's'$. Thus each transporter $\Iso_{R'}(x,y)$ is computed in time polynomial in $b$ and $[H:R']$.

Next we choose the new colors. If some listed object of $X$ has no transporter in $R'$ to a listed object of $Y$, or some listed object of $Y$ has none from a listed object of $X$, then no element of $R'$ carries $X$ to $Y$. Otherwise every listed object $w$ lies in the $R'$-orbit of a listed object of $Y$, and we let $z$ be the first listed object of $Y$ in that orbit. The transporters give an element $u\in R'$ with $z^u=w$. For $w$ in $X$, we take the inverse of an element of $\Iso_{R'}(w,z)$. For $w$ in $Y$, we pick $x$ in $X$ with $\Iso_{R'}(x,w)$ nonempty, and we multiply the inverse of an element of $\Iso_{R'}(x,z)$ by an element of $\Iso_{R'}(x,w)$. Let $H_z$ be the stabilizer of $z$ in $H$. The new object for $w$ is the coset $(R'\cap H_z)u$, with a new color named by $z$. Its group $R'\cap H_z$ contains $N$, because $N\leq R'$ and $N$ fixes every object. This coset is the set of all $v\in R'$ with $z^v=w$. So the new object determines $w$ within the $R'$-orbit of $z$. It also moves with $w$: for $t\in R'$ we have $z^v=w$ exactly when $z^{vt}=w^t$, so the new object for $w^t$ is the new object for $w$ times $t$. Since distinct orbits get distinct colors, the new objects of $X$ and $Y$ have the same transporters in $R'$ as the old ones. Each of $X$ and $Y$ has at most $M$ listed objects, so the rebasing uses at most $M^2$ pairwise transporters. Hence it runs in time polynomial in $b$, $M$ and $[H:R']$.

We can now enumerate the cosets of $R'$ and solve the remaining implicit action in each coset by the preceding profile algorithm. The enumeration bound is in the degree of the given cover, not in the potentially huge coset spaces.

\begin{corollary}[Transitive-kernel multicoset transport]
\label{cor:transitive-kernel-multicoset}
Suppose $N\triangleleft H\leq\Sym(B)$ is transitive on $B$.
In the explicit coset-object model, full multiset transport has
deterministic running time
\begin{equation*}
 \exp\bigl(O(\log^2(b+2)+\log(M+2)\log(b+2))\bigr).
\end{equation*}
\end{corollary}

\begin{proof}
Put $Q=H/N$ and let $R$ be the radical preimage in $H$. Since $N$ is
transitive on $B$, Theorem~1.4 of \cite{GMP} gives
$b(Q)\leq b^{\log_2 b}$. For each nonabelian chief factor $U/V$ of $Q$
in the construction above, $U/V$ is a direct product of isomorphic
nonabelian simple groups, so all of its composition factors are
nonabelian and $[U:V]\leq b(Q)\leq b^{\log_2 b}$. Thus the radical
preimage $R$ is computed in time $b^{O(\log b)}$.
The same bounds hold for the two further runs of the chain construction
in the computation of $R'$, because $R$ and $E_H$ are transitive and
$b(H/E_H)\leq b(H/R)=b(Q)$. The other steps of that computation make
at most $\poly(b)$ passes over lists of at most $b(Q)$ elements, or act
on at most $b(Q)$ cosets. So $R'$ is computed in time $b^{O(\log b)}$,
and Lemma~\ref{lem:radical-free-order} and the bound on $b(Q)$ give
\begin{equation*}
 [H:R']\leq b(Q)^2\leq b^{2\log_2 b}.
\end{equation*}
Enumerate right coset representatives $r$ of $R'$ in $H$ in time
polynomial in $[H:R']$ and $b$, by closing $\{1\}$ under multiplication
by generators of $H$ and testing membership in $R'$ for ratios. For
$v\in H$ we have $X^{vr}=Y$ exactly when $X^v=Y^{r^{-1}}$. So the
elements of the coset $R'r$ that carry $X$ to $Y$ are the products $vr$
with $v\in\Iso_{R'}(X,Y^{r^{-1}})$. In each coset, align the target by
replacing $Y$ with $Y^{r^{-1}}$, rebase the listed objects of $X$ and
$Y^{r^{-1}}$ into $R'$ as described above, and apply
Theorem~\ref{thm:solvable-multicoset} to $(R',N)$. This is allowed
because $N\triangleleft R'$ and $R'/N$ is solvable. Multiply its answer
by $r$ on the right. The union of the valid results over all cosets is
the full transporter, which is compressed to a single coset using
generators of the component stabilizer and ratios of representatives.

There are at most $b^{2\log_2 b}$ cosets. In each one, the rebasing
takes time polynomial in $b$, $M$ and $[H:R']\leq b^{2\log_2 b}$, and
Theorem~\ref{thm:solvable-multicoset} takes time
$\exp(O(\log(b+M+2)\log(b+2)))$. Since
$\log(b+M+2)\leq\log(b+2)+\log(M+2)$, the total time is
$\exp(O(\log^2(b+2)+\log(M+2)\log(b+2)))$.
\end{proof}

\subsubsection{An exact terminal supplied by the row normal form}

Let \(G\leq\Sym(\Omega)\) preserve \(m\) blocks of size \(b\),
with alternating action on those blocks. Assume a clean normal form
as in Theorem~\ref{thm:logarithmic-gat-normal-form} is supplied;
write \(n=mb\). In its identification \(\Omega=B\times[m]\),
write the clean complement as \(R_0\cong\Alt(m)\), the block kernel
as \(K\), and its normal data as
\[
 N^*\triangleleft N\triangleleft H\leq\Sym(B),\qquad
 \mathcal M^*\triangleleft\mathcal M\triangleleft K\triangleleft G,
\]
\[
 \mathcal M^*=(N^*)^m,\qquad
 \mathcal M/\mathcal M^*
   =\{(a_i)\in A^m:\textstyle\sum_i a_i=0\},\qquad A=N/N^*.
\]
Here \(A\) is abelian, \(K/\mathcal M\cong H/N\) is diagonal,
and \(J=\mathcal M R_0\triangleleft G\), with \(G/J\cong H/N\).
All maps and lifts are supplied by that theorem. The following proof
uses these structural identities, without an inequality between \(m\)
and \(b\). A symmetric top action is handled by its two alternating
cosets.

Matching row types leaves a global consistency condition. The independent row corrections must have total zero in the abelian quotient, and the common residual action must preserve the entire multiset of types. Both conditions are imposed before a full answer is returned.

\begin{theorem}[Transitive-\(N\) Giant Action terminal]
\label{thm:transitive-gat-terminal}
In this situation, assume additionally that \(N\) is transitive on
\(B\). Exact string isomorphism under \(G\), including the complete
output coset, reduces to polynomially many ordinary string-isomorphism
calls on \(b\) points and
\[
 \exp\bigl(O(\log^2(b+2)+\log(m+2)\log(b+2))\bigr)
\]
additional time. Thus, if \(F(b)\) bounds those calls, the total is at
most \(\poly(n)F(b)+\exp(O(\log^2(n+2)))\).
\end{theorem}

\begin{proof}
Write a string as \(x=(x_1,\ldots,x_m)\), with each \(x_i\) a string
on \(B\). Using polynomially many degree-\(b\) calls, classify the
local strings into \(N\)-orbits. For each occurring type \(a\), choose
a representative \(z_a\), and find \(n_i\in N\) with
\(z_a^{n_i}=x_i\). Define
\[
 B_a=\operatorname{im}\bigl(\Aut_N(z_a)\longrightarrow A\bigr),
 \qquad \alpha_i=n_iN^* \pmod {B_a},\qquad
 B(x)=\sum_{a\ {\rm occurring}}B_a.
\]
The total \(s(x)=\sum_i\alpha_i\pmod {B(x)}\) is well defined.

Two strings are \(J\)-isomorphic exactly when their type multisets
agree, their total labels agree, and a matching of their fiber types
can be chosen even. Indeed, for a prescribed type matching \(\sigma\),
the possible local transporter images in \(A\) are the cosets
\[
 \alpha_{y,\sigma(i)}-\alpha_{x,i}+B_{a_i}.
\]
Their sum can be zero exactly when the totals agree. A zero-sum
tuple lifts to \(\mathcal M\), since the independent groups \(N^*\)
remain available. If any type repeats, exchanging two equal-type
matches reverses parity without changing the possible sum.
If all types are distinct, the matching is unique and its sign
must be retained.

Solving these finite abelian equations constructs a transporter.
It also constructs \(\Aut_J(x)\): use the local \(N^*\)-stabilizers,
the kernel of the sum map on the product of local stabilizer images,
and lifted generators of the even Young group on equal-type fibers.
The same sum equations correct the latter lifts.

The residual action is \(H/N\). An \(N\)-type \([z]_N\) has
\(H\)-stabilizer \(N\Aut_H(z)\); its transporter to \([w]_N\) is
\(N\Iso_H(z,w)\). Hence degree-\(b\) string calls construct the
explicit coset-object representation required above. These tables
are computed once; their construction is not hidden inside the
additional-time bound.
In particular, store every needed coset \(\Iso_H(z_a,z_c)\).
For a later \(h\) carrying type \(a\) to type \(c\), the known
membership in \(N\Iso_H(z_a,z_c)\), together with constructive
factorization in the normalized product
\(N\Aut_H(z_a)\), supplies the required local \(N\)-transporter
in polynomial time \cite[Theorem~3.1(viii)(c)]{LM}.
The subgroups \(\Aut_N(z_a)\) and \(\Aut_{N^*}(z_a)\) are likewise
normalized-subgroup intersections of the stored stabilizers.
Thus computing offsets in the different cosets below makes no new
string-isomorphism calls.

Compute the subgroup $R'$ of the proof of
Corollary~\ref{cor:transitive-kernel-multicoset} and enumerate its
right cosets in $H$. In each coset, solve the undecorated type-multiset
problem as in that proof. Its
nonempty output is a coset \(D\tau\), with \(D/N\) solvable.
It remains to impose the total label and the possible parity label.
These do not require a return to a larger string domain.

Choose a blockwise lift \(k(\tau)\in K\) of \(\tau N\), and put
\(y'=y^{k(\tau)^{-1}}\). Let \(P\) be the full preimage of
\(D/N\) under \(G\to G/J\cong H/N\). The physical candidates
in this coset are \(Pk(\tau)\), and
\[
 \Iso_{Pk(\tau)}(x,y)=\Iso_P(x,y')k(\tau).
\]
The strings \(x,y'\) have the same type multiset, so their label
torsors use the common group \(A/B(x)\), with the same orientation
convention. We solve this aligned problem under \(P\), whose
residual group is \(D/N\). Each final transporter representative
is multiplied by \(k(\tau)\) on the right; its source stabilizer
is unchanged.

The alignment itself uses the stored tables. Write
\(k_i(\tau)^{-1}=\tau^{-1}e_i\), where \(e_i\in N\).
For a target row \(y_i=z_c^{n_i}\), they give
\[
 y_i'=z_{c^{\tau^{-1}}}^{\,t_c(\tau^{-1})n_i^{\tau^{-1}}e_i}.
\]
Here \(\tau^{-1}\) carries each occurring target type to an
occurring source type, so the required \(t_c(\tau^{-1})\) comes
from the previously stored pairwise transporters. Thus alignment
makes no additional string-isomorphism calls.

On the fixed type multiset, the possible \(J\)-orbits form a torsor
for \(A/B(x)\), with an additional \(C_2\) orientation coordinate
exactly when every type is distinct. An element \(h\in D\), the
subgroup preserving this type multiset, is lifted through \(K\) to a
blockwise permutation \(k(h)\). Its coordinate
actions differ from the common \(h\) by elements of \(N\); their
images in \(A\) give a computable affine offset. Conjugation by \(h\)
permutes the groups \(B_a\), so it induces the linear part on
\(A/B(x)\). Different lifts differ by \(\mathcal M\) and have the
same action on this torsor. Likewise \(h\in N\) acts trivially:
its lift lies in \(\mathcal M\). Thus the affine action factors
through \(D/N\).

Explicitly, for \(h\in D\), write \([n]=nN^*\in A\), let
\(m_a\) be the multiplicity of type \(a\), and choose \(t_a(h)\in N\) with
\(z_a^h=z_{a^h}^{t_a(h)}\), and write the coordinate lifts as
\(k_i(h)=h d_i(h)\), where \(d_i(h)\in N\). On the fixed type
multiset the affine action is
\[
 s\longmapsto s^h+\sum_a m_a[t_a(h)]+
                         \sum_i[d_i(h)]\pmod {B(x)}.
\]
All local cosets here are first mapped to the common quotient
\(A/B(x)\). If \(h\in N\), take \(t_a(h)=h\); the displayed
offset is then \(\sum_i[k_i(h)]=0\), by the defining augmentation
condition for \(\mathcal M\). This also checks the trivial kernel
action without assuming that the diagonal tuple \((h,\ldots,h)\)
itself belongs to \(\mathcal M\).

When the orientation coordinate occurs, its translation is the
sign of the induced permutation of the distinct types. It is a
homomorphism to \(C_2\), independent of the total-sum constraint.
Here $D/N$ is solvable, although $D$ itself need not be solvable. Because the affine label action on $A/B(x)$ and on the optional $C_2$ orientation coordinate factors through $D/N$, the linear image of $D$ on each homogenized primary layer is an extension of a quotient of $D/N$ by the abelian translation kernel and is therefore solvable. The finite-abelian affine transporter procedure in the proof of Theorem~\ref{thm:solvable-multicoset} therefore applies directly to $D$ and imposes both labels in polynomial time within this coset. With the possible orientation factor included, every prime characteristic satisfies $p \leq \max\{b, 2\}$.

Finally lift the surviving quotient representative and correct it
by the constructed \(J\)-transporter. These corrections need no
new string-isomorphism calls. Indeed, for an original row
\(x_i=z_a^{n_i}\) and a coordinate lift \(k_i(h)=h d_i(h)\),
the stored local transporters give
\[
 x_i^{k_i(h)}
   =z_{a^h}^{\,t_a(h)n_i^h d_i(h)}.
\]
Every factor in the exponent lies in \(N\). Hence the type,
local transporter, and abelian label of the transformed row
are obtained from the existing tables, and the same sum equations
construct its \(J\)-correction. For automorphism generators,
lift each generator of the surviving quotient stabilizer and make
the same correction; adjoin \(\Aut_J(x)\). These generators map
onto the exact quotient stabilizer and contain its full kernel,
so the output is the complete isomorphism coset. Combining the
cosets of $R'$ gives the asserted complexity.
\end{proof}

The transitivity assumption on \(N\) is a substantive boundary.
For intransitive \(N\), its normalizer quotient can contain independent
large factors on many \(N\)-orbits. Neither the bound on the radical-free
quotient nor the terminal above applies with the same exponent by merely
replacing \(N\) with its separate projections. The next section treats that case by retaining the normal-orbit column
partition and transposing the large natural action inside its columns.
It does not apply the transitive-kernel bound to an intransitive kernel.

\subsection{Rectangles realized inside inherited permutation actions}
\label{sec:inherited-rectangles}

The following reductions distinguish a new action on ordered pairs from
an ordered-pair action that is already a quotient of an inherited orbit.
Only the latter automatically inherits the marked-chain budget.  All
permutation actions in this section use right point actions.

\subsubsection{The only alternating actions below the triple-subset degree}
\label{subsec:low-index-support}

At large auxiliary degree, the active domain is too small to support three independent label coordinates. The following elementary index argument leaves only single labels and pairs of labels; the constructive part recovers those coordinates.

\begin{lemma}[Alternating subgroups of index below $\binom m3$]
\label{lem:low-index-alt-support}
There is an absolute $m_0$ such that the following holds for $m\geq m_0$.
If $P\leq\operatorname{Alt}(\Gamma)$, $|\Gamma|=m$, and
\[
 [\operatorname{Alt}(\Gamma):P]<\binom m3,
\]
then $P$ is one of the following groups:
\begin{enumerate}
\item $\operatorname{Alt}(\Gamma)$;
\item $\operatorname{Alt}(\Gamma\setminus\{a\})$ for one point $a$;
\item $\operatorname{Alt}(\Gamma\setminus\{a,b\})$ for two distinct points;
\item the setwise stabilizer of $\{a,b\}$ in $\operatorname{Alt}(\Gamma)$.
\end{enumerate}
Their coset actions are respectively the trivial, natural, ordered-pair,
and unordered-pair actions. Given generators of $P$ in its natural
$m$-point action, the case and its one- or two-point support are
computable in polynomial time. No general small-index subgroup
classification is needed.
\end{lemma}

\begin{proof}
We first record a transitive-index gap. For all sufficiently large $s$,
every proper transitive subgroup $U<\operatorname{Alt}(s)$ has
index greater than $s^3$. If $U$ is primitive, Bochert's
theorem~\cite[p.~585]{Bochert1889} (see
also~\cite[Theorem~3.3B, p.~81]{DixonMortimer1996}) gives
$[\operatorname{Sym}(s):U]\geq\lfloor(s+1)/2\rfloor!$, so for all
sufficiently large $s$ we have
\begin{equation*}
 [\operatorname{Alt}(s):U]\geq\frac12\Bigl\lfloor\frac{s+1}{2}\Bigr\rfloor!>s^3.
\end{equation*}
If $U$ is imprimitive, write $s=ab$ for a preserved system of $b$
blocks of size $a$, where $a,b\geq2$. Then
\[
 [\operatorname{Alt}(s):U]
 \geq\frac{s!}{2(a!)^b b!}=\exp(\Omega(s))>s^3.
\]
The exponential lower bound is uniform in $a,b$: for $b\geq16$,
its logarithm is at least
$(s-b)\log b-s-\log2\geq(\tfrac12\log16-1)s-\log2$;
for $2\leq b\leq15$, counting equal-block partitions by one of their
blocks gives a lower bound $\binom sa/(2b)$, whose logarithm is at
least $(s/b)\log b-\log(2b)$. These are the same factorial estimates
used in Lemma~\ref{lem:budget-primitive-valuation}.
The displayed exponential lower bounds exceed $s^3$ for all sufficiently large $s$. Choose $m_0$ so that the gap holds at both $m$ and $m-1$ whenever $m\geq m_0$.

We also use the elementary minimum-index fact
\[
 V<\operatorname{Alt}(r),\quad r\geq5
 \quad\Longrightarrow\quad
 [\operatorname{Alt}(r):V]\geq r.
\]
Indeed, the coset action is faithful by simplicity, and a faithful
action of degree below $r$ would imply
$r!/2\leq(r-1)!$, which is false.

Put $I=[\operatorname{Alt}(m):P]$. If $P=\operatorname{Alt}(m)$,
there is nothing to prove. Otherwise the transitive-index gap, together
with $I<\binom m3<m^3$, shows that $P$ is intransitive on $\Gamma$.
Choose a smallest natural orbit $D$, of size $e\leq m/2$.
Its setwise stabilizer gives $I\geq\binom me$. Therefore $e\leq2$.

Suppose first that $e=2$. Let $S$ be the two-set stabilizer of $D$,
and put $T=\operatorname{Alt}(\Gamma\setminus D)\triangleleft S$.
The quotient $S/T$ has order two, and
\[
 [S:P]=I/\binom m2<(m-2)/3.
\]
Since $[T:T\cap P]\leq[S:P]$, the minimum-index fact forces
$T\leq P$. Hence $P=T$ or $P=S$, giving the third or fourth case.

Suppose next that $e=1$, with $D=\{a\}$. Regard
$P\leq\operatorname{Alt}(\Gamma\setminus\{a\})
 =\operatorname{Alt}(s)$, where $s=m-1$. Its index there is
\[
 j=I/m<s(s-1)/6.
\]
If $P=\operatorname{Alt}(s)$, we obtain the second case. Otherwise
the transitive-index gap forces $P$ to be intransitive on these $s$
points. A smallest orbit has size $e'\leq s/2$. If $e'\geq2$,
then $j\geq\binom{s}{e'}\geq\binom s2$, a contradiction.
Thus $P$ fixes a second point $b$, and
\[
 P\leq\operatorname{Alt}(\Gamma\setminus\{a,b\}),\qquad
 [\operatorname{Alt}(m-2):P]
   =I/(m(m-1))<(m-2)/6.
\]
The minimum-index fact forces equality with $\operatorname{Alt}(m-2)$.
This is the third case and completes the classification.

For the constructive assertion, compute the natural orbits of $P$
and its order. In the nontrivial proper cases its support is either
its unique fixed point, its two fixed points, or its unique orbit of
size two; the other orbit has size $m-1$ or $m-2$. These data identify
the case in polynomial time. The support is intrinsic to $P$; only
an ordering of the two fixed points, when needed, is arbitrary.
\end{proof}

The point-stabilizer image identifies the kernel-orbit action. Its index is at most the number of active positions, so the preceding classification becomes a statement about the actual physical representation.

\begin{corollary}[Constructive physical support thresholds]
\label{cor:physical-support-thresholds}
Let $G\leq\operatorname{Sym}(\Omega)$ be faithful and transitive,
$|\Omega|=n$, and let
$\phi\colon G\twoheadrightarrow\operatorname{Alt}(\Gamma)$ be supplied,
with $m=|\Gamma|\geq m_0$ and $K=\ker\phi$. Suppose
\begin{equation*}
 n<\binom m3,
 \qquad m>\max\{63,12\log_2 n\}.
\end{equation*}
Then the action of $G$ on its $K$-orbits is naturally isomorphic to
the natural, ordered-pair, or unordered-pair action on $\Gamma$.
The equivariant quotient map from $\Omega$ is polynomial-time
computable, and its fibers are exactly the equal-size $K$-orbits.
If in addition $n<\binom m2$, only the natural action is possible;
in particular $K$ is transitive inside each of its $m$ natural blocks.
\end{corollary}

\begin{proof}
Fix $x\in\Omega$ and compute $P=\phi(G_x)$. The stabilizer of the
orbit $x^K$ in $G/K$ is $KG_x/K=P$, so
\begin{equation*}
 [\operatorname{Alt}(m):P]=|\Omega/K|\leq n<\binom m3.
\end{equation*}
Lemma~\ref{lem:low-index-alt-support} applies. Its trivial-action
case would say that $P=\operatorname{Alt}(\Gamma)$, so $G=KG_x$ and
$K$ is transitive on $\Omega$, which is excluded by
Lemma~\ref{lem:transitive-kernel-giant-threshold} since
$m>\max\{63,12\log_2 n\}$. Thus only the
three proper cases remain.

Compute the intrinsic support at $x$ by the preceding lemma and
transport it along a Schreier traversal of the $G$-orbit $\Omega$,
using the supplied images under $\phi$. In the ordered-pair case
choose either ordering at $x$ and transport that ordering. The map
is well defined because $P$ is exactly the corresponding natural,
ordered-pair, or unordered-pair stabilizer. The preceding stabilizer
identity shows that its fibers are precisely the $K$-orbits.
Normality of $K$ and transitivity of $G$ give their equal sizes.
All point stabilizers, orbit computations, image evaluations, and
kernel computations take polynomial time in the stated permutation
representation.

The ordered and unordered pair actions have degrees $m(m-1)$ and
$\binom m2$, respectively. The extra inequality $n<\binom m2$
excludes both, leaving the natural action and its transitive fibers.
\end{proof}

In particular, for a faithful column action of degree $c$, a giant
of degree $t$ above the logarithmic threshold of
Lemma~\ref{lem:transitive-kernel-giant-threshold} and with
$c<\binom t2$ has the natural kernel-orbit quotient. This is the
precise column-kernel transitivity needed by the diagonal-carrier
rectangle construction.

For the pair case, the point-stabilizer image determines whether the pair is ordered. Recovering the map explicitly is necessary: the later matrix query must be pulled back through the actual orbit fibers.

\begin{lemma}[The support-two orbit quotient]
\label{lem:support-two-orbit-quotient}
Let $G\leq\operatorname{Sym}(\Omega)$ be faithful and transitive, let
$\phi:G\twoheadrightarrow\operatorname{Alt}(\Gamma)$, and put
$m=|\Gamma|$, $n=|\Omega|$, and $K=\ker\phi$.  Suppose $m\geq 9$ and,
for some $x\in\Omega$, there is a two-element set $D\subseteq\Gamma$
such that
\[
 \operatorname{Alt}(\Gamma\setminus D)
 \leq \phi(G_x)
 \leq \operatorname{Alt}(\Gamma)_{\{D\}}.
 \tag{IR1}
\]
Then the action of $G$ on its $K$-orbits is equivariantly isomorphic to
either the action on unordered pairs of distinct elements of $\Gamma$
or the action on ordered pairs of distinct elements of $\Gamma$.
The quotient map $f:\Omega\to\Xi$ can be constructed in polynomial
time from the supplied giant map and $D$.  Its fibers are precisely
the $K$-orbits and all have equal size.
\end{lemma}

\begin{proof}
The stabilizer in $G/K$ of $x^K$ is $KG_x/K=\phi(G_x)$.  The quotient
of the upper group in (IR1) by the lower group has order two.  Thus
$\phi(G_x)$ is either the full two-set stabilizer, giving the
unordered-pair action, or the pointwise two-point stabilizer
$\operatorname{Alt}(m-2)$, giving the ordered-pair action.

For the ordered action, choose one ordering of $D$ at the base point
and transport it using $\phi$.  This is well defined because the
base stabilizer fixes both entries.  The other initial ordering
merely interchanges the two coordinate names.  For the unordered
action, transport $D$ itself.  Orbit computation and constructive
membership compute these maps and lift permutations throughout.
Normality of $K$ and transitivity of $G$ give equal fiber sizes.
\end{proof}

\paragraph{When the hypothesis occurs.}
Corollary~\ref{cor:physical-support-thresholds} gives exactly the
support-one and support-two alternatives whenever the physical degree
is below $\binom m3$ and the giant exceeds the logarithmic threshold
of Lemma~\ref{lem:transitive-kernel-giant-threshold}. In the support-two alternative its point-stabilizer image
satisfies (IR1), with the ordered and unordered cases distinguished
by their natural stabilizers. This is the only support classification
needed in the main algorithm.

A rectangle in the pair coordinates is already represented by one orbit of a subgroup of the current group. Pulling back its colors therefore preserves the hereditary analysis parameters.

\begin{lemma}[An exact inherited rectangle query]
\label{lem:inherited-support-two-rectangle}
Assume Lemma~\ref{lem:support-two-orbit-quotient}.  Let $W,R$ be
disjoint nonempty subsets of $\Gamma$, with $w=|W|\geq3$ and $r=|R|$.
Let $Y=\Gamma\setminus(W\cup R)$ and
\[
 P=(\operatorname{Sym}(W)\times\operatorname{Sym}(R)
                   \times\operatorname{Sym}(Y))\cap\operatorname{Alt}(\Gamma),
 \qquad H=\phi^{-1}(P).
\]
Let $\Xi_{W,R}$ consist of the pairs with one entry in $W$ and one
in $R$, using the orientation $(w,r)$ in the ordered-pair case, and
put $\Delta=f^{-1}(\Xi_{W,R})$.  Then $\Delta$ is one $H$-orbit and
\[
 \frac{|\Delta|}{n}=
 \begin{cases}
   wr/(m(m-1)),&\text{ordered pairs},\\
   2wr/(m(m-1)),&\text{unordered pairs}.
 \end{cases}                                                     \tag{IR2}
\]

Given two explicitly colored matrices on $W\times R$, their complete
transporter under the image of $P$ in
$\operatorname{Sym}(W)\times\operatorname{Sym}(R)$ is computable by
one exact string-isomorphism call on $H^\Delta$.  The input strings
are constant on the fibers of $f$.  The call inherits all budgets
and caps valid for subgroup-orbit restrictions of the original action.
The output includes all group generators and a representative, or
certifies emptiness.
\end{lemma}

\begin{proof}
The group $P$ is transitive on $W\times R$.  Indeed, any prescribed
images of one point in each factor can be achieved by permutations
of the factors; their combined sign can be corrected in the
stabilizer of the chosen point of $W$, since $w\geq3$.  The kernel
$K\leq H$ is transitive on each fiber of $f$, proving transitivity
on $\Delta$.  Equal fiber sizes give (IR2).

Pull the matrix color of $(a,b)$ back to every point of
$f^{-1}(a,b)$, with the evident interpretation for unordered pairs.
An element of $H$ preserves or transports these pulled-back strings
exactly when its induced pair permutation preserves or transports
the matrices.  Every permitted pair permutation has a lift in $H$,
so neither direction loses solutions.

The pair action factors through the faithful image $H^\Delta$:
an element fixing $\Delta$ pointwise fixes every pair in
$\Xi_{W,R}$, hence fixes $W$ and $R$ pointwise.  Therefore projecting
the complete string transporter gives exactly the complete permitted
matrix transporter.  To compute this projection, lift its generators
to $H$ and apply $\phi$ followed by restriction to the two sides;
the result is independent of the chosen lifts.  All these maps are
supplied permutation homomorphisms with polynomial-time membership
and lifting operations.

The queried action is literally the action of a subgroup of $G$ on
one of its orbits.  In particular it is not an arbitrary faithful
re-representation of $G$ or of one of its quotients.  The visibility
and inherited-state conditions of the marked-chain budget apply.
\end{proof}

\paragraph{Parity and source--target alignment.}
If source and target rectangles have sides $W_X,R_X$ and $W_Y,R_Y$
of equal sizes, choose an even bijection of $\Gamma$ taking the
three source parts to the three target parts, and lift it through
$\phi$.  Such a bijection exists since $|W_X|\geq3$: the sign of
any alignment can be corrected inside its $W_Y$ part.  This fixes
one alignment coset; the subgroup in
Lemma~\ref{lem:inherited-support-two-rectangle} then supplies all
remaining possibilities.  There is no enumeration of the index of
that set stabilizer.  If corresponding sizes differ, that branch
is empty.

When $|Y|\geq2$, the permitted group on $W\times R$ is the full
$\operatorname{Sym}(W)\times\operatorname{Sym}(R)$: a sign can be
corrected on $Y$.  When $|Y|\leq1$, it is the index-two subgroup
with even product sign.  The latter is sufficient because it is
exactly the restriction of the supplied alternating ideal action.
No unliftable odd matrix automorphism is required.  The full ideal
transporter, if desired, is obtained from the projected lifts by
adjoining the pointwise $(W\cup R)$-stabilizer in the alternating
group.  Parity is imposed on both the resulting group and its coset
representative.

We need only the first bipartite graph of Split-or-Johnson. At that point both sides are still subsets of the old label set, which permits an inherited representation of the rectangle.

\begin{lemma}[The first certified Split-or-Johnson entrance]
\label{lem:first-soj-rectangle}
Suppose the constant-certificate aggregation and Design step has
produced a nontrivial homogeneous coherent configuration on
$C\subseteq\Gamma$, with $|C|\geq\delta m$.  Using only polynomially
many branches and polynomial structural work, one can either obtain
the usual fixed-fraction ideal-cap reduction, or obtain disjoint
actual subsets $W,R\subseteq C$ and a nontrivial semiregular binary
matrix on $W\times R$ such that
\[
 |W|>\eta|C|,\qquad |R|\leq |C|-|W|,\qquad
 |W||R|\leq \eta(1-\eta)m^2,                                      \tag{IR3}
\]
and every left twin class has size at most $|W|/2$.
These are the sides at the first entrance to
\cite[Proposition~5.7]{H}, before any right-side contraction or
replacement by artificial block vertices.
\end{lemma}

\begin{proof}
An imprimitive coherent configuration gives the canonical equal-block
partition from the first paragraph of \cite[Section~5.2]{H}.  Otherwise
choose a source vertex $x$ and branch over its target image.  The
colors $c(x,y)$ are canonical after this choice.  If none of their
classes exceeds $\eta|C|$, they give the required colored partition.

In the remaining case, let $W$ be the unique dominant class.  The
proof of \cite[Theorem~5.3]{H} selects a different nonempty color
class $R$ and an edge color such that the induced graph on
$W\times R$ is nonempty, noncomplete, and semiregular.  These are
still subsets of the original $C$, and $x$ belongs to neither side.
The color choice may be made from the canonical coherent palette;
alternatively all the finitely many vertex/color choices can be
branched over at polynomial cost.  They cover every parent
isomorphism.

Since $w>\eta c$ and $r\leq c-w$, with $c=|C|$, the decreasing
quadratic $w(c-w)$ on $[\eta c,c]$ gives $wr\leq\eta(1-\eta)c^2$.
Finally, the semiregular twin argument of
\cite[Exercise~5.4]{H} bounds every left twin class by $w/2$:
after complementing if necessary the common right degree is at
most $w/2$, and it is at least the size of any left twin class.
\end{proof}

\paragraph{The returned parent cap.}
Apply Lemma~\ref{lem:inherited-support-two-rectangle} at the first
entrance in Lemma~\ref{lem:first-soj-rectangle}.  For sufficiently
large absolute $m$, (IR2)--(IR3) give $|\Delta|\leq\theta n$; the
unordered case is the larger one, with ratio at most
$2\eta(1-\eta)m/(m-1)$.

Let $L$ be the left projection of the exact permitted matrix
automorphism group.  The kernel of the right action of the full
matrix automorphism group is the product of the symmetric groups
on left twin classes; imposing alternating parity only restricts this
kernel to a subgroup.  The quotient is a subgroup of
$\operatorname{Sym}(R)$, so every nonabelian simple section of $L$
comes from a left twin group or a subgroup of
$\operatorname{Sym}(R)$.  Consequently
\[
 \operatorname{aw}(L)\leq \max\{w/2,r\}\leq m/2.
\]
Extend the left transporter by the full symmetric group on
$\Gamma\setminus W$ and impose the alternating sign condition.
Its group has alternating width at most
\[
 \max\{w/2,r,m-w\}\leq\max\{1/2,1-\eta\delta\}m.
 \tag{IR4}
\]
This extension is an overcoset of the ideal images of the parent
isomorphisms in the current branch.  Pull it back through the
supplied full giant map and solve the original strings inside that
preimage.  Thus extra permutations admitted by forgetting the right
side or earlier branch choices are filtered exactly, while every
genuine parent isomorphism remains covered.  No arbitrary group
intersection is used.  Emptiness and full transporter output are
preserved throughout.

\paragraph{Cost at support-two nodes.}
The auxiliary query is a genuine subgroup-orbit restriction of
degree at most $\theta n$.  Its inherited $\rho$, $\sigma$, and cap
moment $Q$ do not increase.  The source/target branch choices have
total size $m^{O(1)}$ because the certificate size is fixed.
The later same-domain continuation has the cap contraction (IR4)
at the selected giant interval.  Thus the auxiliary query is paid
by the decrease of $\log^2(n+2)$, and the continuation by the
corresponding decrease of the cap moment.  These are separate
recursive children; no cap learned in the auxiliary call is
transferred to its parent or to a sibling.  Structural polynomial
work remains additive, as in the ordinary string chain rule.

\subsubsection{Promotion without a budget reset}

After two distinct large natural factors have appeared in the same hierarchy, their product leaves too little degree for a third larger factor. This gives a bound on the entire current group and prevents repeated enlargements of the retained row action.

\begin{lemma}[A single larger-carrier promotion]
\label{lem:one-carrier-promotion}
Fix a phase bound $N$ and $\lambda=N^\alpha$, with $1/3<\alpha<1/2$.  Suppose the row
carrier has degree $m>\lambda$, and a later support-one natural
column giant has degree $q>m$.  Suppose all current physical points
lie in the product grid described in
Lemma~\ref{lem:diagonal-carrier}, with total degree $n\leq N$.
Assume also that the fixed-floor dispatcher has exposed the old row
ratio $m$ and the new natural column ratio $q$ as distinct intervals
of the retained tower, with their valid caps $m$ and $q$.
Then $\operatorname{aw}(G)\leq q$.

After constructing a clean row carrier of degree $q$, no further
carrier promotion is needed in any subsequent inherited subgroup,
orbit, or quotient action that retains this carrier. The normal
transpose retains the inherited tower labels and does not increase
$\rho$ or the cap moment; no budget reset is needed.
\end{lemma}

\begin{proof}
The old row partition and the natural column partition are both
$G$-invariant.  Their intersections have the uniform size
$u=n/(mq)$.  The tower formed first by the $q$ column blocks, then
by their $m$ row intersections, and then by a complete local tower
has top image $\operatorname{Alt}(q)$, middle image
$\operatorname{Alt}(m)$, and all remaining constituent degrees at
most $u$.  The middle image is full because the old clean row
carrier lies in the kernel of the column action.  Since
\[
 u\leq N/(mq)<N/\lambda^2=N^{1-2\alpha}<\lambda<q,
\]
the iterated wreath-product bound for alternating-section width
gives $\operatorname{aw}(G)\leq q$.

Hereditary alternating-section width cannot increase in subgroups,
quotients, or their permutation images.  Consequently no subsequent
giant representation can have degree greater than $q$.  In
particular, once the clean carrier has degree $q$, every further
large column giant has degree at most the carrier and is handled
by Lemma~\ref{lem:diagonal-carrier}.

For the stronger inherited-state assertion, use the normal transpose,
whose factorization proof does not require a numerical inequality
between the row and fiber sizes.  If $P=L_0<\cdots<L_s=H$ is an old
local stabilizer chain and $N\triangleleft H$ is the new variation,
each adjacent factor splits exactly as
\[
 [L_i:L_{i-1}]
 =[L_i\cap N:L_{i-1}\cap N]\,[NL_i:NL_{i-1}].
\]
The lower factor is a subgroup orbit of the old constituent, and
the upper factor is a block quotient of that constituent.  Both
therefore retain its cap label.  Modulo the variation group, the
cell-to-row stabilizers have the form
$\operatorname{Alt}(q-1)\times V$.  Replacing these by
$\operatorname{Alt}(q)\times V$ moves the unchanged natural row
ratio $q$ below the upper local factors.  All ratios are only split
or permuted.  Hence $\rho$ cannot increase, and the cap moment
cannot increase because splitting $d=ab$ with cap $c$ replaces
$\log d\log c$ by at most
$\log a\log c+\log b\log c$.

The actual natural intervals $m$ and $q$ were already exposed
before this refactoring; all other ratios have product at most
$N/(mq)<\lambda$, and their effective caps are no larger than
their ratios.  This also proves the global width bound directly
from the retained tower.  A reset of the tower budgets is deliberately
unnecessary: a certificate child can lose the carrier, and paying a
fresh quadratic reset on every such child would not be justified.
\end{proof}

\subsection{Retaining a natural giant as an auxiliary carrier}
\label{sec:carrier-recurrence}

The action used by an auxiliary oracle need not be the full current
parent group.  It must realize the entire requested ideal action, and
its output must be induced back to that ideal action.  A subgroup chosen
only for this purpose does not impose its index on the original-domain
continuation.  This distinction permits the following construction.

A retained row action supplies the missing first coordinate of a column rectangle. The diagonal subgroup is an auxiliary representation only; its index is never charged to, or imposed on, the original-domain continuation.

\begin{lemma}[Diagonal carrier for a rectangle oracle]
\label{lem:diagonal-carrier}
Suppose a transitive permutation group $G$ on $\Omega$ has an explicit
normal subgroup $M$ and an explicit quotient
\[
 G/M\cong\operatorname{Alt}(m)\times Q.
\]
Assume the $M$-orbits are naturally indexed by
$[m]\times\mathcal C$, with the displayed product action, and that
all $M$-orbits have size $u$.  Let $Q'\le Q$ be transitive on
$\mathcal C'\subseteq\mathcal C$, and let
$\psi:Q'\twoheadrightarrow\operatorname{Alt}(q)$ have a kernel
whose orbits on $\mathcal C'$ are exactly $q$ equal parts of size $r$,
with natural quotient action.  Suppose $9\le q\le m$.

For disjoint nonempty $W,R\subseteq[q]$ with $|W|\ge3$, the full permutation
isomorphism coset of an explicitly colored rectangle $W\times R$
under $\operatorname{Sym}(W)\times\operatorname{Sym}(R)$ can be
computed by at most two string-isomorphism calls in an inherited
subgroup-orbit action of $G$, each on
\[
 u r|W||R|
\]
physical points.  All auxiliary output generators and representatives
induce back to the ideal rectangle.  If the current physical orbit is
$[m]\times\mathcal C'$ with its $M$-fibers, this number is at most
\[
 \frac{q}{4m}|\Omega'|\le\frac14|\Omega'|.
\]
More generally, if $W,R\subseteq C\subseteq[q]$,
$|W|\ge\eta|C|$, and $|R|\le |C|-|W|$, the sharper factor is
$\eta(1-\eta)|C|^2/(mq)\le\eta(1-\eta)q/m$.
\end{lemma}

\begin{proof}
Choose $q$ row labels and identify them with $[q]$.  In the first
factor restrict to the alternating subgroup supported on these labels.
In the second factor restrict to $Q'$.  Inside
$\operatorname{Alt}(q)\times Q'$ take the subgroup
\[
 D=\{(\psi(z),z):z\in Q'\}.
\]
Take its full preimage $\widetilde D$ in $G$.  This is constructive
without an arbitrary subgroup-intersection oracle: generate the common
kernel, lift generators of $\ker\psi$ in the second factor, and lift
the paired generators $(\psi(z),z)$ of the displayed subgroup.
The quotient maps and constructive membership give each lift.

On the $q$ chosen rows and the $q$ column parts, this group has the
diagonal natural alternating action.  Its kernel contains $M$ and the
preimage of $\ker\psi$.  That kernel is transitive on every
row--column-part fiber of size $ur$: the latter subgroup is transitive
on its $r$ column labels, and $M$ is transitive on the $u$ points of
each cell.  Restrict further to the inverse image of the setwise
stabilizer of $W$ and $R$ in $\operatorname{Alt}(q)$, then select the
physical orbit above $W\times R$.  Its ideal image is the
even-total-parity subgroup of
$\operatorname{Sym}(W)\times\operatorname{Sym}(R)$, or the full
product if the omitted ideal labels can absorb parity.  This image
is transitive on $W\times R$: after prescribing the images of one
left and one right point, an odd transposition of two remaining left
points corrects parity whenever needed, since $|W|\ge3$.
The preimage has exactly $ur|W||R|$
points and is a genuine subgroup orbit of the original physical
permutation representation.

Pull the rectangle color string back constantly on the known fibers.
String isomorphisms in this physical action induce precisely the
rectangle isomorphisms in its ideal image.  The kernel fixing the ideal
rectangle fixes the pulled-back string, so inducing the full returned
coset gives the entire ideal coset, not a proper subgroup of it.
If the ideal image has index two, solve its second coset by precomposing
the target rectangle colors with a fixed odd-total-parity rectangle
permutation.  This color substitution requires no lift of that odd
permutation to $G$ and does not change the acting physical group.
Translate the induced ideal output back and combine the two cosets.

There are $mqr$ cells in the current physical orbit.  Since
$|W||R|\le q^2/4$, the asserted size ratio follows.  When the left
set occupies at least an $\eta$ fraction of an active ideal subset $C$, the
maximum product is $\eta(1-\eta)|C|^2\le\eta(1-\eta)q^2$.
\end{proof}

\paragraph{The first certified rectangle and its cap.}
Lemma~\ref{lem:first-soj-rectangle} reaches a certified rectangle after
only $q^{O(1)}$ choices and either returns a balanced partition already
or supplies disjoint ideal sides $W,R$.  The rectangle's exact full
coset imposes the constant-factor hereditary cap on the parent ideal
interval proved in that lemma.  In the support-two case,
Lemma~\ref{lem:inherited-support-two-rectangle} realizes this oracle
on at most $(2\eta(1-\eta)+o(1))n$ inherited physical points.  In the carrier
case Lemma~\ref{lem:diagonal-carrier} does the same on at most
$\eta(1-\eta)qn/m\le\eta(1-\eta)n$ points.  No later Split-or-Johnson contraction
phase is needed.  The parent continuation uses the returned ideal
overcoset and then checks its original string constraints exactly.

\subsubsection{Inheritance of the retained row action}
\label{subsec:carrier-inheritance}
In a clean row normal form, put
$J=M\operatorname{Alt}(m)$.  Then $G/J\cong Q$, and the
$J$-orbits are the columns $[m]\times C$, where $C$ ranges over the
local variation orbits.  While processing the quotient $Q$, use full
preimages of all quotient restrictions.  This keeps the entire kernel
$J$, hence keeps the natural row carrier.  Cameron restrictions and
Design cap restrictions have exactly this form.  Local-certificate
endpoint groups are independent recursive calls; they may lose the
carrier, but inherit the original physical action and its certificate
progress.  They are not enlarged back to the group of the parent.

The fixed $J$-floor is created only when there are at least two
columns.  Entry into one column occurs by a block or orbit descent
from a union of at least two equal columns, so that entry loses at
least half of the preceding physical domain.  At entry switch to the
general inherited dispatcher in place; do not introduce another
same-domain recursive call or a fresh budget.  After one column has
been solved, the conditional subgroup used for the next column may
have lost the carrier.  No retention assertion is needed: each such
query is an inherited subgroup-orbit call on a proper smaller domain.

More generally, an intransitive quotient restriction may have unequal
orbits, each a union of columns. Always process a largest current orbit
first. This first child receives the unchanged group restricted to
that orbit, so it retains the carrier. Every later conditional query
lies in a different original orbit, of size at most half the parent
degree; it may discard the carrier with that physical charge. This
ordering prevents a small first orbit from destroying the carrier
before a nearly full-size later query.

For the faithful-image assertion, write the covering group of the
largest first orbit as $H/M=\Alt(m)\times Q'$ and let
$T=H_{(\Omega')}$ be the kernel of its physical action.
Every pointwise-fixing element fixes every row label, since the orbit
contains all rows above its column orbit. Consequently
$TM/M=1\times Z$ for a normal subgroup $Z\triangleleft Q'$, and
\[
 (H/T)/(TM/T)\cong\Alt(m)\times(Q'/Z).
\]
The images of the $M$-orbits are still transitive cells and the row
complement remains faithful. Each retained column contains all $m$
row cells. Its stabilizer acts on them inside the old $\Alt(m)$,
and the clean carrier supplies every element of $\Alt(m)$.
Consequently the whole inherited constituent on those cells is
still the natural degree-$m$ action, with cap $m$ below the floor.
An element fixing the physical orbit
pointwise also fixes every column label that occurs there. Thus $Z$
lies in the pointwise kernel of the restricted column action, and
all subsequent column-action maps kill $Z$. The same covering maps
can therefore be retained. This proves the
structural assertion used by the largest-orbit-first rule.

\clearpage
\section{Major gaps in Du's algorithm}
\label{app:du-gaps}

Du~\cite{Du17} states an $n^{O(\log n)}$ algorithm for general graph
isomorphism. This appendix explains why the argument in that preprint
does not establish the stated result. All page numbers refer to
arXiv:1710.09526v1, dated October~26, 2017. We give an exact
counterexample to a lemma used by the algorithm and identify two
independent missing algorithmic arguments.

\subsection{The incidence routine fails on an infinite graph family}

The algorithm partitions a graph eigenspace $X$ using the hyperplanes
orthogonal to $P_Xe_i$, where $P_X$ is orthogonal projection onto $X$.
For a chamber containing $x$, Lemma~6 of~\cite{Du17}, p.~10, tests
whether the $i$th hyperplane is incident to that chamber by checking
\begin{equation}
 \sgn(P_Xs_i)=s_i,
 \qquad s_i=\sgn(x)-\sgn(x_i)e_i.
 \label{eq:du-incidence-test}
\end{equation}
The sign function is applied coordinatewise, with $\sgn(0)=0$.
The following counterexample uses the eigenspace and seed explicitly
required by the algorithm, rather than an arbitrary subspace.

\begin{proposition}
For every $m\geq4$, the incidence test~\eqref{eq:du-incidence-test}
returns the empty set on the prescribed seed in the zero eigenspace
of the star $K_{1,m}$, although the chamber has $m-1$ incident
hyperplanes.
\end{proposition}

\begin{proof}
Number the center $0$ and the leaves $1,\ldots,m$. The zero eigenspace is
\begin{equation*}
 X=\{z\in\R^{m+1}:z_0=0,\ \textstyle\sum_{j=1}^m z_j=0\},
 \qquad \dim X=m-1>2.
\end{equation*}
Projection sets the center coordinate to zero and subtracts the mean
of the leaf coordinates. The algorithm's seed $x=P_Xe_1$ has leaf
coordinates $((m-1)/m,-1/m,\ldots,-1/m)$. For every nonzero
$p_i=P_Xe_i$, $\langle x,p_i\rangle=x_i\ne0$, as required by
Lemma~6. Its chamber is $z_1>0$ and $z_i<0$ for $i\geq2$, within $X$.

Each hyperplane $z_i=0$ with $i\geq2$ is a facet: a relative interior
point has $z_1=m-2$, $z_i=0$, and all other leaf coordinates $-1$.
A small displacement along $-p_i$ enters the chamber and projects
orthogonally back to that point. The same holds on a relatively open
set of boundary points, which spans the hyperplane. Thus the example
also satisfies Du's definition of incidence by straightforward projections.
The hyperplane $z_1=0$ meets the chamber's closure only at zero.
The correct incidence set is therefore $\{2,\ldots,m\}$.

For $i\geq2$, the leaf coordinates of $s_i$ consist of one $+1$,
one zero, and $m-2$ copies of $-1$. Their sum is $3-m$, so
\begin{equation*}
 (P_Xs_i)_i=(m-3)/m>0=(s_i)_i.
\end{equation*}
The test rejects every actual facet. It also rejects $i=1$, since
$(P_Xs_1)_1=(m-1)/m>0=(s_1)_1$.
\end{proof}

This failure affects the algorithm, not only the statement of a lemma.
The loop on p.~17 of~\cite{Du17} removes the computed incidence set
and repeats on the remaining indices. Here it removes nothing. The
center has zero projection and is already excluded; the remaining leaf
projections still span $X$. The same state and seed recur, so the
specified loop makes no progress. Replacing this incidence routine
would require a different argument and would not resolve the
independent running-time gap below.

\subsection{The logarithmic exposure bound is not proved}

In Case~2.2B, p.~31 of~\cite{Du17}, a cell is divided into $q$
fibers of size $p$. The argument exposes their associated subspaces
one at a time. From pairwise nonorthogonality it concludes that at most
$\min\{q,\lceil\log p\rceil\}$ exposures make the projection
partition discrete. This is the progress assertion used to bound the
branching by $q^{O(\log p)}$ on p.~35. No decrease by a fixed factor,
rank bound, or other charging argument is given.

Pairwise nonorthogonality alone cannot supply such a bound. The
following projection model shows the missing implication explicitly.
Take $q$ disjoint fibers $F_i$, each split into two halves of size
$r\geq2$, so $p=2r$. Put $b_i=\one_{F_i}$,
$a_i=\one_{F_i^+}-\one_{F_i^-}$, $a=\sum_i a_i$, and
$C=\bigcup_iF_i$. Let $W_i$ consist of the vectors supported on
$F_i$ with sum zero separately on its two halves, and set
\begin{equation*}
 X=\operatorname{span}\{b_i-b_j:i,j\in[q]\},\qquad
 Z=\operatorname{span}\{a\}\ \oplus\!\bigoplus_{i=1}^q W_i.
\end{equation*}
The $X$-projection fibers are exactly the $F_i$. Their residual spans
$Z_i=\operatorname{span}\{P_Ze_w:w\in F_i\}$ equal
$\operatorname{span}\{a\}\oplus W_i$, so every pair is nonorthogonal.
After exposing a nonempty set $K$ of fibers, the exposed space is
$E_K=X\oplus\operatorname{span}\{a\}\oplus\bigoplus_{i\in K}W_i$.
For an unexposed $j$ and $w\in F_j^\varepsilon$, $\varepsilon\in\{+1,-1\}$,
\begin{equation*}
 P_{E_K}e_w=\frac{b_j}{p}-\frac{\one_C}{pq}
                    +\frac{\varepsilon a}{pq}.
\end{equation*}
Thus projection equality leaves every unexposed half unresolved,
and all $q$ exposures are needed in this model; $q$ can exceed
$\lceil\log p\rceil$ arbitrarily. This refutes only the stated
inference from pairwise nonorthogonality. It is not a simple-graph
instance satisfying every earlier splitting step or the equitable
refinement invoked on p.~31. Those additional conditions would need
to enter a quantitative progress proof; the preprint supplies no such
proof for the logarithmic bound used in its final recurrence.

\paragraph{Compatibility of the recursive answers.}
The assembly steps on pp.~23, 30, and~32 also require a bounded-time
computation of compatible actions of recursively obtained groups.
Orbit partitions, even together with all point-stabilizer orbit
partitions, do not determine their intersection. For example,
on $\F_p$ with prime $p\geq5$, the regular translation group $H$ and
$H^\tau$, $\tau=(0\ 1)$, have identical such partitions, whereas
$H\cap H=H$ and $H\cap H^\tau=1$. Both groups have prime order,
and they are distinct because $\tau$ is not affine and does not
normalize $H$. Du also promises correspondences between cells of
stabilizer partitions; genuine transporter maps would distinguish
the groups in this example. The missing step is a specified algorithm
that retains those maps and computes globally compatible lifts, with
its cost charged to the recurrence. Together with the false incidence
test and the missing exposure bound, this prevents the preprint from
establishing its claimed algorithm and running time.

\end{document}